\documentclass[12pt]{sourcebook}
\usepackage{graphicx}
\usepackage{epsfig}
\usepackage{fancyhdr}
\usepackage{multind}

\newcommand{\beq}{\begin{equation}}
\newcommand{\eeq}[1]{\label{#1}\end{equation}}
\newcommand{\eeqn}{\end{equation}}
\newenvironment{Eqnarray}{\arraycolsep 0.14em\begin{eqnarray}}{\end{eqnarray}}
\newcommand{\beqa}{\begin{Eqnarray}}
\newcommand{\eeqa}[1]{\label{#1}\end{Eqnarray}}
\newcommand{\eeqan}{\end{Eqnarray}}
\newcommand{\CR}{\nonumber \\ }
\newcommand{\leqn}[1]{(\ref{#1})}
\renewcommand{\bar}[1]{\overline{#1}}
\newcommand{\etal}{{\it et al.}}
\newcommand{\ie}{{\it i.e.}}
\newcommand{\eg}{{\it e.g.}}

\newcommand{\VEV}[1]{\left\langle{ #1} \right\rangle}
\newcommand{\lsim}{\mathrel{\raise.3ex\hbox{$<$\kern-.75em\lower1ex\hbox{$\sim$}}}}
\newcommand{\gsim}{\mathrel{\raise.3ex\hbox{$>$\kern-.75em\lower1ex\hbox{$\sim$}}}}

\newcommand{\ee}{e^+e^-}
\newcommand{\sstw}{\sin^2\theta_w}

\newcommand{\mz}{m_Z}
\newcommand{\gz}{\Gamma_Z}
\newcommand{\mw}{m_W}
\newcommand{\mt}{m_t}
\newcommand{\mh}{m_h}
\newcommand{\GF}{G_F}
\newcommand{\alphas}{\alpha_s}
\newcommand{\msb}{{\bar{\scriptscriptstyle M \kern -1pt S}}}
\newcommand{\eps}{\epsilon}

\makeatletter
\def\section{\@startsection{section}{0}{\z@}{5.5ex plus .5ex minus
 1.5ex}{2.3ex plus .2ex}{\large\bf}}
\def\subsection{\@startsection{subsection}{1}{\z@}{3.5ex plus .5ex minus
 1.5ex}{1.3ex plus .2ex}{\normalsize\bf}}
\def\subsubsection{\@startsection{subsubsection}{2}{\z@}{-3.5ex plus
-1ex minus  -.2ex}{2.3ex plus .2ex}{\normalsize\sl}}

\renewcommand{\@makecaption}[2]{%
   \vskip 10pt
   \setbox\@tempboxa\hbox{\small #1: #2}
   \ifdim \wd\@tempboxa >\hsize     
       \small #1: #2\par          
     \else                        
       \hbox to\hsize{\hfil\box\@tempboxa\hfil}
   \fi}

 \def\citenum#1{{\def\@cite##1##2{##1}\cite{#1}}}
\def\citea#1{\@cite{#1}{}}
 
\newcount\@tempcntc
\def\@citex[#1]#2{\if@filesw\immediate\write\@auxout{\string\citation{#2}}\fi
  \@tempcnta\z@\@tempcntb\m@ne\def\@citea{}\@cite{\@for\@citeb:=#2\do
    {\@ifundefined
       {b@\@citeb}{\@citeo\@tempcntb\m@ne\@citea\def\@citea{,}{\bf ?}\@warning
       {Citation `\@citeb' on page \thepage \space undefined}}%
    {\setbox\z@\hbox{\global\@tempcntc0\csname b@\@citeb\endcsname\relax}%
     \ifnum\@tempcntc=\z@ \@citeo\@tempcntb\m@ne
       \@citea\def\@citea{,}\hbox{\csname b@\@citeb\endcsname}%
     \else
      \advance\@tempcntb\@ne
      \ifnum\@tempcntb=\@tempcntc
      \else\advance\@tempcntb\m@ne\@citeo
      \@tempcnta\@tempcntc\@tempcntb\@tempcntc\fi\fi}}\@citeo}{#1}}
\def\@citeo{\ifnum\@tempcnta>\@tempcntb\else\@citea\def\@citea{,}%
  \ifnum\@tempcnta=\@tempcntb\the\@tempcnta\else
  {\advance\@tempcnta\@ne\ifnum\@tempcnta=\@tempcntb \else\def\@citea{--}\fi
    \advance\@tempcnta\m@ne\the\@tempcnta\@citea\the\@tempcntb}\fi\fi}
\makeatother

\newcommand{\evqq}     {\mbox{$ e \nu q \bar{q}$}}
\newcommand{\muvqq}     {\mbox{$ \mu \nu q \bar{q}$}}
\newcommand{\tauvqq}     {\mbox{$ \tau \nu q \bar{q}$}}
\newcommand{\qqqq}     {\mbox{$ q \bar{q} q \bar{q}$}}
\newcommand{\lvlv}     {\mbox{$ \ell \nu \ell \nu$}}
\newcommand{\wwa}{W^+W^- }
\newcommand{\dkg}{\Delta\kappa_\gamma}
\newcommand{\ra}{\rightarrow}
\newcommand{\wwl}{W_{\rm L}W_{\rm L}}
\newcommand{\eeff}{e^+e^- \rightarrow f\overline{f}}

\newcommand{\TeV}{{\rm\,TeV}}

\newcommand{\MeV}{{\rm\,MeV}}
\newcommand{\GeV}{{\rm\,GeV}}

\newcommand{\ttb}{\hbox{$t {\bar t}$}}
\newcommand{\dlde}{\hbox{d${\cal L}$/dE}}
\newcommand{\wz}{\sqrt{2}}
\newcommand{\ppl}  {{\cal P}_{\rm{e}^+}}
\newcommand{\pmi}  {{\cal P}_{\rm{e}^-}}
\newcommand{\ppm}  {{\cal P}_{\rm{e}^\pm}}

\newcommand{\sweff}{\sstw^{\mathrm{eff}}}
\newcommand{\cAb}{{\cal A}_b}
\newcommand{\cAe}{{\cal A}_e}
\newcommand{\cAq}{{\cal A}_q}
\newcommand{\Rb}{R_b}
\newcommand{\so}{\sigma_0}
\newcommand{\Ghad}{\Gamma_{\mathrm{had}}}
\newcommand{\ALR}{A_{LR}}
\newcommand{\ww}{$W^+W^-$ }
\newcommand{\invf}{fb$^{-1}$}
\newcommand{\re}{\mbox {Re}\,}
\newcommand{\tb}{\tan \beta}
\newcommand{\Stope}{\tilde{t}_1}
\newcommand{\Stopz}{\tilde{t}_2}
\newcommand{\mste}{m_{\tilde{t}_1}}
\newcommand{\mstz}{m_{\tilde{t}_2}}
\newcommand{\mgl}{m_{\tilde{g}}}
\newcommand{\tst}{\theta_{\tilde{t}}}
\newcommand{\costt}{\cos\tst}
\newcommand{\MA}{M_A}
\newcommand{\Vcb}{V_{cb}}
\newcommand{\Vub}{V_{ub}}
\newcommand{\KL}{\left(}
\newcommand{\KR}{\right)}
\newcommand{\BC}{\begin{center}}
\newcommand{\EC}{\end{center}}
\newcommand{\twol}{two--loop}
\newcommand{\onel}{one--loop}
\newcommand{\fh}{{\em FeynHiggs}}
\newcommand{\epem}{$e^+e^-$}
\newcommand{\refta}[1]{\mbox{Table~\ref{#1}}}
\newcommand{\refse}[1]{\mbox{Section~\ref{#1}}}
\newcommand{\citere}[1]{\mbox{Ref.~\cite{#1}}}
\newcommand{\order}[1]{${\cal O}(#1)$}
\newcommand{\ed}[1]{\frac{1}{#1}}

\newcommand{\emem}{$e^-e^- \,$}

\newcommand{\slashchar}[1]{\setbox0=\hbox{$#1$}           
  \dimen0=\wd0                 
 \setbox1=\hbox{/} \dimen1=\wd1               
 \ifdim\dimen0>\dimen1                        
 \rlap{\hbox to \dimen0{\hfil/\hfil}}      
 #1                                        
 \else                                        
 \rlap{\hbox to \dimen1{\hfil$#1$\hfil}}   
 /                                         
 \fi}

\newcommand{\bit}{\begin{itemize}}         
\newcommand{\eit}{\end{itemize}}

\newcommand{\alpmz}{$\alpha_s(m_Z^2)$}

\def\D0{D\O\ \hskip-0.5mm}

\begin{document}

\pagestyle{fancy}
\thispagestyle{empty}

\setcounter{footnote}{0}
\renewcommand{\thefootnote}{\fnsymbol{footnote}}

\begin{flushright}
{\small
  BNL--52627, 
           CLNS 01/1729, 
           FERMILAB--Pub--01/058-E,  \\
           LBNL--47813,  
           SLAC--R--570,  
           UCRL--ID--143810--DR\\   
           LC--REV--2001--074--US \\
           hep-ex/0106057\\
             June 2001}  
\end{flushright}

\bigskip
\begin{center}
{\bf\LARGE
 Linear Collider Physics Resource Book\\[1ex] for Snowmass 2001 \\[4ex]
Part 3: Studies of Exotic and\\[1.5ex] Standard Model Physics}
\\[6ex]
{\it American Linear Collider Working Group}
\footnote{Work supported in part by the US Department of Energy under
contracts DE--AC02--76CH03000,
DE--AC02--98CH10886, DE--AC03--76SF00098, DE--AC03--76SF00515, and
W--7405--ENG--048, and by the National Science Foundation under
contract PHY-9809799.}
\medskip
\end{center}

\vfill

\begin{center}
{\bf\large
Abstract }
\end{center}

This Resource Book reviews the physics opportunities of a next-generation
$e^+e^-$ linear collider and discusses options for the experimental program.
Part 3 reviews the possible experiments on that can be done at a linear
collider on strongly coupled electroweak symmetry breaking, exotic
particles, and extra dimensions, and on the top quark, QCD, and two-photon
physics. It also discusses the improved precision electroweak measurements
that this collider will make available.

\vfill
\vfill

\newpage
\emptyheads
\blankpage \thispagestyle{empty}
\fancyheads
\emptyheads

 \frontmatter\setcounter{page}{1}

\hbox to\hsize{\null}
\thispagestyle{empty}
\vfill
\begin{figure}[hp]
\begin{center}
\epsfig{file=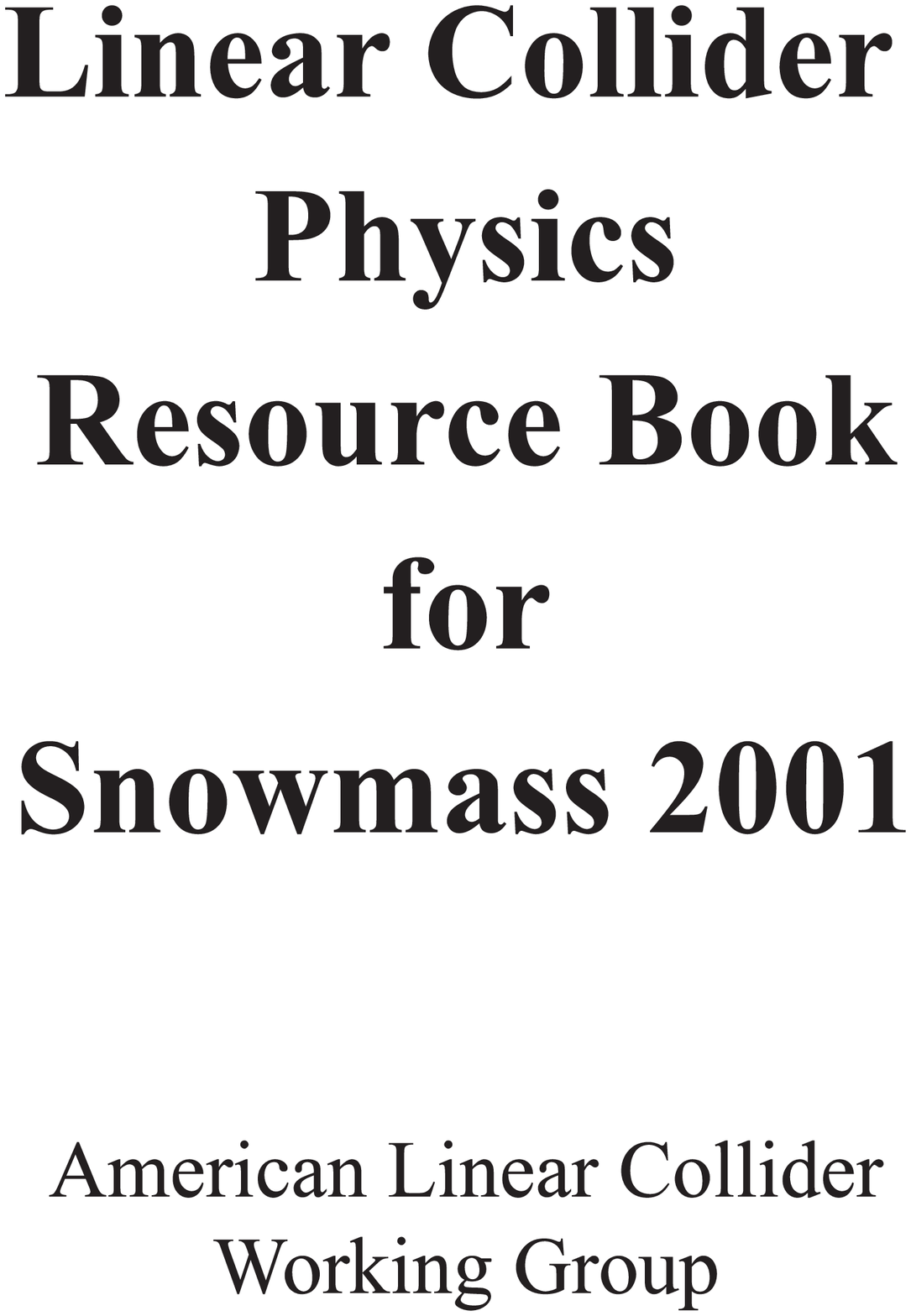,height=6.5in}
\end{center}
\end{figure}
\vfill
\newpage

\begin{center} 
           BNL--52627, 
           CLNS 01/1729, 
           FERMILAB--Pub--01/058-E,  \\
           LBNL--47813,  
           SLAC--R--570,  
           UCRL--ID--143810--DR\\[2ex]   
            LC--REV--2001--074--US \\[2ex]
            June 2001  
\end{center}

\vfill

\begin{center}
\fbox{\parbox{5in}
{This document, and the material and data contained therein, was developed
under sponsorship of the United States Government.  Neither the United
States nor the Department of Energy, nor the Leland Stanford Junior University, 
nor their employees, nor their respective contractors, subcontractors, or
their employees, makes any warranty, express or implied, or assumes any
liability of responsibility for accuracy, completeness or usefulness of any
information, apparatus, product or process disclosed, or represents that its
use will not infringe privately owned rights.  Mention of any product, its
manufacturer, or suppliers shall not, nor is intended to imply approval,
disapproval, or fitness for any particular use.  A royalty-free,
nonexclusive right to use and disseminate same for any purpose
whatsoever, is expressly reserved to the United States and the
University.
}}\end{center}

\vfill

\noindent 
Cover:  Events of $\ee \to Z^0 h^0$, simulated with the Large linear collider
detector \hfill \break described in Chapter 15.
Front cover:  $h^0 \to \tau^+\tau^-$, $Z^0 \to b\bar b$.
Back cover:  $h^0 \to b\bar b$, $Z^0 \to \mu^+\mu^-$.

\vfill

\noindent
Typset in \LaTeX\ by S. Jensen.

\vfill
\noindent
Prepared for the Department of Energy under contract number DE--AC03--76SF00515
by Stanford Linear Accelerator Center, Stanford University, Stanford, California.
Printed in the United State of America.  Available from National Technical
Information Services, US Department of Commerce, 5285 Port Royal Road,
Springfield, Virginia 22161.

  \thispagestyle{empty}

  \begin{center}
{\Large American Linear Collider Working Group}
\end{center}
 
\bigskip\bigskip
\begin{center}

T. Abe$^{52}$,
N.~Arkani-Hamed$^{29}$,
D.~Asner$^{30}$,
H.~Baer$^{22}$,
J.~Bagger$^{26}$,    
C.~Balazs$^{23}$,
C.~Baltay$^{59}$,           
T.~Barker$^{16}$,
T.~Barklow$^{52}$,   
J.~Barron$^{16}$
% ...

\end{center}

\newpage

\centerline{$^{1}$ Argonne National Laboratory, Argonne, IL 60439}
\centerline{$^{2}$ Universitat Autonoma de Barcelona, E-08193 Bellaterra,Spain}
\centerline{$^{3}$ University of British Columbia, Vancouver, BC V6T 1Z1, Canada}
\centerline{$^{4}$ Brookhaven National Laboratory, Upton, NY 11973}
\centerline{$^{5}$ Budker INP, RU-630090 Novosibirsk, Russia}
\centerline{$^{6}$ University of California, Davis, CA 95616}
\centerline{$^{7}$ University of California, Riverside, CA 92521}
\centerline{$^{8}$ University of California at San Diego, La Jolla, CA  92093}
\centerline{$^{9}$ University of California, Santa Cruz, CA 95064}
\centerline{$^{10}$ California Institute of Technology, Pasadena, CA 91125}
\centerline{$^{11}$ Carleton University, Ottawa, ON K1S 5B6, Canada}
\centerline{$^{12}$ Universidad Catolica de Chile, Chile}
\centerline{$^{13}$ CERN, CH-1211 Geneva 23, Switzerland}
\centerline{$^{14}$ University of Chicago, Chicago, IL 60637}
\centerline{$^{15}$ Chonbuk National University, Chonju 561-756, Korea}
\centerline{$^{16}$ University of Colorado, Boulder, CO 80309}
\centerline{$^{17}$ Cornell University, Ithaca, NY  14853}
\centerline{$^{18}$ DESY, D-22063 Hamburg, Germany}
\centerline{$^{19}$ DESY, D-15738 Zeuthen, Germany}
\centerline{$^{20}$ Fairfield University, Fairfield, CT 06430}
\centerline{$^{21}$ Fermi National Accelerator Laboratory, Batavia, IL 60510}
\centerline{$^{22}$ Florida State University, Tallahassee, FL 32306}
\centerline{$^{23}$ University of Hawaii, Honolulu, HI 96822}
\centerline{$^{24}$ Indian Institute of Science, Bangalore, 560 012, India}
\centerline{$^{25}$ Indiana University, Bloomington, IN 47405}
\centerline{$^{26}$ Johns Hopkins University, Baltimore, MD 21218}
\centerline{$^{27}$ Kansas State University, Manhattan, KS 66506}
\centerline{$^{28}$ Kyoto University,  Kyoto 606, Japan}
\centerline{$^{29}$ Lawrence Berkeley National Laboratory, Berkeley, CA 94720}
\centerline{$^{30}$ Lawrence Livermore National Laboratory, Livermore, CA 94551}
\centerline{$^{31}$ Universit\"at Leipzig, D-04109 Leipzig, Germany}
\centerline{$^{32}$ Ludwigs-Maximilians-Universit\"at, M\"unchen, Germany}
\centerline{$^{32a}$ Manchester University, Manchester M13~9PL, UK}
\centerline{$^{33}$ Centre de Physique Theorique, CNRS, F-13288 Marseille, France}
\centerline{$^{34}$ University of Massachusetts, Amherst, MA 01003}
\centerline{$^{35}$ Massachussetts Institute of Technology, Cambridge, MA 02139}
\centerline{$^{36}$ Max-Planck-Institut f\"ur Physik, M\"unchen, Germany}
\centerline{$^{37}$ University of Michigan, Ann Arbor MI 48109}
\centerline{$^{38}$ State University of New York, Buffalo, NY 14260}
\centerline{$^{39}$ State University of New York, Stony Brook, NY 11794}
\centerline{$^{40}$ Northern Illinois University, DeKalb, IL 60115}
\centerline{$^{41}$ University of Notre Dame, Notre Dame, IN 46556}
\centerline{$^{42}$ University of Oregon, Eugene, OR 97403}
\centerline{$^{43}$ Oxford University, Oxford OX1 3RH, UK}
\centerline{$^{44}$ Paul Scherrer Institut, CH-5232 Villigen PSI, Switzerland}
\centerline{$^{45}$ University of Pennsylvania, Philadelphia, PA 19104}
\centerline{$^{46}$ Pennsylvania State University, Mont Alto, PA 17237}
\centerline{$^{47}$ Perkins-Elmer Bioscience, Foster City, CA 94404}
\centerline{$^{48}$ Purdue University, West Lafayette, IN 47907}
\centerline{$^{49}$ RWTH Aachen, D-52056 Aachen, Germany}
\centerline{$^{50}$ University of Rochester, Rochester, NY 14627}
\centerline{$^{51}$ Southern Methodist University, Dallas, TX 75275}
\centerline{$^{52}$ Stanford Linear Accelerator Center, Stanford, CA 94309}
\centerline{$^{53}$ Texas A\&M University, College Station, TX 77843}
\centerline{$^{54}$ Tokoku University, Sendai 980, Japan}
\centerline{$^{55}$ University of Wisconsin, Madison, WI  53706}
\centerline{$^{57}$ Uniwersytet Warszawski, 00681 Warsaw, Poland}
\centerline{$^{57}$ Wayne State University, Detroit, MI 48202}
\centerline{$^{58}$ Universit\"at W\"urzburg, W\"urzburg 97074, Germany}
\centerline{$^{59}$ Yale University, New Haven, CT 06520}

\vfill

\noindent
Work supported in part by the US Department of Energy under
contracts DE--AC02--76CH03000,
DE--AC02--98CH10886, DE--AC03--76SF00098, DE--AC03--76SF00515, and
W--7405--ENG--048, and by the National Science Foundation under
contract PHY-9809799.

  \blankpage  \thispagestyle{empty}

\setcounter{chapter}{4}

\setcounter{page}{185} \thispagestyle{empty}
\renewcommand{\thepage}{\arabic{page}}

\emptyheads

\fancyheads

\chapter{New Physics at the TeV Scale and Beyond}
\fancyhead[RO]{New Physics at the TeV Scale and Beyond}

\section{Introduction}

The impressive amount of data collected in the past several decades
in particle physics experiments is well accommodated by the
Standard Model. This model provides an accurate description
of Nature up to energies of order 100 GeV.
Nonetheless, the Standard Model is an incomplete theory, since  many
key elements are left unexplained:
 (i) the origin of electroweak symmetry breaking,
(ii) the generation and stabilization of the hierarchy, {\it i.e.}, the
large disparity between the electroweak and the Planck scale,
(iii) the connection of elementary particle forces with gravity, and
(iv) the generation of fermion masses and mixings.
 These deficiencies
imply that there is physics beyond the Standard Model and point toward
the principal goal
of particle physics during the next decade: the
elucidation of the electroweak symmetry breaking mechanism
and the new physics that must  necessarily accompany it.
Electroweak symmetry is broken at the TeV scale. In
the absence of highly
unnatural fine-tuning of the parameters in the underlying theory,
the energy scales of the associated new phenomena should
also lie in the TeV range or below.

Numerous theories have been proposed to address these outstanding
issues and embed the Standard Model in a larger framework.  In
this chapter, we demonstrate the ability of a
linear collider operating at 500 GeV and above to make
fundamental progress in the illumination of new phenomena over
the broadest  possible range. The
essential role played by $e^+e^-$ machines in this endeavor has a
strong history.  First, $e^+e^-$ colliders are discovery machines and
are complementary to hadron colliders operating at similar energy
regions.  The discoveries of the gluon, charm, and tau sustain
this assertion.  Here, we show that 500-1000 GeV is a discovery
energy region and that $\ee$ experiments there add
 to the search capability of the LHC in many scenarios.
Second, $e^+e^-$ collisions offer excellent tools for the
intensive study of new phenomena, to precisely determine the
properties of new particles and interactions,
and to unravel the underlying theory.  This claim is chronicled by the
successful program at the $Z$ pole carried out at LEP and the SLC.
The diagnostic tests of new physics scenarios provided by a 500--1000 GeV
linear collider are
detailed in this chapter.  For the new physics discovered
at the LHC or at the LC, the linear collider will provide further information on
what it is and
how it relates to higher energy scales.

Chapter 9 of this book gives a survey of the various possible mechanisms
for electroweak symmetry breaking that motivate the search for new
physics beyond the Standard Model at energies below 1~TeV.
Among these models,
supersymmetry has been the most intensively studied in the past few years.
We have devoted Chapter 4 of this document to a discussion of how
supersymmetry  can be studied at a linear collider.  But supersymmetry is
only one of many proposals that have been made for the nature of the
new physics that will appear at the TeV scale.  In this chapter, we will
discuss how several other classes of models can be tested at the linear
collider. We will  also discuss the general experimental probes of new
physics that the linear collider makes available.

The first few sections of this chapter present the tools that linear
collider experiments bring to models in which electroweak symmetry breaking
is the result of new strong interactions at the TeV energy scale.
We begin this study in Section 2 with a discussion of precision measurements
of the $W$ and $Z$ boson couplings.  New physics at the TeV scale
typically modifies the couplings of the weak gauge bosons, generating, in
particular, anomalous contributions to the triple gauge couplings (TGCs).
These effects appear both in models with strong interactions in the Higgs
sector, where they are essentially nonperturbative,  and in models with
new particles, including supersymmetry, where they arise as perturbative
loop corrections.  We document the special
power of the linear collider to observe these effects.

In Section 3, we discuss the role of linear collider
 experiments in studying models
in which electroweak symmetry breaking arises from new strong interactions.
These include both models with no Higgs boson and models in which the
Higgs boson is a composite of more fundamental fermions.  The general methods
from Section 2 play an important role in this study, but there are also new
features specific to each class of model.

In Section 4, we discuss the related notion that quarks and leptons are
composite states built of more fundamental constituents.  The best tests
for composite structure of quarks and leptons involve the sort of
precision measurements that are a special strength of the linear collider.

In Section 5, we discuss the ability of linear collider experiments to
discover new gauge bosons.  New $Z$ and $W$ bosons arise in many extensions
of the Standard Model.  They may result, for example,
 from extended gauge groups of
grand unification or from new interactions associated with a strongly coupled
Higgs sector.  The linear collider offers many different experimental
probes for these particles, involving their couplings to all Standard
Model species that are pair-produced in $\ee$ annihilation.
  This experimental program neatly complements
the capability of the LHC to discover new gauge bosons as resonances in
dilepton production.  We describe how the LHC and linear collider results
can be put together to obtain a complete phenomenological profile of a
$Z^\prime$.  Grand unified models that lead to $Z^\prime$ bosons often
also lead to exotic fermions, so we also discuss the experiments that
probe for these particles at a linear collider.

It is possible that the new physics at the TeV scale includes the
appearance of new dimensions of space.  In fact, models with extra
spatial dimensions have recently been introduced to address the
outstanding problems of the Standard Model, including the origin of
electroweak symmetry breaking.  In Section 6, we review these models
and explain how they can be tested at a linear collider.

Further new and distinctive
ideas about physics beyond the Standard Model are likely to appear in the
future.  We attempt to explore this unchartered territory  in
Section 7 by discussing
collider tests of some
unconventional possibilities arising from string theory.
More generally, our limited imagination cannot span
the whole range of alternatives for new physics allowed
by the current data.  We must prepare to discover the unexpected!

Finally,  we devote Section 8 to a discussion of the determination
of the
origin of new physics effects.  Many investigations of new phenomena at
colliders focus only on defining the search reach.  But
once a discovery is made, the next step is  to elucidate the
characteristics of the new phenomena.  At the linear collider,
general methods such as the precision study of $W$ pair production
and fermion-antifermion production can give signals in  many different
scenarios for new physics.  However, the specific signals expected in
each class of models are characteristic and can be used to distinguish
the possibilities.  We give an example of this and review the tools that
the linear collider provides to distinguish between possible new physics
sources.

We shall see in this chapter that the reach of the linear collider to
discover new physics and  the ability of the linear collider
 to perform detailed
diagnostic tests combine to provide a facility with very strong
capabilities to study the unknown new phenomena that we will meet
at the next step in energy.

\section{Gauge boson self-couplings}

The measurement of gauge boson self-couplings at a linear collider 
can provide insight
into new physics processes in the presence or absence of new 
particle production.
In the absence of particle resonances, and in particular in the absence of 
a Higgs boson resonance, the measurement of gauge boson self-couplings will
provide a window to the new physics responsible for electroweak
symmetry breaking.   If there are many new particles being 
produced---if, for example, supersymmetric particles abound---then the 
measurement of gauge 
boson self-couplings will prove valuable since the gauge boson 
self-couplings will reflect 
the properties of the new particles through radiative corrections.

\subsection{Triple gauge boson coupling overview}

Gauge boson self-couplings include the triple gauge couplings (TGCs) 
and quartic gauge couplings (QGCs) 
of the photon, $W$ and $Z$.  Of special importance at a linear collider are 
the $WW\gamma$ and $WWZ$ TGCs
since a large sample of fully reconstructed $\ee\ra\wwa$ events will be 
available to 
measure these couplings.           

   The effective Lagrangian for the general $\wwa V$ vertex ($V=\gamma,Z$) 
contains 7 complex TGCs, denoted by $g^V_1$, $\kappa_V$, $\lambda_V$,
         $g^V_4$, $g^V_5$, $\tilde{\kappa}_V$, 
and $\tilde{\lambda}_V$~\cite{Hpzh:1987}.
The magnetic dipole and electric quadrupole moments of the 
$W$ are linear combinations of $\kappa_\gamma$ and $\lambda_\gamma$ while the
magnetic quadrupole and electric dipole moments are linear combinations of 
$\tilde{\kappa}_\gamma$ and $\tilde{\lambda}_\gamma$.  The TGCs $g^V_1$, 
$\kappa_V$, and $\lambda_V$
are C- and P-conserving, $g^V_5$ is  C- and P-violating but conserves 
CP, and
$g^V_4$, $\tilde{\kappa}_V$, and $\tilde{\lambda}_V$ are CP-violating.
In the SM at tree-level all the TGCs are zero except
$g^V_1$=$\kappa_V$=1.  

If there is no Higgs boson resonance below about 800~GeV,  the 
interactions of the $W$ and $Z$ gauge bosons become strong above 1~TeV in the
$WW$, $WZ$ or $ZZ$ center-of-mass system.  In analogy 
with $\pi\pi$ scattering below the $\rho$ resonance,
the interactions of the $W$ and $Z$ bosons below the strong symmetry breaking resonances
can
be described by an effective chiral Lagrangian~\cite{Bagger:1993}.
These interactions induce
anomalous TGC's at tree-level:
\begin{eqnarray}
   \kappa_\gamma &=& 1+\frac{e^2}{32\pi^2s_w^2}\bigl(L_{9L}+L_{9R}\bigr) 
\nonumber
 \\  \kappa_Z  &=& 1+\frac{e^2}{32\pi^2s_w^2}
   \left(L_{9L}-\frac{s_w^2}{c_w^2}L_{9R}\right) \nonumber
 \\  g_1^Z  &=& 1+\frac{e^2}{32\pi^2s_w^2}\frac{L_{9L}}{c_w^2} \  \nonumber ,
\end{eqnarray}
where $s_w^2=\sin^2\theta_w$,
      $c_w^2=\cos^2\theta_w$, and  $L_{9L}$ and  $L_{9R}$
are chiral Lagrangian parameters.   
If we replace $L_{9L}$  and  $L_{9R}$ by the values of these parameters in
QCD, $\kappa_\gamma$  is shifted by  $\dkg \sim -3\times 10^{-3}$.

Standard Model radiative corrections~\cite{Ahn:1988fx} cause shifts in the 
TGCs of 
${\cal{O}}(10^{-4}-10^{-3})$ for CP-conserving
couplings and of ${\cal{O}}(10^{-10}-10^{-8})$ for CP-violating TGC's.  
Radiative corrections in the MSSM can cause shifts of 
${\cal{O}}(10^{-4}-10^{-2})$ 
in both the CP-conserving~\cite{Arhrib:1996dm} and CP-violating 
TGC's~\cite{Kitahara:1998bt}.  

\subsection{Triple gauge boson measurements}

The methods used at LEP2 to measure TGCs provide a useful guide
to the measurement of TGCs at a linear collider.
When measuring TGCs
the kinematics of an $\ee\ra\wwa$ event can be conveniently expressed in
 terms of 
the $\wwa$ center-of-mass energy following initial-state radiation (ISR), 
the masses of the $W^+$ and $W^-$, and five angles:  the angle 
between the  ${W^-}$ and initial
$e^-$ in the ${W^+W^-}$ rest frame, the polar and azimuthal
angles of the fermion in the rest frame of its parent $W^-$, and the polar
 and azimuthal
angles of the anti-fermion in the rest frame of its parent $W^+$.   

In practice not all of these
variables can be reconstructed unambiguously.   
For example, in events with hadronic decays it is often difficult to measure 
the flavor of the quark jet, and so there is usually a two-fold ambiguity 
for quark jet directions.
Also, it can be difficult to measure ISR and
consequently the measured $\wwa$ center-of-mass energy is
often just the nominal $\sqrt{s}$. Monte Carlo simulation is used to account
 for 
detector resolution, quark hadronization, initial- and final-state radiation,
and other  effects.

The TGC measurement error at a linear collider can be estimated to a good
 approximation
by considering \evqq\ and \muvqq\ channels only, and by ignoring
all detector and radiation effects except for the requirement that 
the $\wwa$ fiducial volume be restricted to  $|\cos{\theta_W}|<0.9$.    
Such an approach correctly predicts the TGC
sensitivity of LEP2 experiments and of detailed linear collider 
simulations~\cite{Burgard:1999}.
This rule-of-thumb approximation works because
LEP2 experiments and detailed linear collider simulations also use the 
\tauvqq\ , \lvlv\  and \qqqq\ channels, and the increased sensitivity 
from these 
extra channels makes up for the lost sensitivity due to
detector resolution, initial- and final-state radiation, and systematic errors.

\begin{table}[ht!]
\begin{center}
\begin{tabular}{|l|cc|cc|}\hline\hline
     & \multicolumn{4}{c|}{error $\times 10^{-4}$} \\
\hline
     & \multicolumn{2}{c|}{$\sqrt{s}=500$ GeV} &  
\multicolumn{2}{c|}{$\sqrt{s}=1000$ GeV} \\
 TGC & Re & Im & Re & Im \\
\hline
& & & & \\
$g^\gamma_1$ &  15.5    & 18.9     & 12.8     &  12.5     \\
$\kappa_\gamma$ &  \ 3.5    &  \ 9.8    & \ 1.2     &   \ 4.9    \\
$\lambda_\gamma$ & \ 5.4     &  \ 4.1    &  \ 2.0    &   \ 1.4    \\
$g^Z_1$          &  14.1    & 15.6     & 11.0     &  10.7     \\
$\kappa_Z$        & \ 3.8     &  \ 8.1    &  \ 1.4  &  \ 4.2    \\
$\lambda_Z$        &  \ 4.5     & \ 3.5     &  \ 1.7  &  \ 1.2  \\
\hline\hline
\end{tabular}
\caption{\label{tab:cp-conserving}
Expected errors for the real and imaginary parts of CP-conserving TGCs 
assuming $\sqrt{s}=500$~GeV, 
${\cal L}=500$~fb$^{-1}$ and  $\sqrt{s}=1000$~GeV, ${\cal L}=1000$~fb$^{-1}$.  
The results are for one-parameter fits in which all other 
TGCs are kept fixed at their SM values.}
\end{center}
\vspace{.1in}
\begin{center}
\begin{tabular}{|l|cc|cc|}\hline\hline
     & \multicolumn{4}{c|}{error $\times 10^{-4}$} \\
\hline
     & \multicolumn{2}{c|}{$\sqrt{s}=500$ GeV} &  
\multicolumn{2}{c|}{$\sqrt{s}=1000$ GeV} \\
 TGC & Re & Im & Re & Im \\
\hline
& & & & \\
  $\tilde{\kappa}_\gamma$     & 22.5     &  16.4    &  14.9    &   12.0     \\
  $\tilde{\lambda}_\gamma$    & \ 5.8     & \ 4.0     & \ 2.0     & \ 1.4      \\
  $\tilde{\kappa}_Z$          & 17.3     &  13.8    & 11.8     &  10.3     \\
  $\tilde{\lambda}_Z$         & \ 4.6      & \ 3.4     & \ 1.7     & \ 1.2      \\
  $g^\gamma_4$                & 21.3     & 18.8     & 13.9     & 12.8      \\
  $g^\gamma_5$                & 19.3     & 21.6     & 13.3     & 13.4      \\
  $g^Z_4$                     & 17.9     & 15.2     & 12.0     & 10.4      \\
  $g^Z_5$                     & 16.0     & 16.7     & 11.4     & 10.7      \\
\hline\hline
\end{tabular}
\caption{\label{tab:cp-violating}
Expected errors for the real and imaginary parts of C- and P-violating 
TGCs assuming $\sqrt{s}=500$~GeV, 
${\cal L}=500$~fb$^{-1}$ and  $\sqrt{s}=1000$~GeV, ${\cal L}=1000$~fb$^{-1}$.  
The results are for
one-parameter fits in which all other TGCs are kept fixed at their SM values.}
\end{center}
\end{table}

Table~\ref{tab:cp-conserving} contains the estimates of the TGC
precision that can be obtained at $\sqrt{s}=500$ and 1000~GeV for the
CP-conserving couplings  $g^V_1$, $\kappa_V$, and $\lambda_V$.  
These estimates are derived
from one-parameter fits in which all other TGC parameters are kept fixed 
at their tree-level SM values.  
Table~\ref{tab:cp-violating} contains the corresponding estimates 
for the C- and P-violating couplings $\tilde{\kappa}_V$, 
$\tilde{\lambda}_V$, $g^V_4$, and $g^V_5$.  An alternative method of
measuring the $WW\gamma$ couplings is provided by the channel
$e^+e^-\rightarrow\nu\bar\nu\gamma$~\cite{k1}.

\begin{figure}[htp] 
\centerline{\epsfig{file=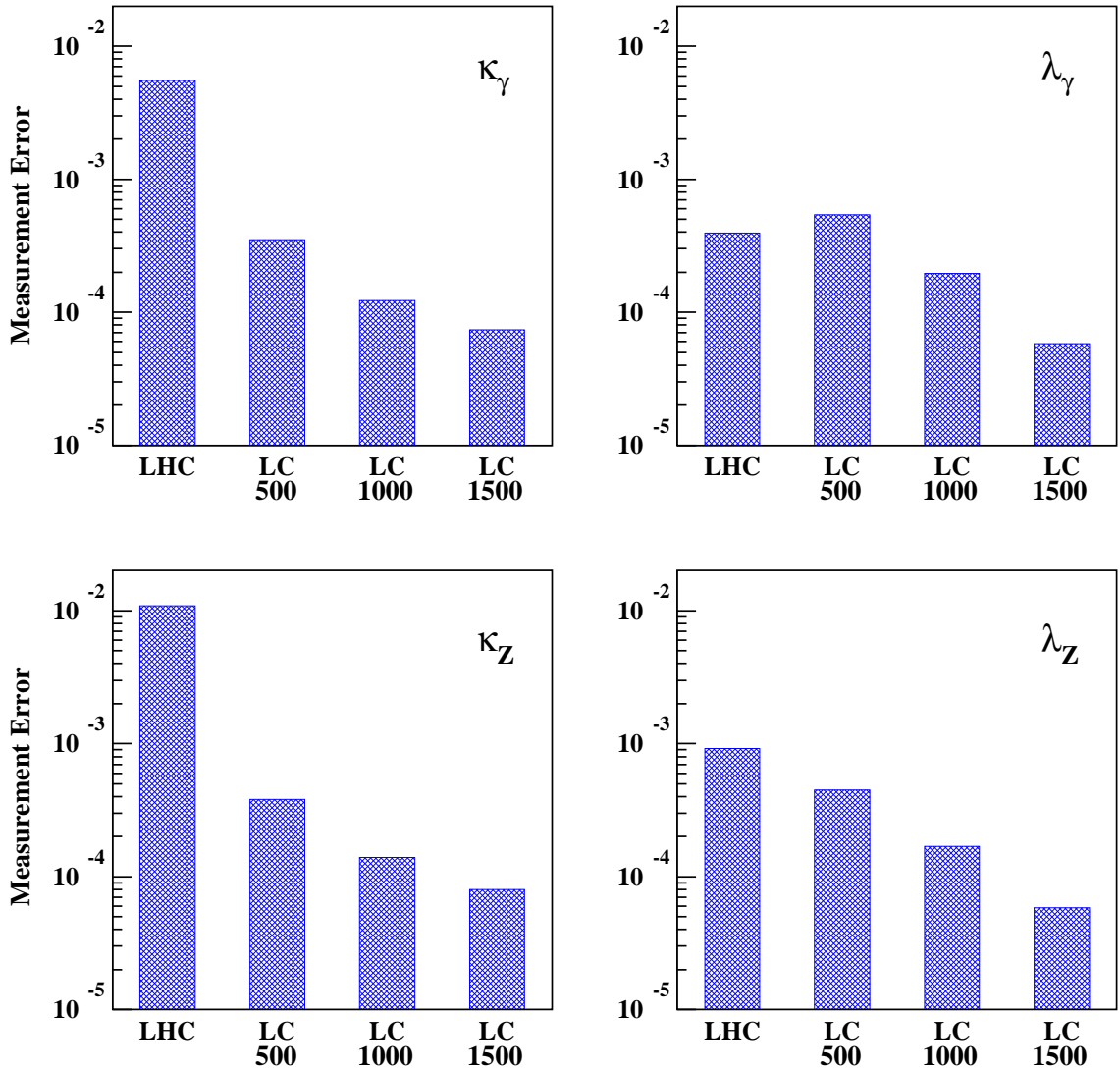,height=17cm}}
\vspace{10pt}
\caption{\label{fig:gauge_lc_lhc}
Expected measurement error for the real part of four different TGCs. 
The numbers below the ``LC'' labels refer to the
center-of-mass energy of the linear collider in GeV.
The luminosity of the LHC is assumed to be 300~fb$^{-1}$, while the 
luminosities
of the linear colliders are assumed to be  500, 1000, and 1000~fb$^{-1}$ for 
$\sqrt{s}$=500, 1000, and 1500~GeV respectively.}
\end{figure}

The difference in TGC precision between the LHC and a linear collider depends
 on the TGC, 
but typically the TGC precision at the linear collider will be substantially
 better,
even at $\sqrt{s}=500$~GeV.
Figure~\ref{fig:gauge_lc_lhc} shows the measurement
precision expected for the LHC~\cite{atlas:1999} and for linear colliders of
 three different
energies for four different TGCs.

\begin{figure}[htb] 
\centerline{\epsfig{file=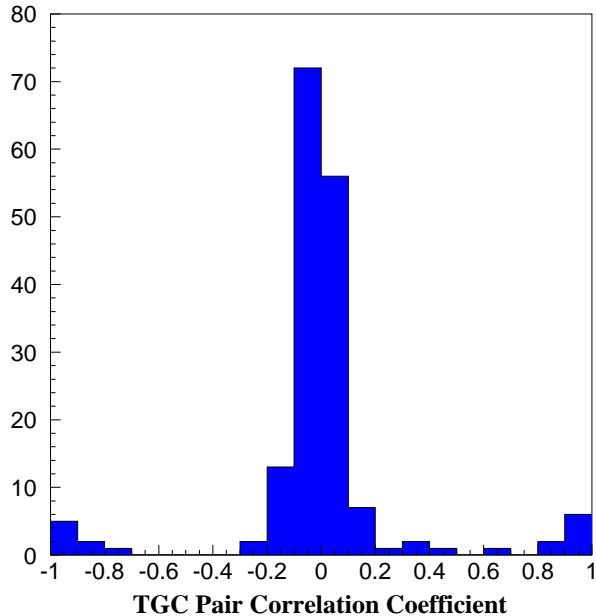,height=9cm}}
\vspace{10pt}
\caption{Histogram of correlation coefficients
for all 171 pairs of TGCs when 19 different TGCs are measured using
one-parameter fits at LEP2 (unpolarized beams).  The 19 TGCs are
made up of the real and imaginary
parts of the 8 C- and P-violating couplings along with the real parts of
the three CP-conserving couplings
$g^Z_1$, $\kappa_\gamma$,  $\lambda_\gamma$.}
\label{fig:tgc_pair_corr}
\end{figure}

If the goal of a TGC measurement program is to search for the  first sign of 
deviation from 
the SM, one-parameter fits in which all other TGCs are kept fixed at their
 tree-level SM values
are certainly appropriate.  But what if the goal is to survey a large number
 TGCs, all of 
which seem to deviate from their SM value?  Is a 28-parameter fit required? 
 The answer is probably no,
as illustrated in Fig.~\ref{fig:tgc_pair_corr}.   

Figure~\ref{fig:tgc_pair_corr} shows the histogram of the correlation
 coefficients
for all 171 pairs of TGCs when 19 different TGCs are measured at LEP2 using 
one-parameter fits.  
The entries in Fig.~\ref{fig:tgc_pair_corr} with large positive correlations 
are pairs of TGCs 
that are related  to each 
other by the interchange
of $\gamma$ and $Z$.  The correlation between 
the two TGCs of each pair can be removed  using the dependence on electron beam polarization.
The entries in Fig.~\ref{fig:tgc_pair_corr} with large negative correlations 
are TGC pairs of the type 
$Re(\tilde{\kappa}_\gamma)/Re(\tilde{\lambda}_\gamma)$,
  $Re(\tilde{\kappa}_Z)/Re(\tilde{\lambda}_Z)$, {\it etc.}  Half of the TGC pairs 
with large negative
correlations will become uncorrelated once polarized electron beams are used,
  leaving only a small
number of TGC pairs with large negative or positive correlation coefficients.

\boldmath
\subsection{Electroweak radiative corrections to $e^+ e^- \to 4$ fermions}
\unboldmath

We have seen that the experimental accuracy at a linear collider 
for the basic electroweak cross section measurements is
expected to be at the level of $0.1-0.01 \%$, requiring the
inclusion of electroweak radiative corrections to the predictions for
the underlying production processes such as $e^+e^- \to WW \to 4f$.

The full treatment of the processes $e^+e^- \to 4f$ at the one-loop
level is of enormous complexity.  Nevertheless there is ongoing work
in this direction \cite{Vicini:1998iy}.  While the real bremsstrahlung
contribution is known exactly, there are severe theoretical problems
with the virtual order-$\alpha$ corrections.  A detailed description of
the status of predictions for $e^+e^- \to 4f (\gamma)$ processes can be found
in \cite{Grunewald:2000ju}.  A suitable approach to include
order-$\alpha$ corrections to gauge-boson pair production is a
double-pole approximation (DPA), keeping only those terms in an 
expansion about the gauge-boson
resonance poles  that are enhanced by two
resonant gauge bosons. All present calculations of order-$\alpha$
corrections to $e^+e^-\to WW \to 4f$ rely on a DPA
\cite{Beenakker:1999gr,Denner:2000bj,Jadach:1998hi,Kurihara:2000ii}.  
Different versions of a DPA have been implemented in the Monte
Carlo (MC) generators RacoonWW \cite{Denner:2000bj} and YFSWW3
\cite{Jadach:1998hi}.  The intrinsic DPA error is estimated to be
$\alpha \Gamma_W/(\pi M_W) \sim 0.5\%$ whenever the cross section
is dominated by doubly resonant contributions. This is the case at
LEP2 energies sufficiently above threshold. The DPA is not a valid
approximation close to the $W$-pair production threshold. At higher
energies diagrams without two resonant $W$ bosons become sizable,
especially single $W$ production, and appropriate cuts must be applied to
extract the $WW$ signal.

The theoretical uncertainty of present predictions for the total
$W$-pair production cross section, $\sigma_{WW}$, is of the order of 
$0.5\%$ for energies between 170~GeV and 500~GeV \cite{Grunewald:2000ju},
which is within the expected DPA uncertainty.  This is a result of a
tuned numerical comparison between the state-of-the-art MC generators
RacoonWW and YFSWW3, supported by a comparison with a semi-analytical
calculation \cite{Beenakker:1999gr} and a study of the intrinsic DPA
ambiguity with RacoonWW \cite{Grunewald:2000ju,Denner:2000bj}.  In the
threshold region $\sigma_{WW}$ is known only to about $2\%$, since
predictions are based on an improved Born approximation
\cite{Grunewald:2000ju} that neglects non-universal electroweak
corrections.  Further improvements of the theoretical uncertainty on
$\sigma_{WW}$ are anticipated only when the full order-$\alpha$
calculation becomes available.  Above 500 GeV, large electroweak
logarithms of Sudakov type become increasingly important and
contributions of higher orders need to be taken into account.

A tuned comparison has also been performed of RacoonWW and YFSWW3
predictions for the $W$ invariant mass and the  $W$ production angle 
distributions, as well as for several photon observables
such as photon energy and production angle distributions,
at 200 GeV \cite{Grunewald:2000ju,Denner:2001aa} and 
500 GeV \cite{Denner:2001aa}.
Taking the observed differences between the RacoonWW and YFSWW3
predictions as a guideline, a theoretical uncertainty of the order of
$1\%$ can be assigned to the $W$ production angle distribution and the $W$
invariant mass distribution in the $W$ resonance region.  A recent
comparison of RacoonWW predictions for photon observables including
leading higher-order initial-state radiation
\cite{Denner:2001aa} with YFSWW3 predictions yields relative differences of less
than $5\%$ at 200 GeV and about $10 \%$ at 500 GeV.  These differences might be
attributed to the different treatment of visible photons in the
two MC generators: in RacoonWW the real order-$\alpha$ corrections are
based on the full $4f+\gamma$ matrix element, while in YFSWW3 
multi-photon radiation in $W$-pair production is combined with
order-$\alpha^2$ LL photon radiation in $W$ decays.

\subsection{Quartic gauge boson couplings} 

The potential for directly probing anomalous quartic gauge boson
couplings \break (AQGCs) via triple gauge-boson production at LEP2, at a future
high-energy LC, and at hadron colliders has been investigated in
\cite{Denner:2001aa,Stirling:1999xa,Belanger:2000aw,Dervan:2000as,Montagna:2001uk},
\cite{Denner:2001aa,Stirling:1999xa,Belanger:1992qh,AbuLeil:1995jg,Dawson:1996aw} and
\cite{Dervan:2000as,Belyaev:1999ih,Eboli:2001ad}, respectively.  The
AQGCs under study arise from genuine 4- and 6-dimensional operators,
{\it i.e.},  they have no connection to the parametrization of the anomalous
TGCs.  It is conceivable that there are extensions of the SM that 
leave the SM TGCs unchanged but modify quartic self-interactions of
the electroweak gauge bosons
\cite{Belanger:1992qh}.  The possible number of
operators is considerably reduced by imposing a global custodial
${\rm SU}(2)$ symmetry to protect the $\rho$ parameter from large
contributions, {\it i.e.}, to keep $\rho$ close to 1, and by the local
${\rm U}(1)_{{\rm QED}}$ symmetry whenever a photon is involved.

The sensitivity of triple-gauge-boson cross sections to dimension-4
operators, which only involve massive gauge bosons,  has been studied
for  a high-energy LC and the LHC in~\cite{Belanger:1992qh,%
Dawson:1996aw} 
and \cite{Belyaev:1999ih},
respectively.  Only weak constraints are expected from
$WWW,WWZ,WZZ$ and $ZZZ$ productions at the LHC \cite{Belyaev:1999ih},
but these processes may provide complementary information if non-zero AQGCs are
found.  The genuine dimension-4 AQGCs may be best probed in a
multi-TeV LC. The sensitivity to the two ${\rm SU}(2)_c$-conserving AQGCs in
the processes $e^+e^- \to 6 f$ at a 1 TeV LC with a luminosity
of 1000 ${\rm fb}^{-1}$ can be expected to be between $10^{-3}$ and
$10^{-2}$ \cite{Dawson:1996aw}.

The following discussion is restricted to AQGCs involving at least one
photon, which can be probed in $WW\gamma, ZZ\gamma, Z\gamma \gamma$ and
$W\gamma\gamma$ production.  The lowest-dimension operators that
lead to the photonic AQGCs $a_0$, $a_{\rm c}$, $a_{\rm n}$, $\tilde a_0$, and $\tilde a_{\rm n}$
are of dimension-6 \cite{Denner:2001aa,Belanger:1992qh,AbuLeil:1995jg,Eboli:2001ad}
and yield anomalous contributions to the SM
$WW\gamma\gamma, WWZ\gamma$ vertices, and a non-standard
$ZZ\gamma\gamma$ interaction at the tree level.  Most studies of AQGCs
consider the separately P- and C-conserving couplings $a_0, a_{\rm c}$ and
the CP-violating coupling $a_{\rm n}$.  Recently the P-violating AQGCs 
$\tilde a_0$, and $\tilde a_{\rm n}$
have also been considered \cite{Denner:2001aa}.  More general AQGCs that 
have been embedded in
manifestly ${\rm SU}(2)_{\rm L} \times {\rm U}(1)_{\rm Y}$ gauge invariant 
operators are discussed
in \cite{Belanger:2000aw,Montagna:2001uk}.  The AQGCs depend on a mass
scale $\Lambda$ characterizing the scale of new physics.  The choice
for $\Lambda$ is arbitrary as long as no underlying model is specified
which gives rise to the AQGCs.  For instance, anomalous quartic
interactions may be interpreted as contact interactions, which might
be the manifestation of the exchange of heavy particles with a mass scale $\Lambda$.

Recently, at LEP2, the  first direct bounds on the AQGCs $a_0, a_{\rm c}, a_{\rm n}$
have been imposed by investigating the total cross sections and photon
energy distributions for the processes $e^+ e^- \to WW\gamma,
Z\gamma\gamma, Z\nu \bar \nu$ \cite{LEPEWWG}.
The results, in units of GeV$^{-2}$, are
\begin{equation}
-0.037 < \frac{a_0}{\Lambda^2} < 0.036 \qquad
-0.077 < \frac{a_{\rm c}}{\Lambda^2} < 0.095 \qquad
-0.45  < \frac{a_{\rm n}}{\Lambda^2} < 0.41 \; ,
\end{equation}
for 95\%\ CL intervals.
These limits are expected to  improve considerably as the energy
increases.  It has been found that a 500 GeV LC with a total
integrated luminosity of 500 ${\rm fb}^{-1}$ can improve the LEP2
limits by as much as three orders of magnitude \cite{Belanger:2000aw}.
 
At hadron colliders the search for AQGCs is complicated by 
an arbitrary form factor that is introduced to suppress
unitarity-violating contributions at large parton center-of-mass
energies.  At the LHC, however, the dependence of a measurement of
AQGCs on the form-factor parametrization may be avoided by measuring
energy-dependent AQGCs \cite{Belyaev:1999ih}.  At Run II of the
Tevatron at 2 TeV, with 2 ${\rm fb}^{-1}$, AQGC limits
comparable to the LEP2 limits are expected
\cite{Dervan:2000as,Eboli:2001ad}.
  
Numerical studies of AQGCs are not yet as sophisticated as the ones
for TGCs.  For instance, most studies of AQGCs have not yet included
gauge boson decays, and MC
generators for the process $e^+e^- \to 4f+\gamma$ including photon
AQGCs have only recently become available \cite{Denner:2001aa,Montagna:2001uk}.  
To illustrate the typical size of the limits
that can be obtained for the AQGCs at a 500 GeV
LC with 50 ${\rm fb}^{-1}$, the following
$1\sigma$ bounds have been extracted from the total cross section 
measurement of
$e^+ e^-\to u \bar d \mu^- \bar \nu_{\mu}+\gamma$,
with all bounds in units of  $10^{-3} \; {\rm GeV}^{-2}$~\cite{Denner:2001aa}:
\begin{eqnarray}
-0.12 < \frac{a_0}{\Lambda^2} < 0.14  &\qquad&
-0.31 < \frac{a_{\rm c}}{\Lambda^2} < 0.16  \qquad
-0.82 < \frac{a_{\rm n}}{\Lambda^2} < 0.79  \qquad
\nonumber \\
-0.10 < \frac{\tilde a_0}{\Lambda^2} < 0.10  &\qquad&
-0.69 < \frac{\tilde a_{\rm n}}{\Lambda^2} < 0.90  \; .
\end{eqnarray}
The availability of MC programs
\cite{Denner:2001aa,Montagna:2001uk,Dawson:1996aw} 
will allow more
detailed studies to be performed.  For example, longitudinally polarized
gauge bosons have the greatest effect on AQGCs, and gauge bosons 
with this polarization can be isolated through an
analysis 
of gauge boson production and decay angles \cite{Belanger:1992qh}.

\section{Strongly coupled theories}

The Standard Model with a light Higgs boson provides a good fit to 
the electroweak data. Nevertheless, the electroweak observables depend
only logarithmically on the Higgs mass, so that the effects of the 
light Higgs could be mimicked by new particles with masses
as large as several TeV. A recent review of such scenarios is given in~\cite{Peskin:2001rw}. 
One can even imagine that no Higgs boson 
exists. In that case, the electroweak symmetry should be broken by some 
other interactions, and gauge boson scattering should become strong at a
scale of order 1~TeV.  An often discussed class of theories 
of this kind is called technicolor~\cite{technicolor}, which is discussed in the
next subsection.

Electroweak symmetry is often assumed to be
either connected to supersymmetry or driven by some strong dynamics, such as 
technicolor, without a Higgs boson.
There is, however, a distinctive alternative where a strong interaction
gives rise to bound states that include a Higgs boson. The latter could
be light and weakly coupled at the electroweak scale.
At sub-TeV energies these scenarios are described by
a (possibly extended) Higgs sector, while the strong dynamics 
manifests itself only above a TeV or so.

\boldmath
\subsection{Strong $WW$  scattering and technicolor}
\unboldmath

The generic idea of technicolor theories is that 
a new gauge interaction, which is asymptotically free,
becomes strong at a scale of order 1 TeV, such that the 
new fermions (``technifermions'') 
that feel this interaction form condensates that break
 the electroweak symmetry. This idea is based on 
the observed dynamics of QCD, but arguments
involving the fits to the electroweak data and the generation of 
quark masses suggest that the technicolor interactions
should be described by a  strongly coupled gauge theory that
has a different behavior from QCD (see, {\it e.g.}, \cite{Appelquist:1997fp}). 

A generic prediction of technicolor theories
is that there is a vector resonance with 
mass below about 2 TeV which unitarizes the $WW$ scattering cross section.
In what follows we will concentrate on the capability of a
linear $e^+e^-$ collider of studying $WW$ scattering, but first we
briefly mention other potential signatures associated with various 
technicolor models.  The chiral symmetry  of the technifermions
may be large enough that its dynamical breaking 
leads to pseudo-Goldstone bosons, which are pseudoscalar bound states
that can be light enough to be produced at a linear $e^+e^-$ collider
(for a recent study, see \cite{bCasalbuoni:1999fs}).
The large top-quark mass typically requires a special dynamics associated 
with the third generation. A thoroughly studied model along these lines
is called Topcolor Assisted Technicolor \cite{Hill:1995hp}, 
and leads to a rich phenomenology. This model predicts the existence 
of spinless bound states with large couplings to the top quark, 
called  top-pions and top-Higgs, which may be studied 
at a linear $e^+e^-$ collider \cite{Yue:2000xa}.

Strong $\wwa$ scattering is an essential test not only of technicolor
theories, but in fact of any model that does not include a Higgs 
boson with large couplings to gauge boson pairs.
 It can be studied at a linear collider with the
reactions $\ee\ra\nu \bar{\nu}\wwa$, $\nu \bar{\nu}ZZ$, $\nu \bar{\nu} t\bar{t}$, and $\wwa$~\cite{Barklow:1997nf}.
The final states $\nu \bar{\nu}\wwa$, $\nu \bar{\nu}ZZ$ are used to study 
the I=J=0 channel in $\wwa$ scattering, while the final state $\wwa$ is best suited
for studying the I=J=1 channel.   The $\nu \bar{\nu} t\bar{t}$ final state can be used to
investigate strong electroweak symmetry breaking in the fermion sector through the process
$\wwa\ra t\bar{t}$.

The first step in studying strong $\wwa$ scattering is
to separate the scattering of a pair of longitudinally polarized $W$'s, denoted by $\wwl$,
from transversely polarized $W$'s, and from background such as 
$\ee\ra\ee\wwa$ and $e^- \bar{\nu}W^+Z$.
 Studies have shown that simple cuts can be used to achieve this separation 
in $\ee\ra\nu \bar{\nu}\wwa$, $\nu \bar{\nu}ZZ$
at $\sqrt{s}=1000$~GeV, and that the
signals are comparable to those obtained at the LHC~\cite{Barger:1995cn}. 
Furthermore, by analyzing the gauge boson production
and decay angles it is possible to 
use these reactions to measure chiral Lagrangian parameters with an accuracy 
greater than that which can be achieved at the LHC~\cite{Chierici:2001}.

The reaction $\ee\ra \nu \bar{\nu} t\bar{t}$ provides unique access to 
$\wwa\ra t\bar t$, since this process is overwhelmed 
by the background $gg\ra t\bar{t}$ at the LHC.  Techniques similar to those employed
to isolate $\wwl\ra \wwa, ZZ$ can be used to measure
the enhancement in  $\wwl\ra t\bar{t}$ production~\cite{Barklow:1997wwtt}.
       Even in the absence
of a resonance it will be possible to establish a clear
signal.  The ratio $S/\sqrt{B}$ is expected to be 12 for a linear
collider with $\sqrt{s} = 1$ TeV and 1000~fb$^{-1}$ and 80\%/0\%
electron/positron beam polarization,
increasing to
28 for the same data sample at 1500~GeV.  

There are two approaches to studying strong $\wwa$ scattering with
the process $\ee\ra \wwa$.  
The first approach was discussed
in Section 2:  a strongly coupled gauge
boson sector induces anomalous TGCs that could be measured in 
$\ee\ra\wwa$.  The precision of $4\times 10^{-4}$ 
 for the TGCs $\kappa_\gamma$ and $\kappa_Z$ at $\sqrt{s}=500$~GeV 
can be interpreted as a precision of $0.26$ for the chiral Lagrangian parameters
$L_{9L}$ and $L_{9R}$.
Assuming naive dimensional analysis~\cite{Manohar:1984md}, 
such a measurement 
would provide a $8\sigma$ ($5\sigma$) signal for $L_{9L}$ and $L_{9R}$
if the strong symmetry breaking energy scale were 3~TeV (4~TeV).
The only drawback to this approach is that  the detection of anomalous TGCs does
not by itself provide unambiguous proof of strong electroweak symmetry breaking.   

The second approach involves an effect unique to 
strong $\wwa$ scattering.
When $\wwa$ scattering becomes strong
the amplitude for $\ee\ra\wwl$ develops a complex form factor $F_T$
in analogy with the pion form factor in $\ee\ra\pi^+\pi^-$~\cite{peskinomnes}.  
To evaluate the size of this effect the 
following expression for $F_T$ has been suggested:
\[
 F_T =
         \exp\left[{1\over \pi} \int_0^\infty
          ds'\delta(s',M_\rho,\Gamma_\rho)
          \left\{ {1\over s'-s-i\epsilon}-{1\over s'}\right\}
         \right]
\]
where
\[
\delta(s,M_\rho,\Gamma_\rho) = {1\over 96\pi} {s\over v^2}
+ {3\pi\over 8} \left[ \
\tanh (
{
s-M_\rho^2
\over
M_\rho\Gamma_\rho
}
)+1\right] \ .
\]
Here $M_\rho,\Gamma_\rho$ are the mass and width, respectively, of a vector 
resonance in $\wwl$ scattering.
The term 
\[
\delta(s) = {1\over 96\pi} {s\over v^2}
\]
is the Low Energy Theorem (LET) amplitude for $\wwl$ scattering at energies below a resonance.
Below the resonance, the real part of $F_T$ is proportional to $L_{9L} +
L_{9R}$ and can therefore be interpreted as a TGC.
The imaginary part, however,  is a distinct new effect.  

The real and imaginary parts of $F_T$ are measured~\cite{Barklow:2000let} in the 
same manner as the TGCs.
The $\wwa$ production and decay angles are analyzed, and an electron beam 
polarization of 80\% is assumed.
In contrast to TGCs, the analysis of $F_T$ seems to benefit from
even small amounts of jet flavor tagging. We therefore assume
that charm jets can be tagged with a
purity/efficiency of  100/33\%.
These purity/efficiency numbers are based on research~\cite{Damerell:1997} 
that indicates that
it may be possible to tag
charm jets with a 
purity/efficiency as high as 100/65\%,  given that
$b$-jet contamination is not a significant factor in $\wwa$ pair production and decay.

\begin{figure}[tbh] 
\centerline{\epsfig{file=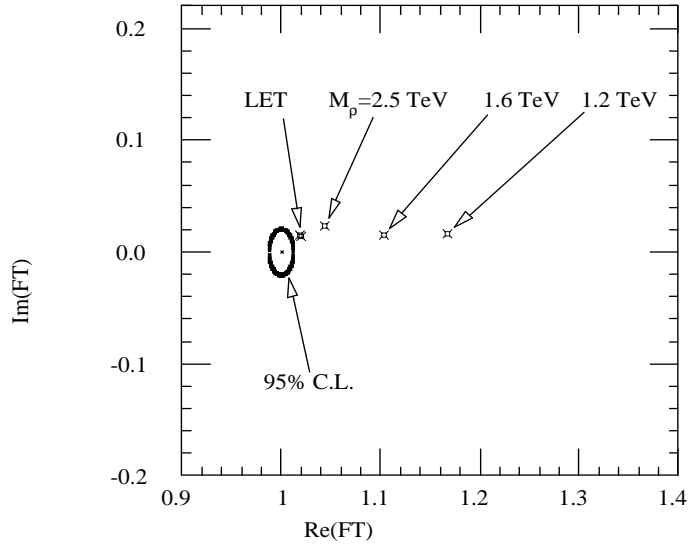,angle=-90,clip=,width=9cm}}
\vspace{10pt}
\caption{\label{fig:fteight}
95\% C.L. contour for $F_T$  for $\sqrt{s}=500$~GeV
and 500~fb$^{-1}$. Values of $F_T$  for various masses $M_\rho$ of a vector 
resonance in $\wwl$ scattering
are also shown. The $F_T$ point ``LET'' refers to the
case where no vector resonance exists at any mass in strong $\wwl$ scattering.}
\end{figure}

The expected 95\% confidence level limits for $F_T$ for $\sqrt{s}=500$~GeV
and a luminosity of 500~fb$^{-1}$ are shown in Fig. \ref{fig:fteight}, 
along with the predicted values of $F_T$ for various  masses $M_\rho$ of a vector 
resonance in $\wwl$ scattering.
The masses and widths of the vector resonances are chosen to coincide with 
those used in the ATLAS TDR~\cite{atlas:1999}.
The technipion form factor $F_T$ affects only the amplitude for $\ee\ra\wwl$, whereas TGCs
affect all amplitudes.   Through the use of electron beam polarization and the rich
angular information in $\wwa$ production and decay, 
it will be possible to disentangle anomalous values of $F_T$ from 
other anomalous TGC values and to deduce the mass of a strong vector resonance 
well below threshold, as suggested by Fig.~\ref{fig:fteight}.

\begin{figure}[] 
\centerline{\epsfig{file=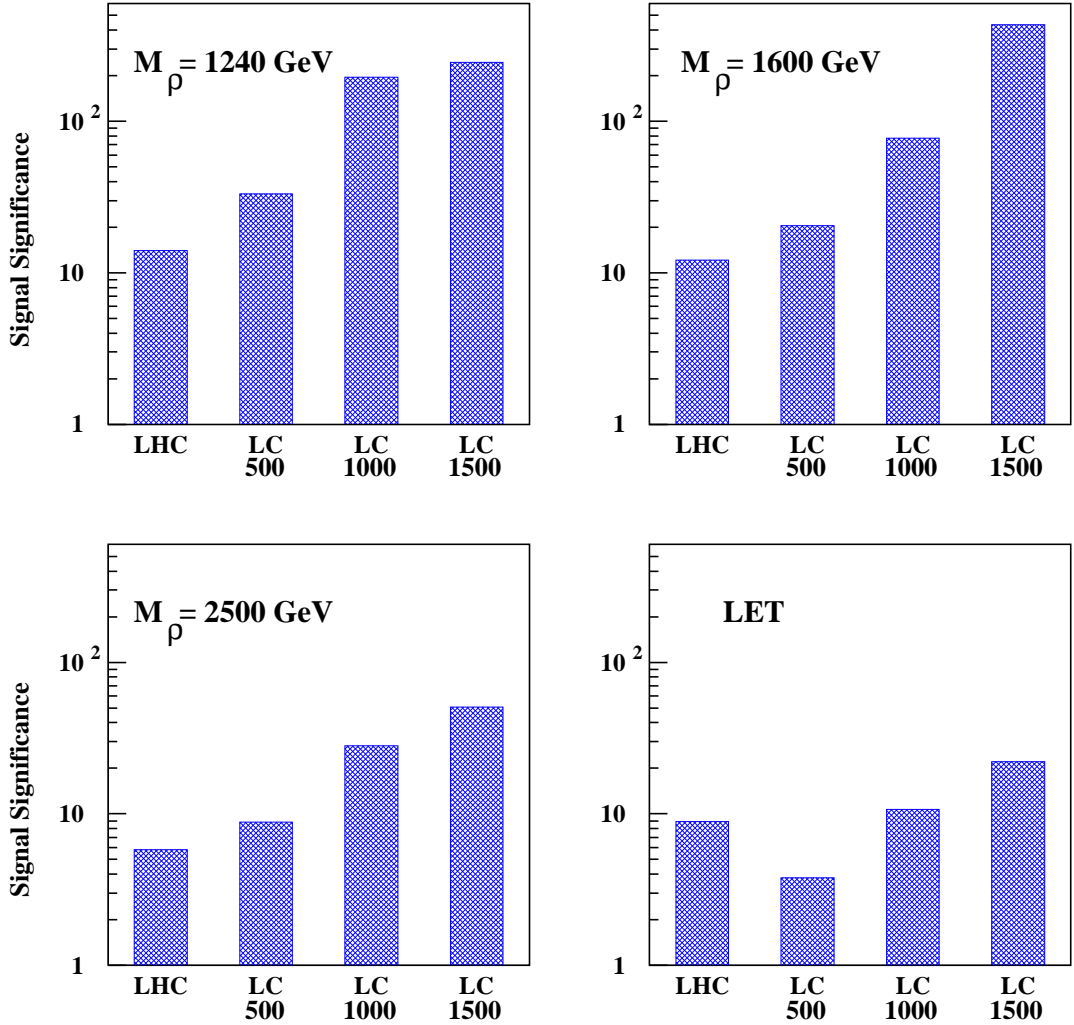,height=16cm,width=16cm}}
\vspace{10pt}
\caption{\label{fig:strong_lc_lhc}
Direct strong symmetry breaking signal significance in $\sigma$'s 
for various masses $M_\rho$ of a vector resonance in $\wwl$ scattering.
In the first three plots the signal at the LHC is a bump in the $WW$ cross
section; in the LET plot, the LHC signal is an enhancement over the SM
cross section.  The various LC signals are for enhancements of the
amplitude for pair production of longitudinally polarized $W$ bosons.
The numbers below the ``LC'' labels refer to the
center-of-mass energy of the linear collider in GeV.
The luminosity of the LHC is assumed to be 300~fb$^{-1}$, while the luminosities
of the linear colliders are assumed to be  500, 1000, and 1000~fb$^{-1}$ for 
$\sqrt{s}$=500, 1000, and 1500~GeV respectively.  The lower right hand plot ``LET'' refers to the
case where no vector resonance exists at any mass in strong $\wwl$ scattering.}
\end{figure}

The signal significances obtained by combining the results for 
$\ee\ra\nu \bar{\nu}\wwa$, $\nu \bar{\nu}ZZ$~\cite{Barger:1995cn} 
with the $F_T$ analysis of $\wwa$~\cite{Barklow:2000let}
are displayed in Fig.~\ref{fig:strong_lc_lhc} along with the
results expected from the LHC~\cite{atlas:1999}.   
The LHC signal is a mass bump in $W^+W^-$; the LC signal is less
direct.  Nevertheless, the  signals at the LC are strong, particularly in $\ee\to W^+W^-$, where
the technirho effect gives a large enhancement of a very well-understood
Standard Model process.
Since the technipion 
form factor includes an integral
over the technirho resonance region, the linear collider signal 
significance is 
relatively insensitive to the technirho width.   (The real part of 
$F_T$ remains fixed 
as the width is varied, while the imaginary part grows as the width grows.)  
The LHC signal significance will drop
as the technirho width increases.  The large linear collider signals can 
be utilized to study a
vector resonance in detail; for example, 
the evolution of $F_T$ with $\hat{s}$ can be determined by measuring
the initial-state radiation in $\ee\ra\wwa$.

    Only when the vector resonance
disappears altogether (the LET case in the lower right-hand panel in Fig.~\ref{fig:strong_lc_lhc} ) 
does the  direct strong symmetry breaking signal from the $\sqrt{s}=500$~GeV linear collider
drop below the LHC signal.   At higher $\ee$ center-of-mass energies the linear collider 
signal exceeds the LHC signal.  

\subsection{Composite Higgs models}

The good fit of the Standard Model to the electroweak data
suggests that the new physics has a decoupling limit
in which the new particles carrying $SU(2)_W\times U(1)_Y$ 
charges can be much heavier than the electroweak scale 
without affecting the Standard Model.
This is the reason why the Minimal Supersymmetric Standard Model
is viable: all the superpartners and the states associated with a
second Higgs doublet can be taken to be heavier than the electroweak scale,
leaving a low-energy theory given by the Standard Model.
At the same time, it is hard to construct viable technicolor 
models because they do not have a decoupling limit: the new fermions 
that condense and give the $W$ 
and $Z$ masses are chiral, {\it i.e.}, their
masses break the electroweak symmetry.

There is a class of models of electroweak symmetry breaking
that have a decoupling limit given by the Standard Model, so they 
are phenomenologically viable, and yet the Higgs field arises 
as a bound state due to some strong interactions.
An example of such a composite Higgs model is the Top Quark Seesaw Theory, 
in which a Higgs field appears as a bound state of the top quark 
with a new heavy quark. This has proven phenomenologically viable and
free of excessive fine-tuning~\cite{Dobrescu:1998nm}.
Furthermore, the top quark is naturally the heaviest Standard Model
fermion in this framework, because it participates directly in the 
breaking of the electroweak symmetry. 

The interaction responsible for binding the Higgs field is
provided by a spontaneously broken gauge symmetry, such
as topcolor~\cite{Hill:1991at}, or some flavor or family
symmetry~\cite{Burdman:1999vw}.
Such an interaction is asymptotically
free, allowing for a solution to the hierarchy problem. At the
same time the interaction is non-confining, and therefore has a very
different behavior from the technicolor interaction discussed in the first 
part of this section.

Typically, in the top quark seesaw theory, the Higgs boson is
heavy, with a mass of order 500 GeV \cite{Chivukula:2000px}. 
However, the effective theory
below the compositeness scale may include an extended Higgs sector,
in which case the mixing between the CP-even scalars could bring the
mass of the Standard Model-like Higgs boson down to the current LEP 
limit \cite{Dobrescu:1998nm,Dobrescu:1999gv}. 
One interesting possibility in this context is that there is a light 
Higgs boson with nearly standard couplings to fermions and gauge
bosons, but whose decay modes are completely non-standard.
This happens whenever
a CP-odd scalar has a mass less than half the Higgs
mass and the coupling of the Higgs to a pair of CP-odd scalars is not
suppressed.  The Higgs boson decays in this case into a pair of CP-odd
scalars, each of them subsequently decaying into a pair of
Standard Model particles with model-dependent branching
fractions \cite{Dobrescu:2001jt}.
If the Higgs boson has Standard Model branching fractions, then the
capability of an $e^+e^-$ linear collider depends on $M_H$, 
as discussed in \cite{fermilab-report}.
On the other hand, if the Higgs boson has non-standard decays, 
an $e^+ e^-$ collider may prove very useful in disentangling the
composite nature of the Higgs boson, by measuring its width and
branching fractions.

The heavy-quark constituent of the Higgs has a mass of a few TeV, 
and the gauge bosons associated with the strong interactions that
bind the Higgs are expected to be even heavier.
Above the compositeness scale there must be some additional 
physics that leads to the spontaneous breaking of the
gauge symmetry responsible for binding the Higgs. This may involve 
new gauge dynamics~\cite{Collins:2000rz}, or fundamental scalars and
supersymmetry. For studying  these
interesting strongly interacting particles, the 
$e^+ e^-$ collider should operate at the highest energy achievable.

Other models of Higgs compositeness have been proposed
recently~\cite{Georgi:2000wt}, and more are likely to be constructed
in the future. Another framework in which a composite Higgs boson
arises from a strong interaction is provided by extra spatial 
dimensions accessible to the Standard Model particles; this is discussed in Section 6.

\section{Contact interactions and compositeness}

There is a strong historical basis for the consideration of composite
models that is currently mirrored in the proliferation of fundamental
particles.  If the fermions have substructure, then their constituents
are bound by a confining force at the mass scale $\Lambda$, which 
characterizes the radius of the bound states.  At energies above $\Lambda$,
the composite nature of fermions would be revealed by the break-up
of the bound states in hard scattering processes.  At lower energies,
deviations from the Standard Model may be observed via form factors
or residual effective interactions induced by the binding force.  These
composite remnants are usually parameterized by the introduction of
contact terms in the low-energy Lagrangian.  More generally, four-fermion
contact interactions represent a useful parametrization of many types
of new physics originating at high energy scales, and specific cases 
will be discussed throughout this chapter.

The lowest-order four-fermion contact terms are of dimension 6.  A 
general helicity-conserving, flavor-diagonal, Standard Model-invariant
parameterization can be written as \cite{contact_par}
\begin{equation}
{\cal L} = {g^2_{\rm eff}\eta\over\Lambda^2}\left( \bar q\gamma^\mu q+
{\cal F}_\ell\bar\ell\gamma^\mu\ell\right)_{L/R} \left( \bar q\gamma_\mu q
+ {\cal F}_\ell\bar\ell\gamma_\mu\ell\right)_{L/R} \,,
\end{equation}
where the generation and color indices have been suppressed, $\eta=\pm 1$,
and ${\cal F}_\ell$ is inserted to allow for different quark and lepton
couplings but is anticipated to be ${\cal O}(1)$.  Since the binding
force is expected to be strong when $Q^2$ approaches $\Lambda^2$, it
is conventional to define $g^2_{\rm eff}=4\pi$.

Interference between the contact terms and the usual gauge interactions
can lead to observable deviations from Standard Model predictions at
energies lower than $\Lambda$.  Currents limits from various processes
at the Tevatron and LEP II place $\Lambda$ above the few-TeV range.
At the LHC \cite{atlas:1999}, $\Lambda_{\ell q}$ terms can be probed  
to $\sim 20-30$ TeV for integrated luminosities of $10-100$ fb$^{-1}$,
while the $\Lambda_{qq}$ case is more problematic because of uncertainties
in the parton distributions and the extrapolation of the calorimeter
energy calibration to very high values of the jet $p_T$.

At a LC, the use of polarized beams, combined with angular distributions,
allows for a clear determination of the helicity of the contact term.
An examination of contact effects in $e^+e^-\to f\bar f$, where $f=\mu\,,c\,,
b$ was performed for LC energies in \cite{CGH}.  This study concentrated 
on tagged final states, since contact effects are diluted when all quark
flavors are summed  because of cancellations between the up- and down-type
quarks.  Here, both polarized and unpolarized angular distributions
were examined with tagging efficiencies of 60\% and 35\% for $b$- and
$c$-quarks, respectively, and the detector acceptance was taken to be
$|\cos\theta|< 0.985$.  The resulting 95\% CL sensitivity for
${\cal L}=500$ fb$^{-1}$ to $\Lambda$
from the polarized distributions with 90\% electron beam polarization
is listed in Table \ref{jlh_contact}.

\begin{table}
\centering
\begin{tabular}{|l|c|c|c|c|} \hline\hline
 & $\Lambda_{LL}$ & $\Lambda_{LR}$ & $\Lambda_{RL}$ & $\Lambda_{RR}$ \\ \hline
$\sqrt s=0.5$ TeV & & & &  \\ \hline
$e^-_Le^+\to \mu^+\mu^-$ & 57 & 52 & 18 & 18 \\
$e^-_Re^+\to \mu^+\mu^-$ & 20 & 18 & 52 & 55 \\
$e^-_Le^+\to c\bar c$ & 59 & 50 & 9 & 15 \\
$e^-_Re^+\to c\bar c$ & 21 & 20 & 43 & 57 \\
$e^-_Le^+\to b\bar b$ & 68 & 53 & 9 & 16 \\
$e^-_Re^+\to b\bar b$ & 30 & 21 & 59 & 59\\ \hline
$\sqrt s=1.0$ TeV & & & &  \\  \hline
$e^-_Le^+\to \mu^+\mu^-$ & 79 & 72 & 25 & 26 \\
$e^-_Re^+\to \mu^+\mu^-$ & 28 & 25 & 73 & 78 \\
$e^-_Le^+\to c\bar c$ & 82 & 72 & 12 & 21 \\
$e^-_Re^+\to c\bar c$ & 30 & 28 & 62 & 78 \\
$e^-_Le^+\to b\bar b$ & 94 & 77 & 14 & 23 \\
$e^-_Re^+\to b\bar b$ & 43 & 30 & 82 & 84\\ \hline\hline
\end{tabular}
\caption{\label{jlh_contact}
95\% CL search reach in TeV for contact interaction scales with
various helicities.}  
\end{table}

\begin{figure}[]
\centerline{\epsfig{file=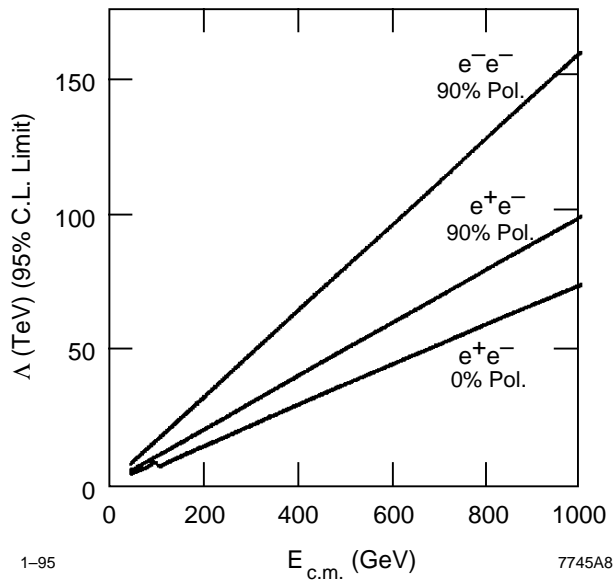,height=8cm}}
\caption[*]{ \label{nlcbhapol}
The 95\% confidence level limits for the compositeness scale $ \Lambda^+_{\rm LL}$ 
from M{\o}ller and Bhabha scattering as a function of the \emem or $\ee$ 
center-of-mass energy.  The luminosity is given by 
$ {\cal L}=680\ {\rm pb}^{-1}\cdot s/M_Z^2$.  The polarization
of the electron beam(s) is indicated in the figure.}
\end{figure}

Compositeness limits for $\Lambda^+_{\rm LL}$ from M{\o}ller and Bhabha 
scattering \cite{Barklow:1996ut} are 
summarized in Fig.~\ref{nlcbhapol}.  For equal luminosities the limits from  M{\o}ller
scattering are significantly better than those from  Bhabha scattering.  This is due not only to the 
polarization of both beams, 
but also to the  M{\o}ller/Bhabha crossing relation in the
central region of the detector.  Limits on  $\Lambda^+_{\rm LL}$ for different
energies and luminosities can be calculated under the assumption that the 
compositeness limit scales as ${\cal{L}}^{1/4}s^{1/2}$.

\section{New particles in extended gauge sectors and GUTs}

\subsection{Extended gauge sectors}

New gauge bosons are a feature of many extensions of the 
Standard Model.  They arise naturally in grand unified theories, such as 
$SO(10)$ and $E_6$, where the GUT group gives rise to extra $U(1)$  
or $SU(2)$ subgroups after decomposition.   There are also numerous non-unified 
extensions, such as the Left-Right Symmetric model and Topcolor.
More recently, there has been renewed interest in 
Kaluza-Klein excitations of the SM gauge bosons, which are realized in 
theories of extra space dimensions at semi-macroscopic scales.  All of these 
extensions of the SM predict the existence of new gauge bosons, 
generically denoted as $Z'$ or $W'$.  The search for extra gauge bosons 
thus provides a common coin in the quest for new physics
at high-energy colliders.
Here, we concentrate on the most recent developments on the
subject, and refer the interested to recent 
reviews \cite{leike}.

\subsubsection{$Z^\prime$\ discovery limits and identification}

The signal for the existence of a new neutral gauge boson at linear collider
energies arises through the indirect effects of $s$-channel $Z^\prime$ exchange.
Through its interference with the SM $\gamma$ and $Z$
exchange in $e^+e^- \to f\bar{f}$,  significant deviations from SM
predictions can occur even when  $M_{Z^\prime}$ is much
larger than $ \sqrt{s}$.   This sensitivity to the $Z^\prime$ nicely
complements the ability of the LHC to discover a $Z^\prime$ as a
resonance in lepton pair production.   The combination of
many LC observables such as the cross sections for $f\bar f$ final states,
forward-backward asymmetries, $A_{FB}^f$, and left-right asymmetries,
$A_{LR}^f$, where $f=\mu$, $\tau$, $c$, $b$, and light quarks,
can fill in the detailed picture of the $Z^\prime$ couplings.

\begin{figure}[t]
\leavevmode
\centerline{\epsfig{file=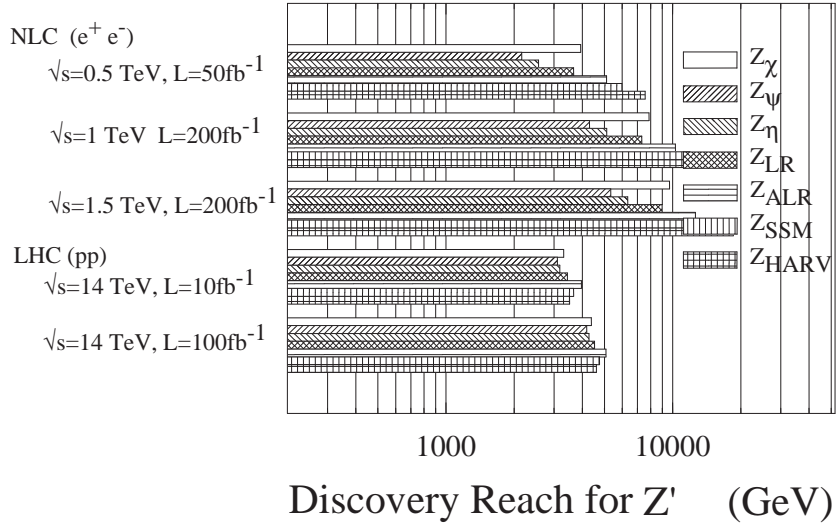,width=12cm,clip=}}
\caption{\label{stevezp}
$95\%$ CL search limits for extra neutral gauge bosons, for
various models,  at high-energy linear colliders, by observation of corrections to $e^+e^-
\to f\bar f$ processes, and at the LHC, by observation of a peak
in dilepton pairs.}
\end{figure}

\begin{figure}[t]
\begin{minipage}[t]{6.8cm} {
\centerline{\epsfig{file=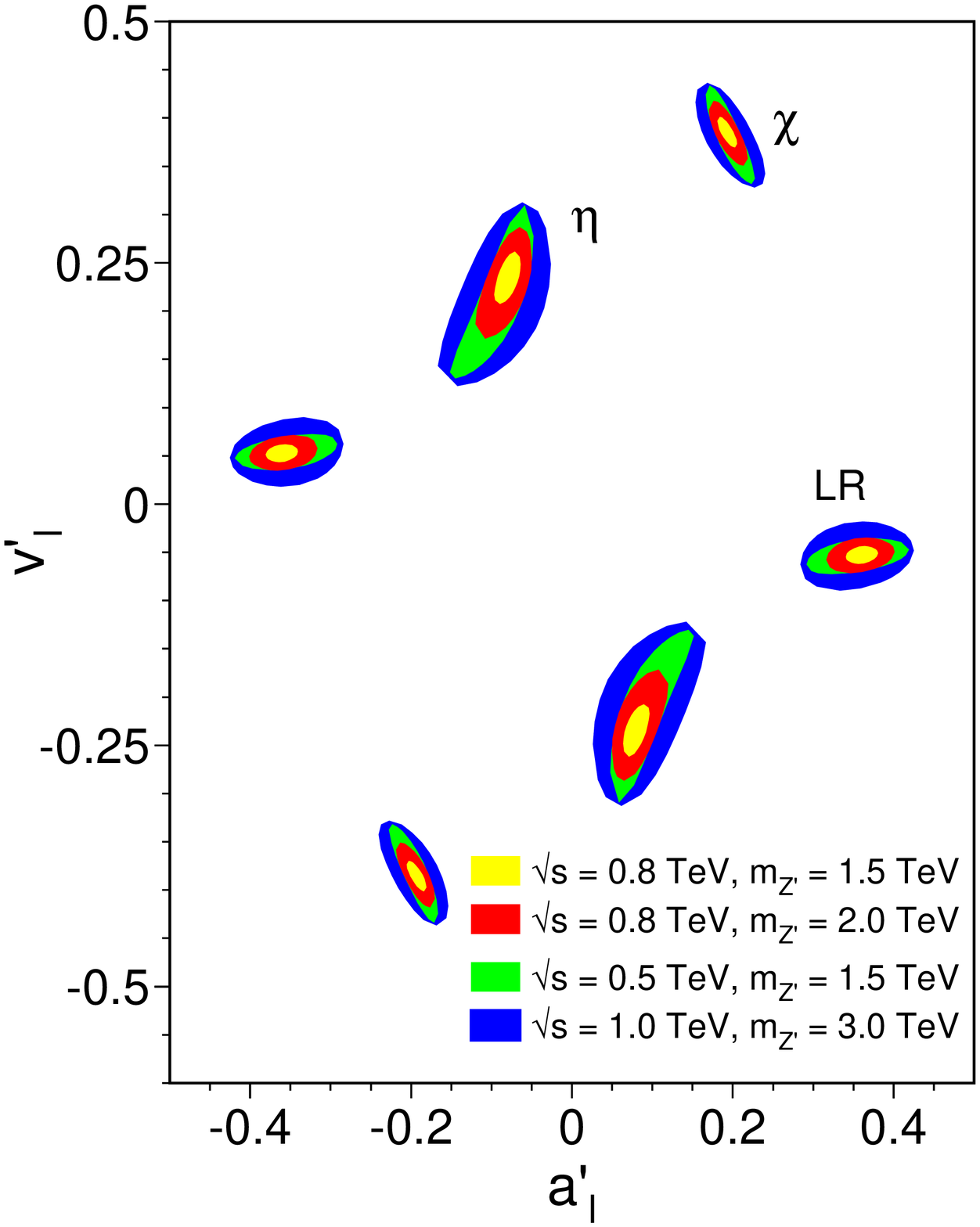,width=6.0cm,clip=}  }    }
\end{minipage}       
\hspace*{0.5cm}
\begin{minipage}[t]{6.8cm} {
\centerline{\epsfig{file=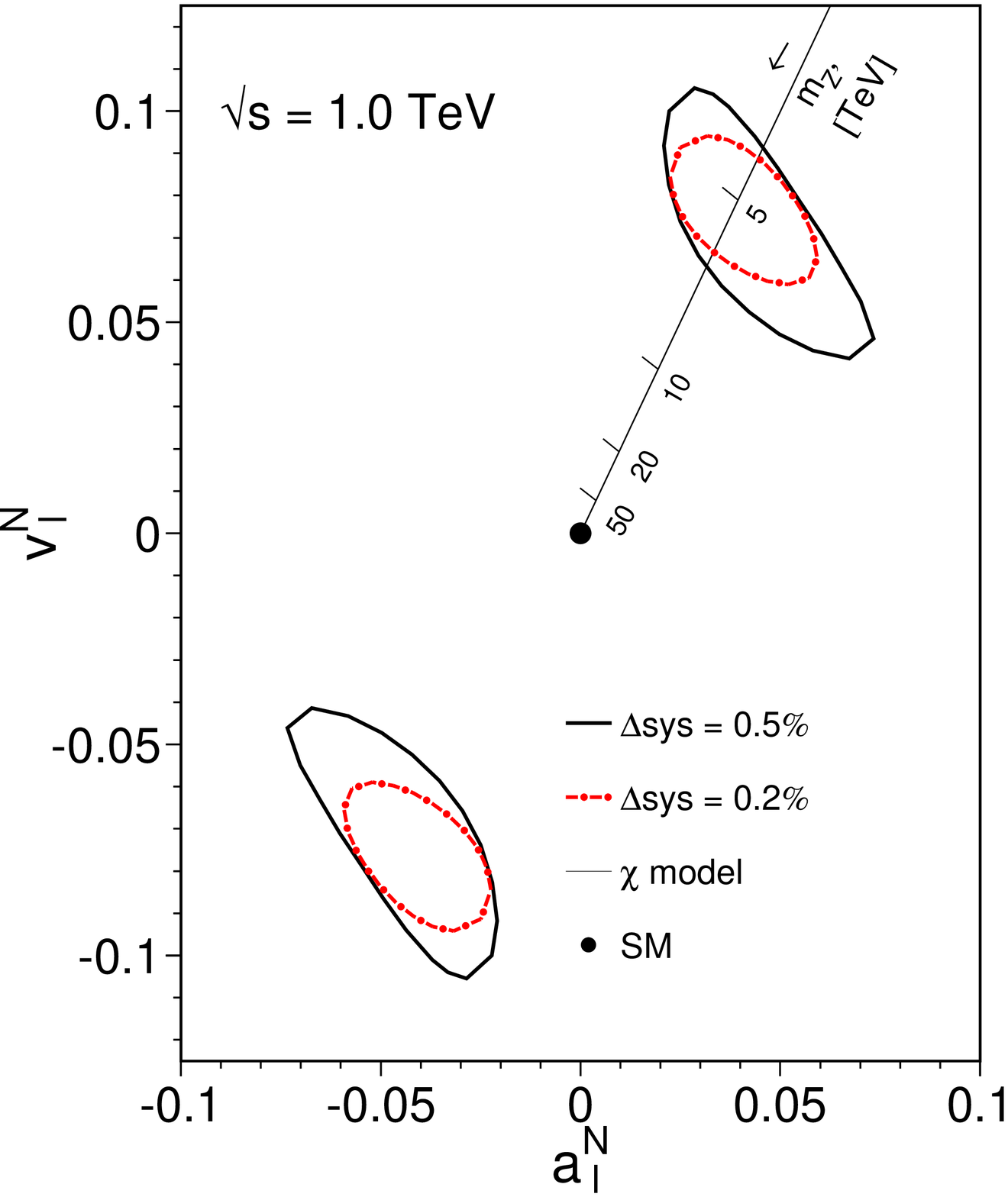,width=6.0cm,clip=} }     }
\end{minipage}
\caption[*]{\label{zpcoupl}
Left Panel: Resolution power (95\% C.L.) for different $M_{Z'_\chi}$ based on 
measurements of leptonic observables at $\sqrt{s}=1$~~TeV 
with a luminosity of $L_{\rm int}=1 \; $ab$^{-1}$ \cite{sabine}.
Right Panel:
Resolution power (95\% C.L.) for different $M_{Z'}$ based on 
measurements of leptonic observables at $\sqrt{s}=500$~GeV, 800~GeV, 
1 TeV with a luminosity of $L_{\rm int}=1 \; $ab$^{-1}$. The leptonic 
couplings of the $Z'$ correspond to the $\chi$, $\eta$, or $LR$ model
\cite{sabine}.}
\end{figure}

The combined sensitivity of the LC measurements for various $Z^\prime$ models
is shown in  Fig. \ref{stevezp} \cite{leike}.  We see that 
if a $Z'$ is  detected at the LHC, precision measurements at the LC 
could be used to measure its properties and determine the 
underlying theory.  Figure~\ref{zpcoupl}
displays the resolving power between $Z'$ models 
assuming that the mass of the $Z'$ was measured previously at the 
LHC.  This study only considers leptonic final states and assumes 
lepton universality.  If $M_{Z'}$ were beyond the LHC discovery reach 
or if the $Z'$ does not couple to quarks  then no  prior knowledge of it
would be obtained before the LC turns on.  However, in this case, the LC
can still yield some information on the $Z'$ couplings and mass.
Instead of extracting $Z'$ couplings directly, ``normalized'' couplings, 
defined by
\begin{equation}
a_f^N = a_f' \sqrt{{s\over {M_{Z'}^2 -s}}}; \quad
v_f^N = v_f' \sqrt{{s\over {M_{Z'}^2 -s}}}
\end{equation}
could be measured.  For a demonstration of this case, the diagnostic power 
of a 1 TeV LC for a
$Z'$ with couplings of the $E_6$ model $\chi$ and mass $M_{Z'}=5$~TeV
is displayed in Fig.  \ref{zpcoupl} 
for $f=\ell$.  An additional 
determination of the $Z'$ mass and couplings could be 
performed \cite{leike} in this case from cross section and 
asymmetry measurements at several different values of $\sqrt s$.

A recent study of the process $e^+e^-\to \nu\bar{\nu}\gamma$ has 
demonstrated that the process can also be used to obtain information on 
$Z' - \nu\bar{\nu}$ couplings \cite{wp1}.

\subsubsection{$W^\prime$\  discovery limits and identification}

While considerable effort has been devoted to the study of $Z'$ bosons 
at $e^+e^-$ colliders, a corresponding endeavor for the $W'$ sector
has only recently been undertaken.  A preliminary investigation \cite{hewett} 
of the sensitivity of $e^+e^-\to \nu\bar{\nu}\gamma$ to $W'$ bosons
was performed at Snowmass 1996, and more detailed examinations 
\cite{wp1,wp2}
have recently been performed.
The models with extra $SU(2)$ factors considered 
in these studies are the 
Left-Right symmetric model (LRM) based on the gauge group $SU(2)_L 
\times SU(2)_R \times U(1)_{B-L}$, the Un-Unified model (UUM) based on 
$SU(2)_q\times SU(2)_l \times U(1)_Y$ where the quarks 
and leptons each transform under their own $SU(2)$, a Third Family 
Model (3FM) based on the group $SU(2)_h\times SU(2)_l \times U(1)_Y$ 
where the quarks and leptons of the third (heavy) family transform 
under a separate group, and the KK model which contains the 
Kaluza-Klein excitations of the SM gauge bosons that are a possible 
consequence  of theories with large extra dimensions. 

\begin{figure}[h]
\leavevmode
\centerline{\epsfig{file=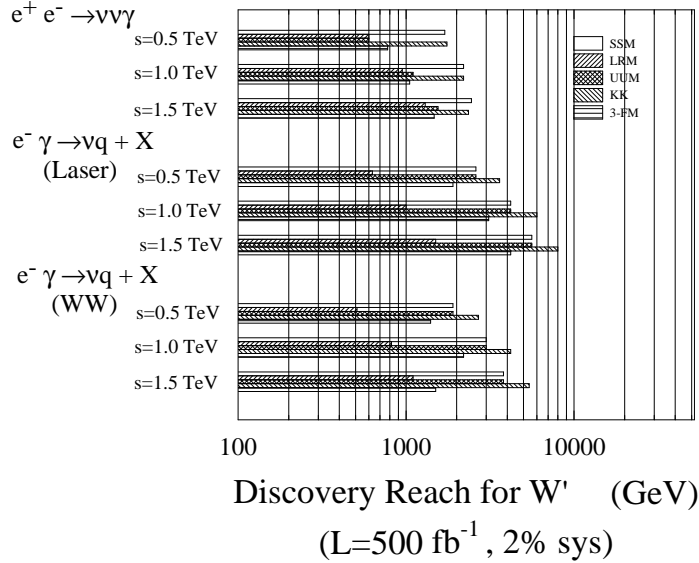,width=9.2cm,clip=}}
\caption{\label{stevewp}
95\% CL search limits for $W'$ bosons at the LC.}
\end{figure}

In the process $e^+e^-\rightarrow \nu\bar\nu\gamma$,
both charged and neutral extra gauge bosons can contribute.
In the analysis of \cite{wp1}, 
the photon energy and angle with respect to the beam axis are
restricted to $E_\gamma \geq 10$ GeV and 
$10^\circ\leq \theta_\gamma \leq 170^\circ$, to take into account detector acceptance.
The most serious background,
radiative Bhabha scattering in which the scattered $e^+$ and $e^-$
go undetected down the beam pipe, is suppressed by restricting
the photon's transverse momentum  to 
$p_T^\gamma > \sqrt{s} \sin\theta_\gamma \sin\theta_v / 
(\sin\theta_\gamma + \sin\theta_v)$,
where $\theta_v$ is the minimum angle at which the veto detectors
may observe electrons or positrons; here, $\theta_v = 25$ mrad.  The observable
$d\sigma/dE_\gamma$ was found to provide the
most statistically significant search reach.  The $95\%$ CL reach
is displayed graphically in Fig. \ref{stevewp} and in Table \ref{steve_tab},
which shows the degradation when a
2\% systematic error is added in quadrature with the statistical error.
The corresponding $W'$ search reach at the LHC is in the range 5--6 TeV
\cite{leike}.

\begin{figure}[t]
\centerline{
\epsfig{file=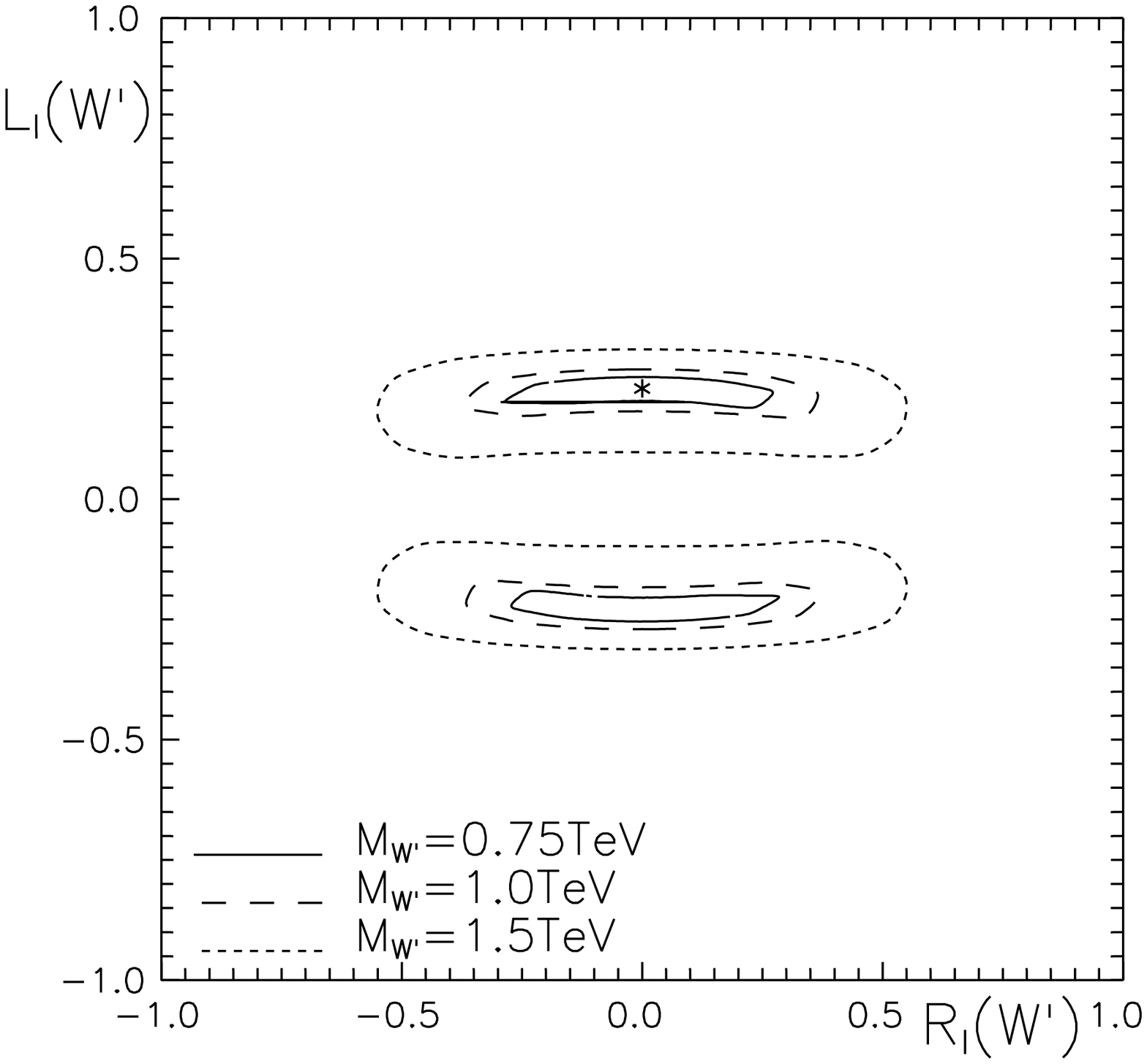,width=7.0cm,clip=}
\hspace*{0.5cm}
\epsfig{file=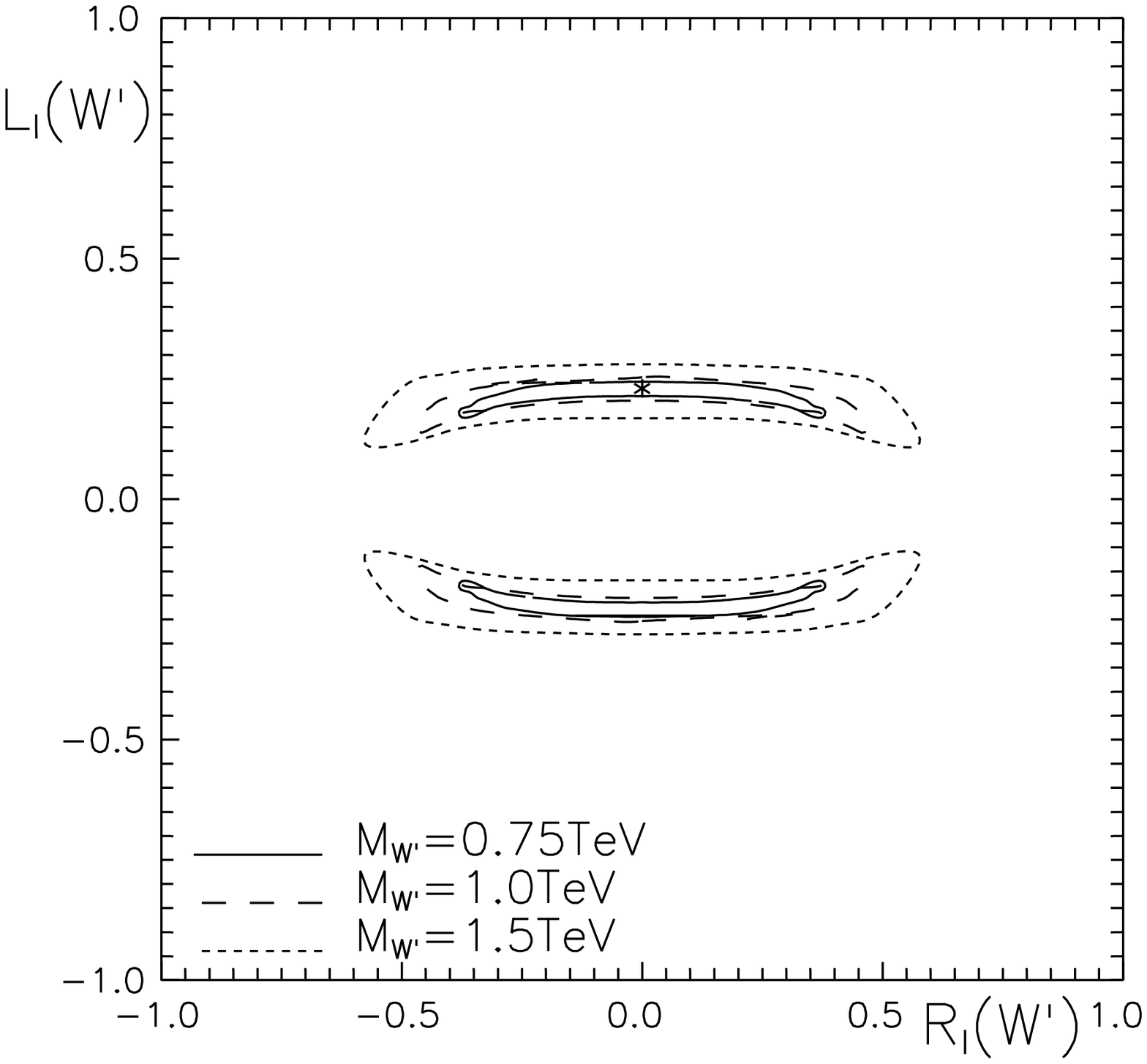,width=7.0cm,clip=}}
\caption{\label{wpcoupl}
Left Panel: 
95\% CL constraints from $e^+e^-\to \nu\bar{\nu}\gamma$
on  couplings of the SSM $W'$ indicated by a star 
for $\sqrt{s}=0.5$\, TeV  and
$L_{\rm int}=1000$\, fb$^{-1}$ with a systematic error of 0.5\% (0.25\%) for 
$\sigma (A_{LR})$
for different $W'$ masses.
Right Panel: 95\% C.L. constraints from $e\gamma \to \bar{\nu} q + X$ 
on couplings of the SSM $W'$ for $\sqrt{s}=0.5$\, TeV and
$L_{\rm int}=1000\ 
{\rm fb}^{-1}$ with a 2\% systematic error for different $W'$ masses.
}
\end{figure}

The 95\% CL constraints that can be placed 
on the right- and left-handed couplings of a $W'$ to fermions,
assuming that the $W'$ has Standard Model-like couplings, and that there
is no corresponding $Z'$ contribution to $e^+e^-\to\nu\bar\nu\gamma$,
are shown in Fig.   \ref{wpcoupl}.  
Here, 
the total cross section $\sigma$ and the left-right asymmetry $A_{LR}$
are used as observables, with
the systematic errors for $\sigma (A_{LR})$ taken as 2\%(1\%) and
80\% electron and 60\% positron polarization are assumed.
The axes in this figure correspond to couplings 
normalized as $L_f(W) =C_L^{W'} g/(2\sqrt{2})$ and similarly for 
$R_f(W)$. It is found that 2\%
systematic errors dominate the coupling determination.  In addition, we note
that the $W'$ couplings can only be constrained up to a two-fold ambiguity,
which could be resolved by reactions in which the $W'$ couples to a
triple gauge vertex.

Additional sensitivity to the existence of a $W'$ can be gained from 
$e\gamma\rightarrow \nu q + X$ \cite{wp2}. This process 
receives contributions only
from charged and not from neutral gauge bosons.  The $W'$ contribution can be 
isolated by imposing a kinematic cut requiring either the $q$ or 
$\bar{q}$ to be collinear to the beam axis.  
In  order to take into account detector acceptance, the angle $\theta_q$ 
of the detected quark relative to the beam axis is restricted to
$10^\circ \leq \theta_q \leq 170^\circ$. The kinematic variable that is most sensitive 
to a $W'$ is the $p_{T_q}$ distribution.
The quark's transverse momentum relative to the beam is restricted to
$p_T^q > 40 (75)$ GeV for $\sqrt{s}= 0.5(1.0)$\, TeV, to
suppress various Standard Model backgrounds.  
Figure \ref{wpcoupl}  and Table \ref{steve_tab} 
show the resulting 95\% CL constraints on the $W'$ fermionic couplings 
for the case of backscattered laser photons. 
As seen above, the assumed systematic error of 2\% again
dominates the statistical error,
thus eliminating the potential gain from high luminosities.
$W'$ coupling determination 
from backscattered laser photons are considerably better than 
those from Weizs\"acker-Williams photons or from $e^+e^-$ collisions.
Polarized beams give only a minor improvement to these results after the 
inclusion of systematic errors.

If a $W'$ were discovered elsewhere, measurements of its couplings in 
both $e^+e^-\to \nu\bar{\nu}\gamma$ and $e\gamma \to \nu q 
+X$ could provide valuable information regarding the underlying model, 
with the latter process serving to isolate the $W'$ couplings from 
those of the $Z'$.

\begin{table}
\centering
\begin{tabular}{|l|ll|ll|ll|ll|}
\hline\hline
&\multicolumn{4}{c|}{$\sqrt{s}=0.5$ TeV, $L_{\rm int}=500$ fb$^{-1}$} &
\multicolumn{4}{c|}{$\sqrt{s}=1$ TeV, $L_{\rm int}=500$ fb$^{-1}$}\\ \hline
&\multicolumn{2}{c|}{$e^+e^-\rightarrow \nu\bar\nu\gamma$} 
&\multicolumn{2}{c|}{$e\gamma\rightarrow \nu q + X$} &
 \multicolumn{2}{c|}{$e^+e^-\rightarrow \nu\bar\nu\gamma$}
&\multicolumn{2}{c|}{$e\gamma\rightarrow \nu q + X$}\\
Model &no syst.&syst.&no syst&syst.&no syst.&syst.&no syst.&syst.\\ \hline
SSM $W'$  & 4.3 & 1.7 & 4.1 & 2.6 & 5.3 & 2.2 & 5.8 & 4.2 \\
LRM       & 1.2 & 0.6 & 0.8 & 0.6 & 1.6 & 1.1 & 1.2 & 1.1 \\
UUM       & 2.1 & 0.6 & 4.1 & 2.6 & 2.5 & 1.1 & 5.8 & 4.2 \\
KK        & 4.6 & 1.8 & 5.7 & 3.6 & 5.8 & 2.2 & 8.3 & 6.0 \\ 
3FM	& 2.3 & 0.8 & 3.1 & 1.9 & 2.7 & 1.1 & 4.4 & 3.1 \\ \hline\hline
\end{tabular}
\caption{\label{steve_tab}
95\% CL search limits for $W'$ bosons, in TeV, for various reactions.}
\end{table}

\subsection{Leptoquarks}

Leptoquarks are natural in theories that relate leptons and quarks
at a more fundamental level.  These spin-0 or -1 particles carry
both baryon and lepton number and are color triplets under SU(3)$_C$.  They
can be present at the electroweak scale in models where baryon and lepton 
number are separately conserved, thus avoiding conflicts with rapid proton
decay.  Their remaining properties depend on the model in which they appear,
and would need to be determined in order to ascertain the framework of
the underlying theory.
Given the structure of the Standard Model fermions, there are 14 different
possible types of leptoquarks; their classification can be found in 
\cite{brw}.  Their fermionic couplings proceed through a Yukawa interaction
of unknown strength, while their gauge couplings are specified for a
particular leptoquark.  Low-energy data place tight constraints on 
intergenerational leptoquark Yukawa couplings and also require that
these couplings be chiral.  A summary of the 
current state of experimental searches for leptoquarks is given in
\cite{zarnecki1}.  

\begin{figure}[t]
\centerline{
\epsfig{figure=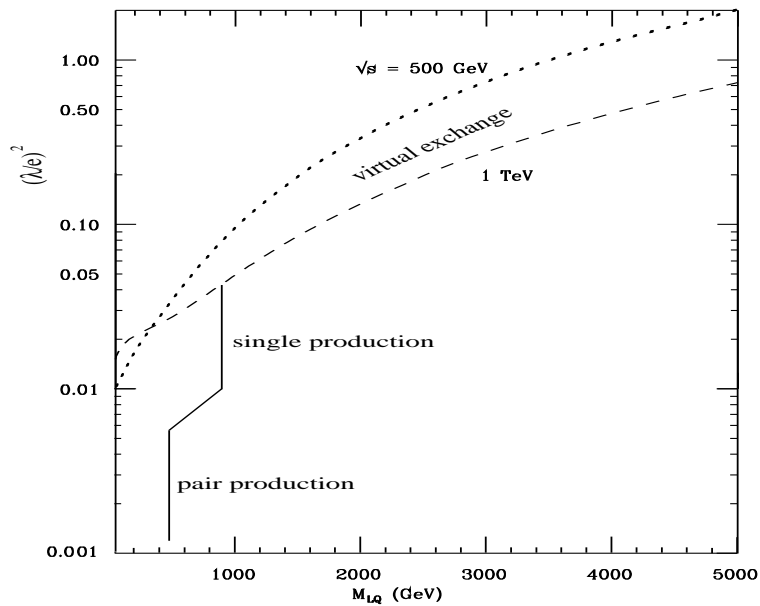,height=8cm,width=10cm,angle=0}}
\caption{\label{leptosearch}
Leptoquark search limits at a LC from the three processes
discussed in the text.  The Yukawa coupling is scaled to $e$.  The
pair- and single-production reaches are shown for $\sqrt s=1$ TeV,
while the indirect reach is displayed for $\sqrt s=0.5$ and 1~TeV.}
\end{figure}

At a linear collider, leptoquarks may be produced in pairs or as single
particles, while virtual leptoquark
exchange may be present in $e^+e^-\to$ hadrons.   Pair production
receives a $t$-channel quark-exchange contribution whose magnitude depends
on the size of the Yukawa coupling.  This only competes with the usual
$s$-channel exchange, which depends on the leptoquark's gauge couplings,
if the Yukawa coupling is of order electromagnetic
strength.  The possible signatures are $e^+e^-$, $e^\pm$ plus missing
$E_T$, or missing $E_T$ alone, combined with two jets.  The 
observation is straightforward essentially up to the
kinematic limit.  A thorough study of the background and resulting 
search reach for each type of leptoquark can be found in \cite{ruckl}.
Single leptoquark production is most easily studied
in terms of the quark content of the photon \cite{mikesteve}.
In this case a lepton fuses with a quark from 
a Weisz\"acker-Williams photon (in $e^+ e^-$ mode) or a laser-backscattered 
photon (in  $e \gamma$ mode) to produce a leptoquark.   
The cross section is a  convolution of the parton-level process with 
distribution functions for the photon in the electron and the quark in the 
photon, and is directly proportional to the $eqLQ$ Yukawa coupling.
The kinematic advantage of single production is lost if the
Yukawa coupling is too small.  For Yukawa couplings of 
electromagnetic strength, leptoquarks with mass up to 
about 90\% of $\sqrt{s}$ can be discovered at a LC \cite{mikesteve}.    
If the Yukawa couplings are sizable enough, then virtual leptoquark
exchange \cite{joatgr} will lead to observable deviations in the hadronic 
production cross section for leptoquark masses in excess of $\sqrt s$.
A summary of the search reach from these three processes is shown in 
Fig. \ref{leptosearch} from \cite{ruckl} in the leptoquark mass-coupling 
plane.  In comparison, leptoquarks are produced strongly at the LHC,
with search reaches in the 1.5 TeV range \cite{lhclepto} independent 
of the Yukawa couplings.

The strength of the LC is in the determination of the leptoquark's
electroweak quantum numbers and the strength of its Yukawa couplings once it is
discovered.  Together, the production rate and polarized left-right
asymmetry can completely determine the leptoquark's electroweak properties
and identify its type \cite{muchado} in both the pair and single production
channels, up to the kinematic limit.  In addition, the Yukawa coupling
strength can be measured via the forward-backward asymmetry in leptoquark
pair production (which is non-vanishing for significant Yukawa couplings),
deviations in the hadronic cross sections, and the comparison of pair
and single production rates.

\subsection{Exotic fermions}

Fermions beyond the ordinary Standard Model content arise in many
extensions of the Standard Model, 
notably in grand unified theories. They are
referred to as exotic fermions if they do not have the usual SU(2)$_L
\times$U(1)$_Y$ quantum numbers. For a review, we refer the 
reader to \cite{djrev}.  Examples of new fermions
are the following: ($i$) The sequential repetition of a Standard Model 
generation (of course, in this case the fermions maintain their 
usual SU(2)$ _L \times$U(1)$_Y$ assignments).
($ii$)  Mirror fermions, which have
chiral properties opposite to those of their Standard 
Model counterparts
\cite{maa}. The restoration of left-right symmetry is a motivating
factor for this possibility.
($iii$) Vector-like fermions that arise when a particular weak isospin representation is
present for both left and right handed components. For instance, in E$_6$
grand unified theories, with each fermion generation in the representation
of dimension {27}, there are two additional isodoublets of leptons,
one sequential (left-handed) and one mirror
(right-handed). This sort of additional content is referred to as a
vector doublet model (VDM) \cite{e6}, whereas the addition of weak
isosinglets in
both chiralities is referred to as a vector singlet model
(VSM) \cite{Val}.  

Exotic fermions can mix with the Standard Model 
fermions; in principle, the
mixing pattern may be complicated and is model-independent. One
simplifying factor is that intergenerational mixing is severely limited by
the constraints on flavor-changing neutral currents, as such mixing is
induced at the tree level \cite{e6}. Thus most analyses neglect
intergenerational
mixing. Global fits of low-energy electroweak data and the high-precision
measurements of the $Z$ properties provide upper limits for the remaining
mixing angles of the order of $\sin^2\theta_{\rm mix} \leq 10^{-2} - 10^{-3}$
\cite{ll}. 

Exotic fermions may be produced in $e^+e^-$ collisions either in pairs or
singly
in association with their Standard Model partners as a result of mixing. The
cross section for pair production of exotic quarks via gluon fusion and
the Drell-Yan process at the LHC is large enough that the reach of the LC
is unlikely to be competitive \cite{quark}.  On the other
hand, the backgrounds to exotic lepton production are large in $pp$
collisions, with production in $e^+e^-$ collisions providing a promising
alternative. Generally, the search reach for exotic leptons is up to the
kinematic limit of the $e^+e^-$ machine, for allowed mixings  \cite{djaz}. 
The experimental signature requires knowledge of the $L^{\pm}$ decay
mode, which is model-dependent and also depends on the mass difference of
the charged and neutral exotic leptons. 
Studies indicate that the signals for
exotic lepton production are
clear and easy to separate from Standard Model 
backgrounds \cite{djrev,djaz,almnew}, and that the use of polarized beams
is important
in determining the electroweak quantum numbers \cite{almnew}.

\begin{figure}[htb]
\leavevmode
\centerline{\epsfig{file=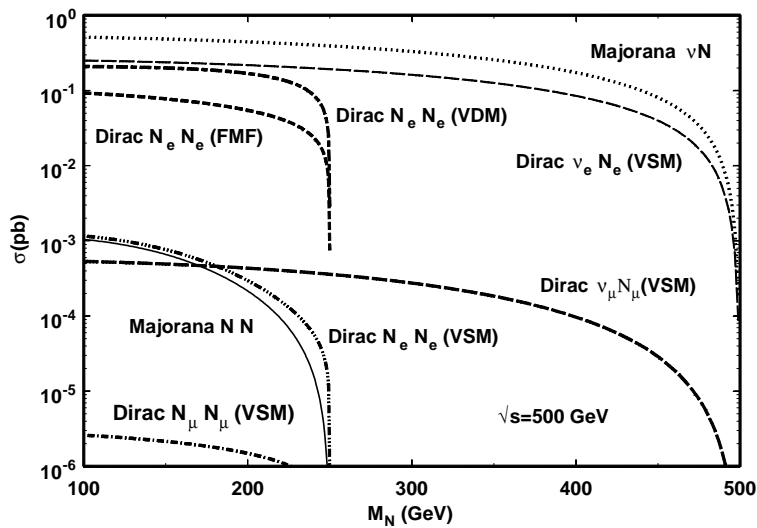,width=12.0cm,clip=}}
\caption{\label{pat3}
Single and pair production cross sections of on-shell heavy
Dirac and Majorana neutrinos at $\sqrt{s} = 500$ GeV for $e^+e^-$
colliders \cite{almneut}.}
\end{figure}

Almeida {\it et al.} have recently presented a detailed study of neutral
heavy lepton production at high-energy $e^+e^-$ colliders
\cite{almneut}. They find
single heavy neutrino production to be more important than pair production
and have calculated the process $e^+e^- \rightarrow \nu e^{\pm} W^{\mp}$
including on-shell and off-shell heavy neutrinos. They conclude that
$e^+e^-$ colliders can test the existence of heavy Dirac and Majorana
neutrinos up to $\sqrt{s}$ in the $\nu e^{\pm}+$ hadrons channel. Single
heavy neutrino production can be clearly separated from Standard Model 
backgrounds,
particularly with the  application of angular cuts on the final-state particle
distributions. Figure \ref{pat3} 
shows the on-shell approximation cross sections
for various pair- and single-production processes, with all mixing angles
such that $\sin^2\theta_{\rm mix} = 0.0052$ \cite{ll}.

\section{Extra dimensions}

The possibility has recently been proposed of utilizing the geometry of extra spatial dimensions to
address the hierarchy problem, {\it i.e.}, the disparity between the 
electroweak and Planck scales~\cite{add,ars}.
This idea exploits the fact that gravity has yet to be probed at energy
scales much above $10^{-3}$ eV in laboratory experiments, implying that the
Planck scale (of order $10^{19}$ GeV), 
where gravity becomes strong, may not be fundamental  but simply an artifact 
of the properties of the higher-dimensional space.  In one
such scenario \cite{add}, the apparent hierarchy is generated by a large
volume for the extra dimensions, while in a second case \cite{ars}, the 
observed hierarchy is created
by an exponential function of the compactification radius of the
extra dimension.  An exciting
feature of these theories is that they afford concrete and
distinctive experimental tests both in high energy physics and
in astrophysics.  Furthermore, if they truly describe the source of the observed
hierarchy, then their signatures should appear in high-energy experiments at
the TeV scale.  

Another possibility is the existence of TeV$^{-1}$-sized extra dimensions
accessible to Standard Model fields. Although these theories do not explicitly
address the hierarchy between the Electroweak and Planck scales,
they are not ruled out experimentally and
may arise naturally from string theory \cite{ant}. Furthermore, they serve as a
mechanism for suppressing proton decay and generating the
hierarchical pattern of
fermion masses \cite{nam}.  Models with TeV-scale extra dimensions
provide a context for new approaches to the problem of explaining
electroweak symmetry breaking
\cite{Cheng:2000bg,Arkani-Hamed:2000hv} and the existence of three
generations of quarks and leptons \cite{Dobrescu:2001ae}.
These theories also give rise to interesting phenomenology at the TeV scale.

We first describe some common features of these theories.  In all the
above scenarios, our universe lies on a 3+1-dimensional brane
(sometimes called a wall) that is embedded in the higher 
$4+\delta$-dimensional space, 
known as the bulk.  The field content that is allowed to propagate in the
bulk varies between the different models.  Upon compactification
of the additional dimensions, all bulk fields expand into a Kaluza-Klein (KK)
tower of states on the $3+1$-dimensional brane, where the masses of the KK states
are related to the $\delta$-dimensional kinetic motion of the bulk field.
It is the direct observation or indirect effects of the KK states
that signal the existence of extra dimensions at colliders.

\subsection{Large extra dimensions}

In this scenario \cite{add},  gravitational fields propagate in the 
$\delta$ new large spatial dimensions, as well as in the usual $3+1$
dimensions.  It is postulated that their 
interactions become strong at the TeV scale.  The volume of
the compactified dimensions, $V_\delta$, relates the scale
where gravity becomes strong in the $4+\delta$-dimensional spaces
to the apparent Planck scale via Gauss' Law 
\begin{equation}
M^2_{Pl}=V_\delta M_*^{2+\delta}\,,
\end{equation}
where $M_*$ denotes the fundamental Planck scale in the 
higher-dimensional space.  Setting $M_*$ to be of order 1 
$\sim$ TeV thus determines the compactification
radius $r_c$ ($V_\delta\sim r_c^\delta$)
of the extra dimensions, which ranges from
a sub-millimeter to a few fermi for $\delta=$ 2--6, assuming that all radii are
of equal size.  The compactification 
scale ($M_c=1/r_c$) associated with these parameters then ranges
from $10^{-4}$ eV to a few MeV.  The case of $\delta=1$ (which yields
$r_c\approx 10^{11}$ m) is immediately
excluded by astronomical data.  Cavendish-type experiments, which search
for departures from the inverse-square law gravitational force,
exclude \cite{adel} $r_c>$ 190 $\mu$m for $\delta=2$, which translates to the
bound $M_*>1.6$ TeV using the convention in \cite{grw}.  In addition, 
astrophysical and cosmological considerations \cite{astro}, such 
as the rate of supernova cooling and the diffuse $\gamma$ ray spectrum, 
disfavor a value of $M_*$ near the TeV scale for $\delta=2$.
Precision electroweak 
data \cite{bunch} do not allow the Standard Model fields to propagate in
extra dimensions with $M_c <$ a few  TeV, and hence they are constrained to the 
$3+1$-dimensional brane in this model.  

The Feynman rules for this scenario \cite{grw,hlz} are obtained by 
considering a linearized theory of gravity in the bulk.  
The bulk  field strength tensor can be decomposed into spin-0, 1, 
and 2 states, each of which expands into KK towers upon compactification.  
These KK states are equally spaced and 
have masses of $n/r_c$ where $n$ labels the KK excitation level.  
Taking $M_*=1$ TeV, we see that the KK state mass splittings are
equal to $5\times 10^{-4}$ eV, 20 keV, and 7 MeV for $\delta=2\,, 4$,
and 6, respectively.  The interactions of the KK gravitons with the Standard
Model fields on the wall are governed by the conserved stress-energy tensor
of the wall fields.  
The spin-1 KK states do not interact with the wall fields because of
the form of the wall stress-energy tensor.  The non-decoupling
scalar KK states couple to the trace of the stress-energy tensor, and are 
phenomenologically irrelevant for most collider processes. 
Each state in the spin-2 KK tower, $G_n$, couples identically 
to the Standard Model wall fields via their stress-energy tensor with the 
strength proportional to the inverse 4-dimensional Planck scale, $M_{Pl}^{-1}$.
It is important to note that this description is an effective 4-dimensional
theory, valid only for energies below $M_*$.  The full theory above 
$M_*$ is unknown.

Two classes of collider signatures arise in this model.  The first is emission 
of the graviton KK tower states in scattering
processes \cite{grw,emit}.  The relevant process at a linear collider 
is $e^+e^-\to\gamma/Z+G_n$, where the graviton appears as missing
energy in the detector, behaving as if it were a massive, 
non-interacting, stable particle.  The cross section is computed for
the production of a single massive graviton excitation and then summed
over the full tower of KK states.  Since the mass splittings of the
KK excitations are quite small compared to the collider center-of-mass 
energy, this sum can be replaced by an integral weighted by the density 
of KK states which is cut off by the specific process kinematics.
The cross section for this process scales as simple powers of $\sqrt s/M_*$.
It is important to note that because of the integral over the effective density 
of states, the emitted graviton appears to have a continuous mass distribution.
This corresponds to the probability of emitting gravitons with different 
extra-dimensional momenta.  The observables for graviton
production, such as the $\gamma/Z$ angular and energy distributions,
are thus distinct from those of other new physics processes,
such as supersymmetric particle production, since the latter corresponds to
a fixed invisible particle mass.  The Standard Model
background transition $e^+e^-\to\nu\bar\nu\gamma$ also has different
characteristics, since it is a three-body process.

\begin{figure}[hp]
\centerline{
\psfig{figure=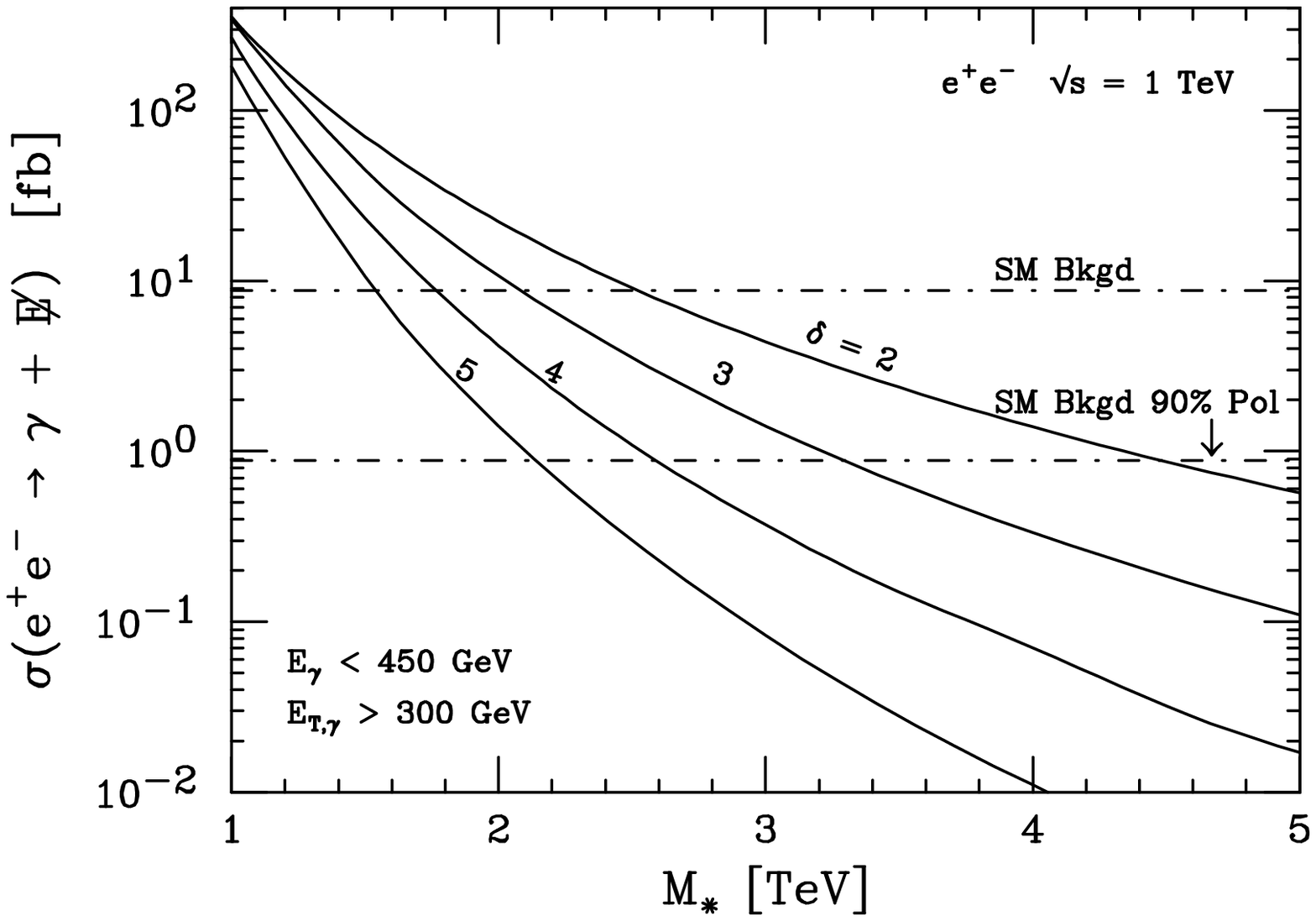,height=7cm,width=12cm,angle=0}}
\caption[*]{\label{emitt}
The cross section for $e^+e^-\to\gamma G_n$ for $\sqrt s=1$ TeV 
as a function of the fundamental Planck scale for various values of $\delta$ 
as indicated.  The cross sections for the Standard Model background, with
and without $90\%$ beam polarization, correspond to the horizontal lines
as labeled.  The signal and background are computed with the requirement
$E_\gamma<450$ GeV in order to eliminate the $\gamma Z\to \nu\bar\nu\gamma$
contribution to the background.  From \cite{grw}.}
 \vspace{.1in}  
\centerline{
\epsfig{figure=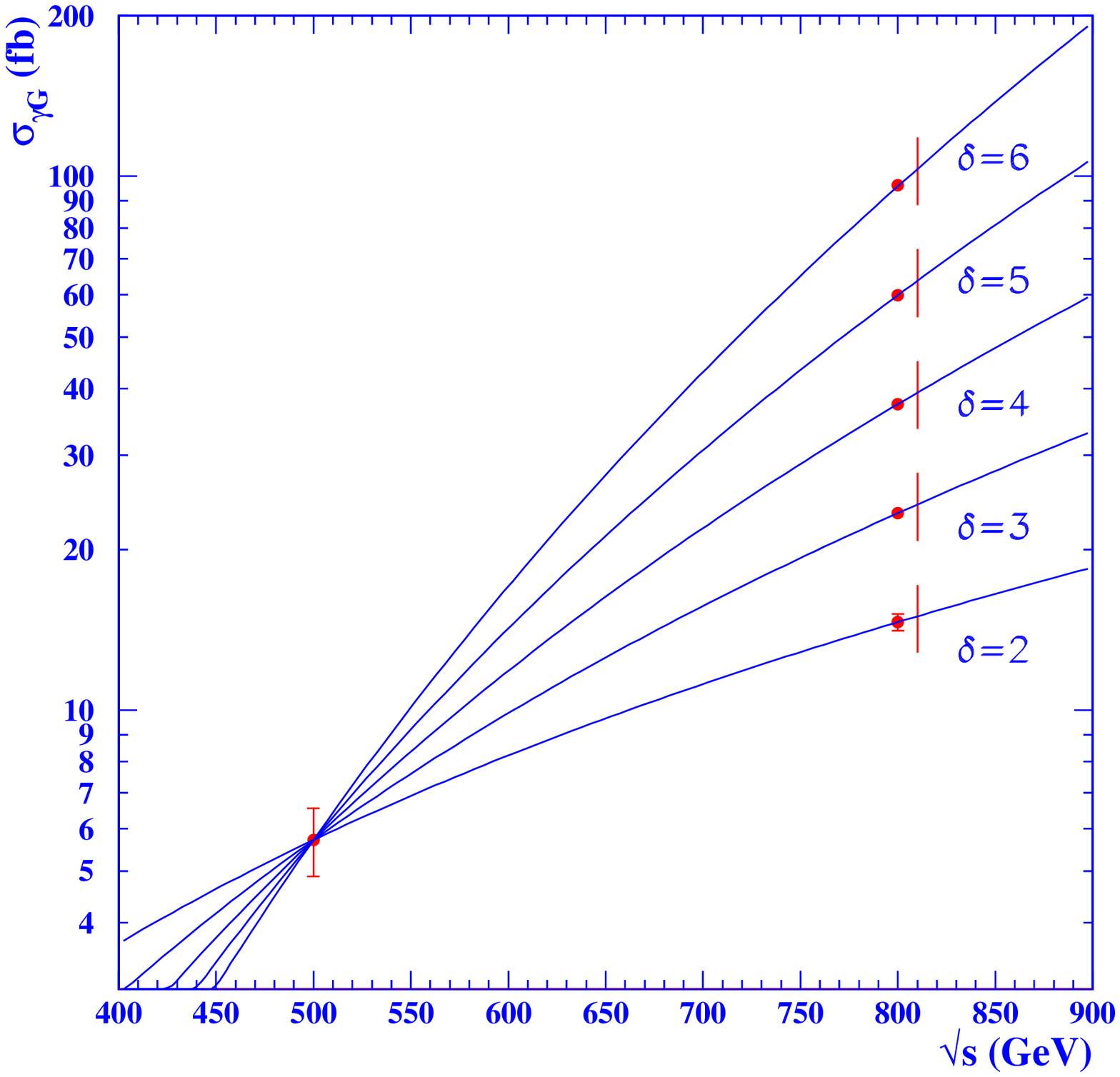,height=7cm,width=12cm,angle=0}}
\caption{\label{numdim}
The determination of $\delta$ from cross section measurements
of $e^+e^-\to\gamma G_n$ at $\sqrt s = 500$ and 800 GeV with 500
fb$^{-1}$ and 1 ab$^{-1}$, respectively,  taking $P_-=80\%$ and
$P_+=60\%$.  The 500 GeV cross section has been normalized for the
case $M_*=5$ TeV and $\delta = 2$.  From [82].}
\end{figure}

The cross section for $e^+e^-\to\gamma G_n$ as a function of the 
fundamental Planck scale is presented in Fig. \ref{emitt} for $\sqrt s= 1$
TeV.  The level of
Standard Model background is also shown, with and without electron beam 
polarization set at $90\%$.  We note that the signal (background) increases
(decreases) with increasing $\sqrt s$.  Details of the various distributions
associated with this process can be found in Cheung and Keung \cite{emit}.
The discovery reach from this process
has been computed in  \cite{teslatdr}, with
$\sqrt s=800$ GeV, 1000 fb$^{-1}$ of integrated luminosity, including various
beam polarizations  and kinematic acceptance cuts, ISR, and 
beamstrahlung.  The results are displayed in Table \ref{emit_tab}.  In this
table, we have also included the $95\%$ CL bounds obtained \cite{moriond_ed}
at LEP for $\sqrt s > 200$ GeV. 

The associated emission process at hadron colliders, $q\bar q\to g+G_n$,
results in a mono-jet signal.  In this case, the effective low-energy
theory breaks down for some regions of the parameter space, as the parton-level 
center-of-mass energy can exceed the value of $M_*$.  The experiment is then
sensitive to the new physics appearing above $M_*$.
An ATLAS simulation \cite{ian} of the missing 
transverse energy in signal and background events at the LHC with 100
fb$^{-1}$ results in the discovery range for the effective theory 
displayed in Table \ref{emit_tab}.  
The lower end of the range corresponds to values at which  the 
ultraviolet physics sets in and the effective theory fails, while the upper 
end represents the boundary where the signal is no longer observable above background.

\begin{table}
\centering
\begin{tabular}{|c|l|c|c|c|} \hline\hline
$e^+e^-\to\gamma+G_n$ & & 2 & 4 & 6 \\ \hline
LC & $P_{-,+}=0$ & 5.9 & 3.5 & 2.5 \\
LC & $P_{-}=0.8$ & 8.3 & 4.4 & 2.9 \\
LC  & $P_{-}=0.8$, $P_{+}=0.6$  & 10.4 & 5.1 & 3.3\\
LEP II & & 1.45 & 0.87 & 0.61 \\ \hline\hline
$ pp\to g+G_n$ & & 2 & 3 & 4\\ \hline
LHC  & & $4.0 - 8.9$ & $4.5-6.8$ & $5.0-5.8$ \\ \hline\hline
\end{tabular}
\caption{\label{emit_tab}
$95\%$ CL sensitivity to the fundamental Planck scale $M_*$ in TeV
for different values of $\delta$,
from the emission process for various polarization configurations and
different colliders as discussed in the text. $\sqrt s= 800$ GeV and
1 ab$^{-1}$ has been assumed for the LC and 100 fb$^{-1}$ for the LHC.}
\end{table}

If an emission signal is observed, one would like to determine the values of 
the fundamental parameters, $M_*$ and $\delta$.  In this case, measurement
of the cross section at a linear collider at two different values of
$\sqrt s$ can be used to determine $\delta$ \cite{teslatdr} and test the
consistency of the data with the hypothesis of large extra dimensions.
 This is displayed for a LC
in Fig. \ref{numdim}.

\begin{figure}[htbp]
\centerline{
\psfig{figure=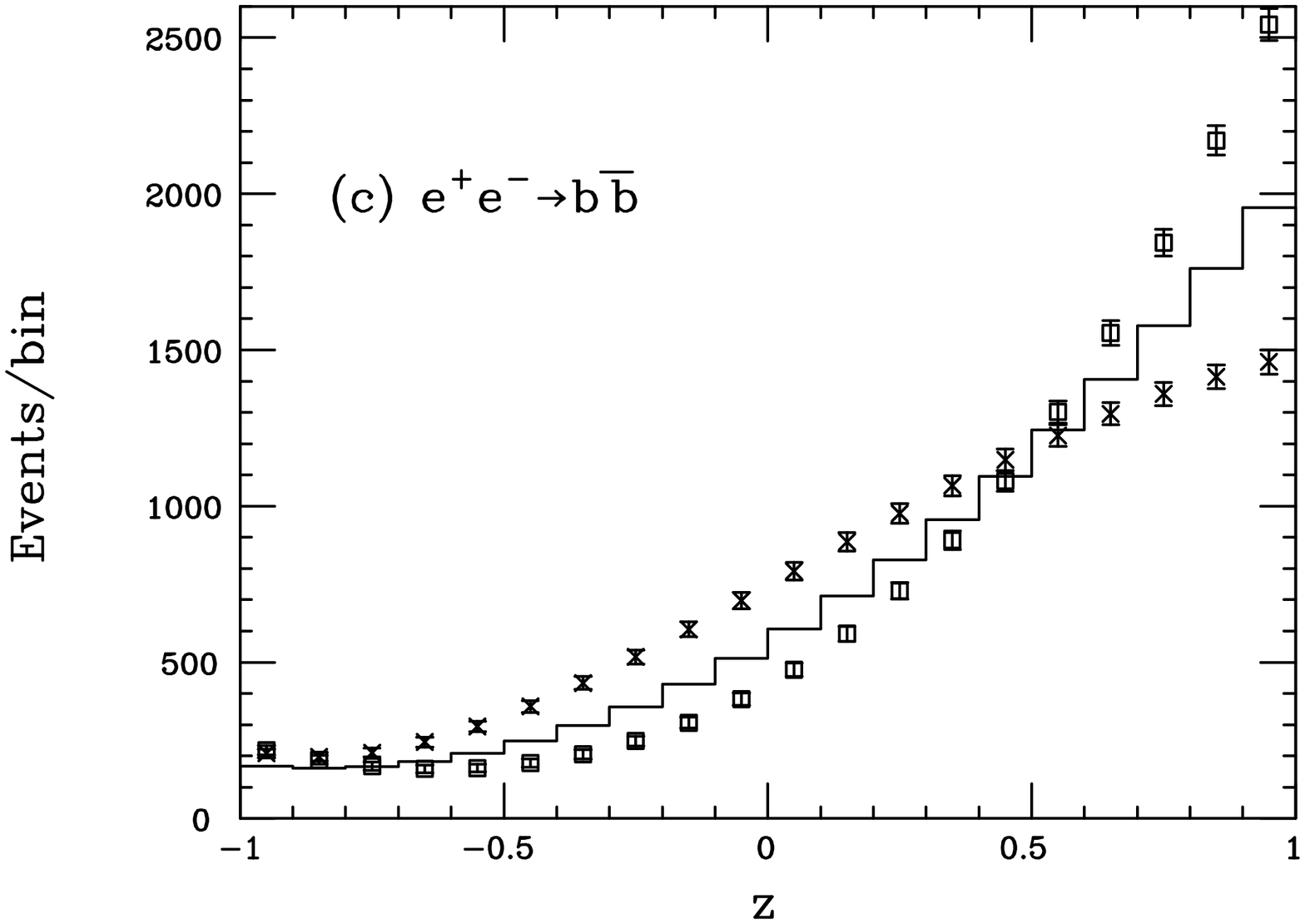,height=8.cm,width=8cm,angle=-90}
\hspace*{-5mm}
\psfig{figure=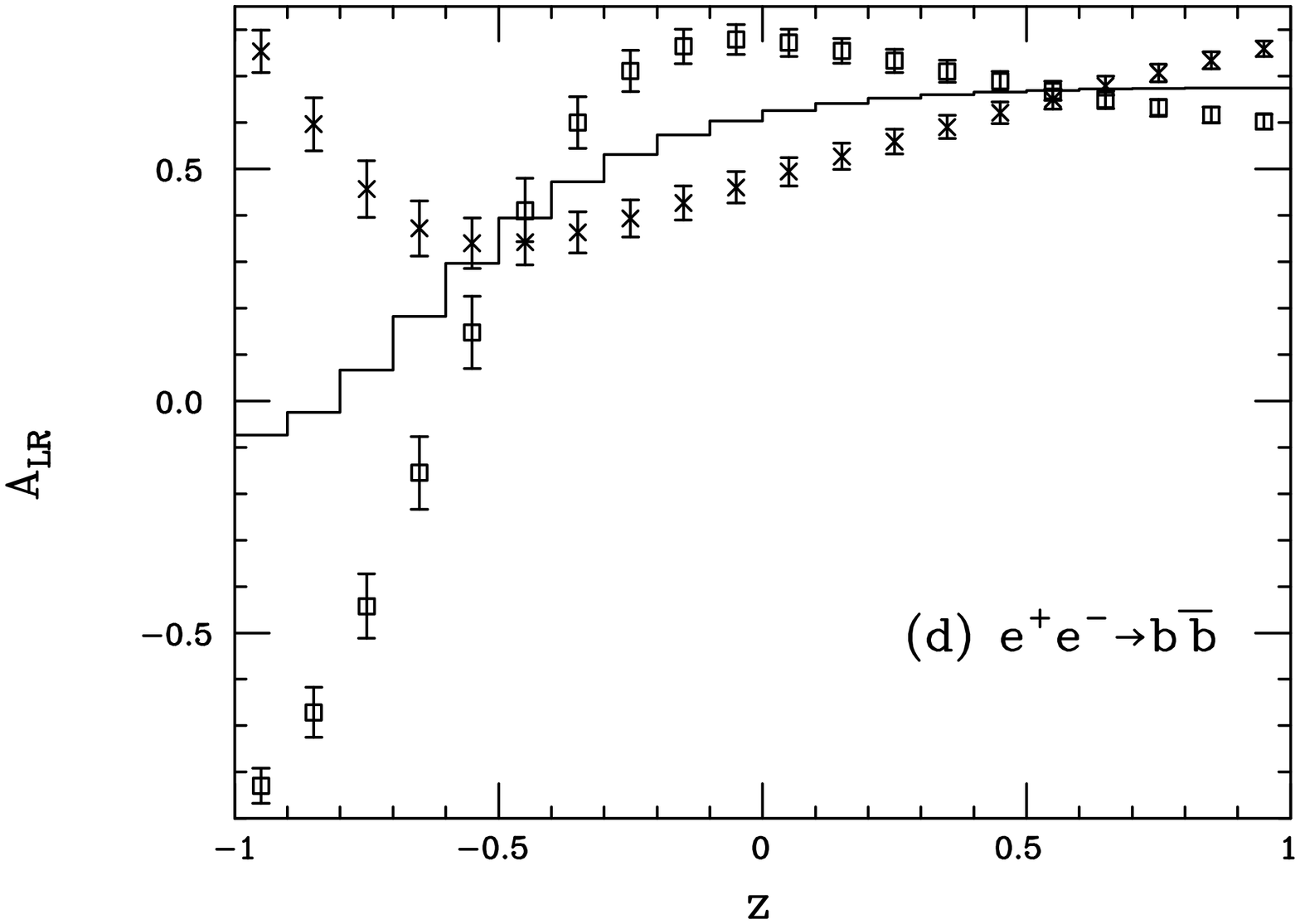,height=8.cm,width=8cm,angle=-90}}
\vspace*{-1cm}
\caption{\label{exchng}
Bin-integrated angular distribution and $z$-dependent 
($z=\cos\theta$) left-right
asymmetry for $e^+e^-\to b\bar b$ at $\sqrt s=500$ GeV.  The solid histogram 
represents the Standard Model while the `data' points are for $M_*=1.5$ with
$\lambda=\pm 1$.  The error bars indicate the statistics in each bin.}
\vspace{.1in}
\centerline{
\psfig{figure=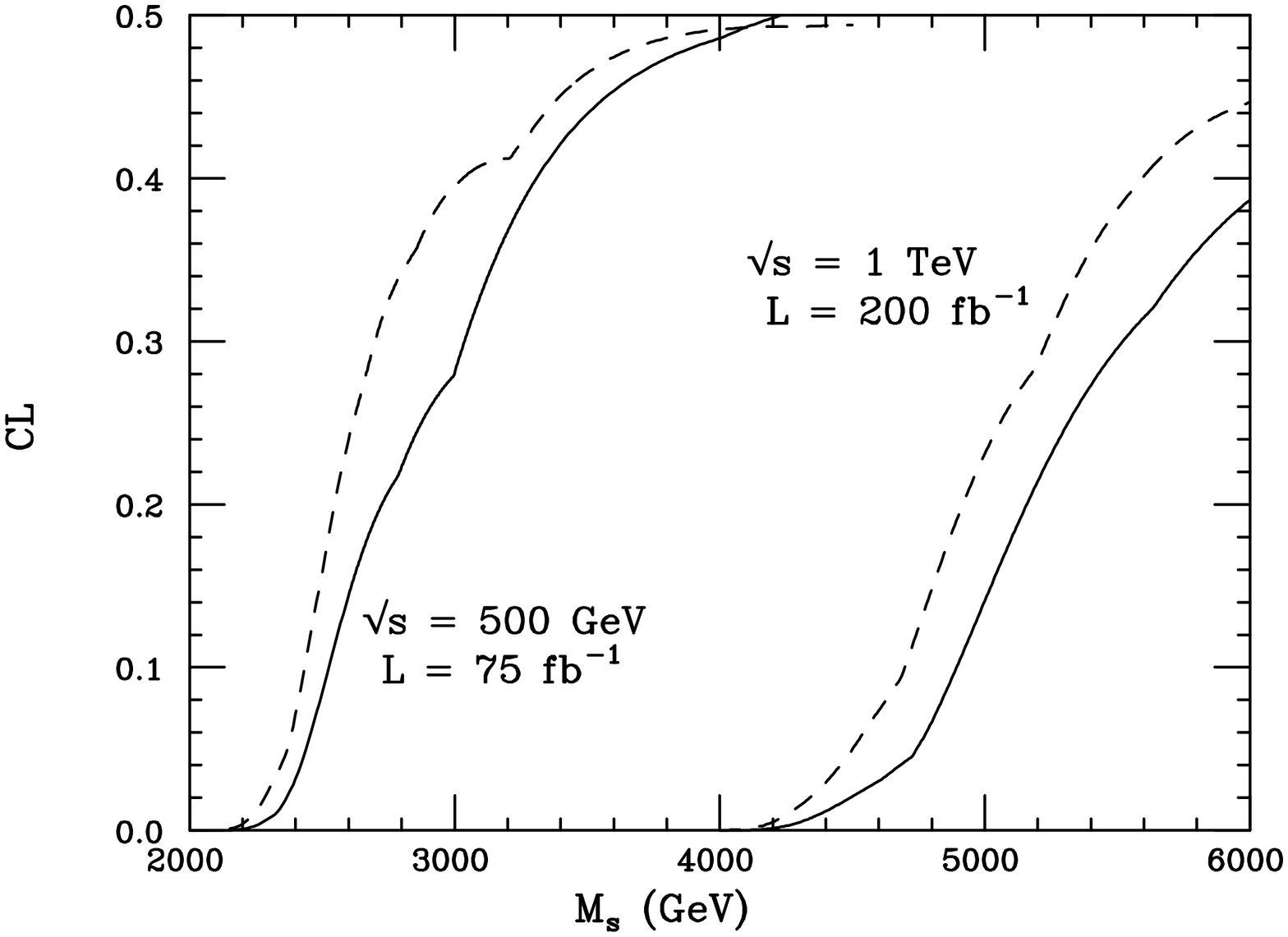,height=9cm,angle=-90}}
\caption{\label{clee}
The percentage confidence level as a function of $M_*$ for a
fit of spin-2 data under a spin-1 hypothesis.  The dashed and solid
curves correspond to the choice $\lambda=\pm 1$.} 
\end{figure}

The second class of collider signals for large extra dimensions is that of 
graviton exchange \cite{grw,hlz,exch} in $2\to 2$ scattering.  This 
leads to deviations in cross sections and asymmetries
in Standard Model processes  such as $e^+e^-\to f\bar f$, and may also
give rise to new production processes that are not otherwise 
present at tree-level, such as $e^+e^-\to hh,$ or $\tilde g\tilde g$.
The exchange amplitude is proportional to the sum over the
propagators for the graviton KK tower states which, as before, may be
converted to an integral over the density of states.  However, in this 
case the integral is divergent for $\delta>1$ and thus introduces a 
sensitivity to the unknown ultraviolet physics.  Several approaches have
been proposed to regulate this integral:  (i) a naive 
cut-off scheme \cite{grw,hlz,exch}, (ii) an exponential damping due to the
brane tension \cite{bando},
(iii) restrictions from unitarity \cite{eboli_99},
or (iv) the inclusion of full weakly coupled
TeV-scale string theory in the scattering process \cite{dudas}.
Here, we adopt the most model-independent approach, that of a naive
cut-off, and set the cut-off equal to $M_*/\lambda^{1/4}$, where $\lambda$
accounts for the effects of the unknown ultraviolet physics.
Assuming that the integral is dominated by
the lowest-dimensional local operator, which is dimension-8, this
results in a contact-type interaction limit for graviton
exchange, which can be described via
\begin{equation}
i{4\lambda\over M_*^4}T^{\mu\nu}T_{\mu\nu}\,,
\end{equation}
where $T^{\mu\nu}$ is the stress-energy tensor.
This is described in the matrix element for $s$-channel $2\to 2$ 
scattering by the replacement
\begin{equation}
{i^2\pi\over {M}_{Pl}^2}\, \sum^\infty_{n=1} 
{1\over s-m_n^2}\to {\lambda\over M_*^4}
\end{equation}
with corresponding substitutions for $t$- and $u$-channel scattering.
Here $m_n$ represents the mass of $G_n$, the $n^{th}$ graviton KK 
excitation.  This substitution is universal for any $2\to 2$ process.
The resulting angular distributions for fermion pair production
are quartic in $\cos\theta$ and thus provide a signal for
spin-2 exchange.  An illustration of this is given in Fig. \ref{exchng}
from \cite{exch}, which displays the unpolarized angular distribution as 
well as the angular dependence of the left-right asymmetry
in $e^+e^-\to b\bar b$, taking $M_*=3\sqrt s=1.5$ TeV and $\lambda=\pm 1$.  
The two sets of
data points correspond to the two choices of sign for $\lambda$, and
the error bars represent the statistics in each bin for an integrated
luminosity of 75 fb$^{-1}$.  Here, a $60\%$
$b$-tagging efficiency, $90\%$ electron beam polarization, 
$10^\circ$ angular cut, and ISR have been included.  The resulting
$95\%$ CL search reach with 500 fb$^{-1}$ of integrated
luminosity is given in Table \ref{tab_exch} from 
summing over the unpolarized and $A_{LR}$ angular distributions for
fermion ($e\,, \mu\,, \tau\,, c\,, b$\,, and $t$) final states.  For 
comparison, we also present the current bounds \cite{moriond_ed} from LEP II,
HERA, and the D\O\  Collaboration
at the Tevatron, as well as estimates for the LHC with 100
fb$^{-1}$ \cite{exch,miagkov}
and $\gamma\gamma$ colliders \cite{tgr_ed}.  Note that the $\gamma\gamma\to
WW$ process has the highest sensitivity to graviton exchange.  This is due
to the large $W$ pair cross section and the multitude of observables
that can be formed utilizing polarized beams and $W$ decays.

The ability of the LC to determine that a spin-2 exchange has taken
place in $e^+e^-\to f\bar f$ is demonstrated in Fig. \ref{clee} from
\cite{exch}.  Here, the confidence level of a fit of spin-2 exchange data
to a spin-1 exchange hypothesis is displayed; the quality of such a
fit is quite poor almost up to the $M_*$ discovery limit, indicating
that the spin-2 nature is discernable.

\begin{table}
\centering
\begin{tabular}{|c|c|c|c|} \hline\hline
 & & $\sqrt s$ (TeV)  & $M_*$ (TeV) \\ \hline
LEPII & $e^+e^-\to\ell^+\ell^-,\gamma\gamma,ZZ$ & 0.2 & 1.03-1.17 \\
LC & $e^+e^-\to f \bar f$ & 0.5 & 4.1 \\
LC & $e^+e^-\to f\bar f$    & 1.0 & 7.2 \\
LC & $\gamma\gamma\to WW$ & 1.0 & 13.0 \\
LC & $\gamma\gamma\to\gamma\gamma$ & 1.0 & 3.5 \\
LC & $e\gamma\to e\gamma$ & 1.0 & 8 \\
HERA &$ep\to e+$ jet & 0.314 & 0.81-0.93 \\
Tevatron Run I & $p\bar p\to\ell^+\ell^-,\gamma\gamma$ & 1.8 & 1.01-1.08 \\
LHC & $ pp\to \ell^+\ell^-$ & 14.0 & 7.5\\
LHC & $pp\to \gamma\gamma$ & 14.0 & 7.1 \\ \hline\hline
\end{tabular}
\caption{\label{tab_exch}
$95\%$ CL search reach for $M_*$ from graviton exchange in
various processes as indicated
and discussed in the text.  In the bounds from present data, a range is 
indicated to account for $\lambda=\pm 1$.}
\end{table}

\begin{figure}[t]
\centerline{
\psfig{figure=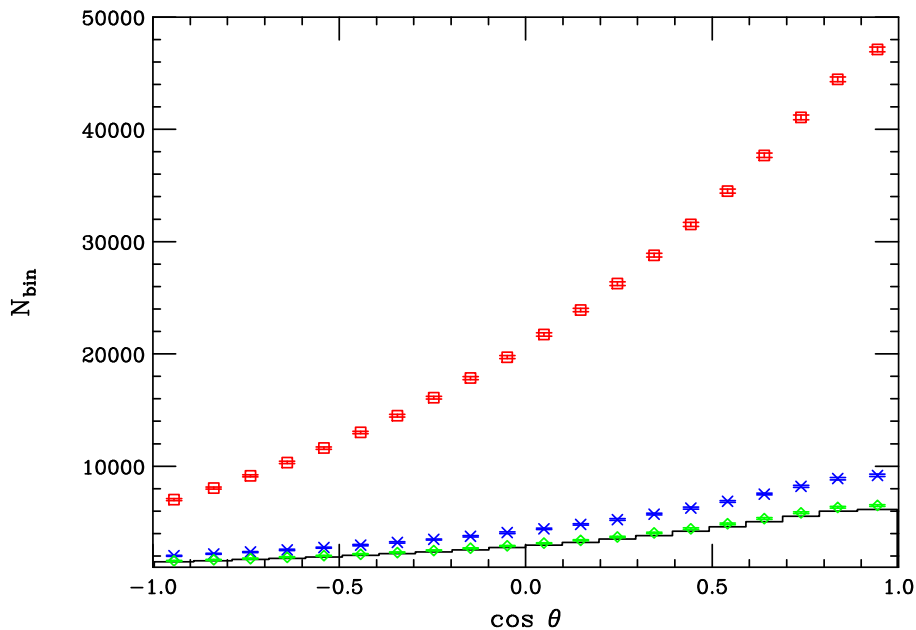,height=12cm,angle=90}}
\vspace*{0.1cm}
\caption{\label{boxb}
The number of events per bin in $e^-_{L,R}e^+\to
\tilde e_L^\mp\tilde e_R^\pm$ for $\sqrt s=500$ GeV with 500 fb$^{-1}$ of
integrated luminosity and $P_- = 80\%$.  The curves correspond to 
$M_*=1.5\,, 3\,, 6$ TeV
from top to bottom with the solid histogram representing the minimal
supersymmetric case.  The error bars correspond to the statistics in
each bin.  Here the values $m_{\tilde e_L}= 220$ GeV and
$m_{\tilde e_R}= 117$ GeV are assumed.}
\end{figure}

The scenario with large extra dimensions resolves the hierarchy 
problem without invoking supersymmetry.  However, if this mechanism
is embedded in a string theory, 
then supersymmetry may also be present at the weak scale.  A supersymmetric
bulk then results in a KK tower of gravitinos, in
addition to the KK gravitons.  In supersymmetric models that expect
a light gravitino, such as gauge-mediated supersymmetry breaking, the
gravitino KK tower can yield interesting phenomenological effects.  An
example of this is in the process $e^+e^-\to\tilde e^+\tilde e^-$, which
would now also receive contributions from  $t$-channel KK gravitino exchange
and $s$-channel KK graviton exchange.  This has been studied in \cite{darius},
which considered an $N=2$ supersymmetry in the bulk, and after compactifying
the gravitino sector, derived the KK gravitino couplings to $N=1$
supersymmetric matter on the brane.  The resulting dramatic effect on
selectron pair production is highlighted by the ability to select various
production channels via the use of electron beam polarization.  This is
displayed in Fig. \ref{boxb}, which shows the binned angular distribution
for $e^-_{L,R}e^+\to\tilde e_L^\mp e_R^\pm$ for various values of $M_*$; 
this choice of polarization isolates the $t$-channel neutralino and KK 
gravitino contributions.  The search reach for this process at $\sqrt s=
500$ GeV with $80\%$ beam polarization and 500 fb$^{-1}$ of integrated
luminosity is $M_* \sim 12$ TeV for the case $\delta=6$.

\subsection{Localized gravity}

We now turn to the scenario where the
hierarchy is generated by an exponential function of the 
compactification radius.  In its simplest form, gravity 
propagates in the bulk, while the Standard Model fields are constrained
to a 3-brane.  This model contains
a non-factorizable geometry embedded in a slice of 5-dimensional Anti-de~Sitter
space (AdS$_5$), which is a space of negative curvature.  Two 3-branes reside rigidly at 
fixed points at the boundaries of the AdS$_5$ slice, located at
$|\phi|=0,\pi$ where $\phi$ parameterizes the fifth dimension.  
The 5-dimensional Einstein 
equations permit a solution that preserves 4-d Poincar\' e
invariance with the metric
\begin{equation}
ds^2=e^{-2kr_c|\phi|}\eta_{\mu\nu}dx^\mu dx^\nu-r_c^2d\phi^2\,,
\end{equation}
where  $\pi r_c$ is the length of the fifth dimenion.
The exponential function, or warp factor, multiplying the usual 4-d
Minkowski term  curves space away from the branes.  The constant
 $k$ is the AdS$_5$ curvature scale,
which is of order the Planck scale and is determined by the bulk
cosmological constant.   The scale of
physical phenomena as realized by the 4-d flat metric transverse to
the fifth dimension is specified by the exponential
warp factor.  If the gravitational wavefunction is localized on the brane at
$\phi=0$ (called the `Planck brane'), then TeV scales can naturally be 
attained \cite{ars} on the 3-brane at $\phi=\pi$ (the `TeV brane', where the
Standard Model fields reside) if 
$kr_c\simeq$ 11--12.  The scale $\Lambda_\pi\equiv\overline{M}_{Pl}
e^{-kr_c\pi}\sim 1$~TeV, where $\overline M_{Pl}=M_{Pl}/\sqrt 8\pi$ is the
reduced Planck scale, then describes the scale of all physical processes
on the TeV-brane.  We note that it has been demonstrated \cite{gw} that this 
value of $kr_c$ can be stabilized  without fine tuning of 
parameters. 

Two parameters govern the 4-d phenomenology of this model,
$\Lambda_\pi$ and the ratio
$k/\overline M_{Pl}$.  Constraints on the curvature of the AdS$_5$ space
suggest that $k/\overline M_{Pl}\lsim 0.1$.  The
Feynman rules are obtained by a linear expansion of the flat
metric, including the warp factor.
After compactification, a KK tower of gravitons appears on the TeV-brane
and has masses $m_n=x_nk
e^{-kr_c\pi}=x_n\Lambda_\pi k/\overline M_{Pl}$ with the $x_n$ being
the roots of the first-order Bessel function, {\it i.e.}, $J_1(x_n)
=0$.  Note that the first excitation is naturally of order a few
hundred GeV and that the KK states are not evenly spaced.  The
interactions of the graviton KK tower with the Standard Model fields on
the TeV brane are \cite{dhr1} 
\begin{equation}
{\cal L}= - {1\over\overline M_{Pl}}T^{\mu\nu}(x)h^{(0)}_{\mu\nu}(x)
-{1\over\Lambda_\pi}T^{\mu\nu}(x)\sum^\infty_{n=1}h^{(n)}_{\mu\nu}(x)
\,,
\end{equation}
where $T^{\mu\nu}$ is the stress-energy tensor.
Note that the zero-mode decouples and that the couplings of the higher
states have inverse-TeV strength.  This results in a
strikingly different phenomenology from the case of large 
extra dimensions.  Here, the graviton KK tower states are directly produced 
as single resonances if kinematically allowed.

If the KK gravitons are too 
massive to be produced directly, their contributions to fermion
pair production may still be felt via virtual exchange.  In this
case, the uncertainties associated with the
introduction of a cut-off are avoided, since there is only one additional 
dimension and the KK states may be neatly summed.  The sensitivity \cite{dhr1} 
to $\Lambda_\pi$ at a linear collider for various values 
of $k/\overline M_{Pl}$ is listed in Table \ref{rscont} for 500 
fb$^{-1}$ of integrated luminosity.  For purposes of comparison, the 
corresponding reach at LEP II, Tevatron Run II, and the LHC is also displayed. 

\begin{figure}[t]
\centerline{
\psfig{figure=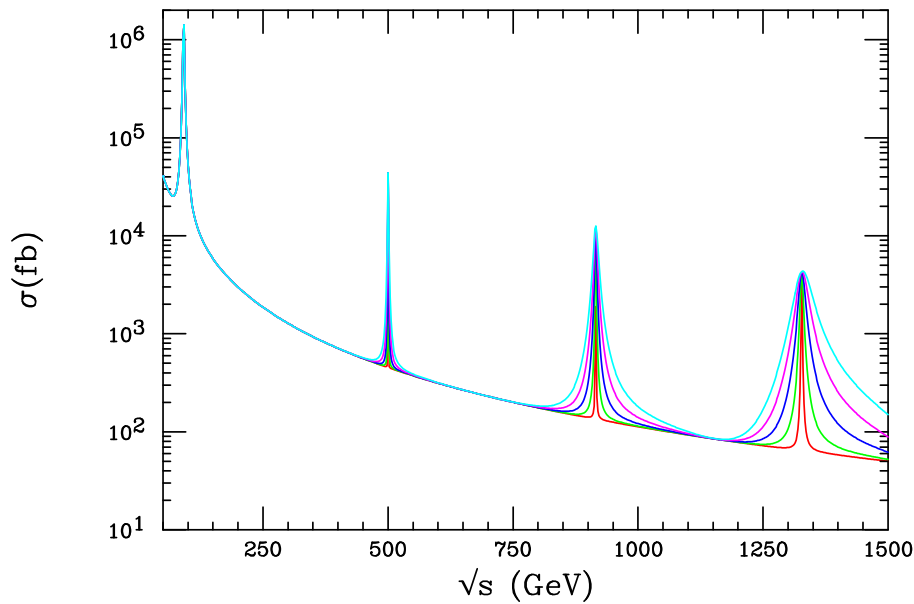,height=12cm,angle=90}}
\vspace*{0.1cm}
\caption{\label{kkspect}
The cross section for $e^+e^-\to\mu^+\mu^-$ including the
exchange of a KK tower of gravitons with $m_1=500$ GeV.  The curves 
correspond to $k/\overline m_{Pl}=$ in the range 0.01--0.05.}
\end{figure}

With sufficient center-of-mass energy the graviton KK states may be
produced as resonances.  To exhibit how this may appear at a linear 
collider, Fig. \ref{kkspect} displays the cross section for $e^+e^-\to
\mu^+\mu^-$ as a function of $\sqrt s$, assuming $m_1=500$ GeV and taking
$k/\overline M_{Pl}= $0.01--0.05.  The height of the third resonance is somewhat
reduced, because  the higher KK excitations decay to the lighter 
graviton states once it is kinematically allowed \cite{dr}.  In this case
one can study graviton self-couplings, and
higher-energy $e^+e^-$ colliders may become graviton factories!

Searches for the first graviton KK resonance in Drell-Yan and di-jet data
at the Tevatron already place non-trivial restrictions \cite{dhr1}
on the parameter space of this model, given roughly by $m_1\gsim$ 175,
550, 1100 GeV for $k/\overline M_{Pl} = 0.01\,, 0.1\,, 1.0$.  Precision
electroweak data extend \cite{dhr3} this search reach for smaller values of
$k$.  These constraints, taken together with the theoretical prejudices that
(i) $\Lambda_\pi\lsim 10 $ TeV, {\it i.e.}, the scale of physics on the 
TeV brane is not far above the electroweak scale and (ii) $k/\overline M_{Pl}
\lsim 0.1$ from the above-mentioned AdS$_5$ curvature considerations, 
result in a closed allowed region in the 2-dimensional parameter space,
which can be completely explored at the LHC \cite{dhr3,brits} via the 
Drell-Yan mechanism.

\begin{table}
\centering
\begin{tabular}{|c|c|c|c|} \hline\hline
 & \multicolumn{3}{c|}{$k/\overline {M_{Pl}}$ } \\ \hline
 & 0.01 & 0.1 & 1.0\\ \hline
LC $\sqrt s=0.5$ TeV & 20.0 & 5.0 & 1.5 \\
LC $\sqrt s=1.0$ TeV & 40.0& 10.0 & 3.0 \\
LEP II & 4.0 & 1.5 & 0.4 \\
Tevatron Run II & 5.0 & 1.5 & 0.5 \\
LHC & 20.0 & 7.0 & 3.0 \\ \hline\hline
\end{tabular}
\caption{\label{rscont}
$95\%$ CL search reach for $\Lambda_\pi$ in TeV in the contact 
interaction
regime taking 500, 2.5, 2, and 100 fb$^{-1}$ of integrated luminosity at
the LC, LEP II, Tevatron, and LHC, respectively.  From \cite{dhr1}.}
\end{table}

Lastly, we note that if the Standard Model fields are also allowed to
propagate in the bulk \cite{dhr3,bulkph}, the phenomenology can 
be markedly different, and is highly dependent 
on the value of the 5-dimensional fermion mass.
For various phenomenological 
reasons, it is least problematic to keep the Higgs field on the
TeV brane \cite{bulkph}.  As a first step, one can 
study the effect of placing the Standard Model gauge
fields in the bulk and keeping the fermions 
on the TeV-brane.  In this case, the
fermions on the wall couple to the KK gauge fields 
 a factor of $\sqrt {2 k r_c \pi} \sim 9$ times more strongly than they
couple to the 
($\gamma, g, W^{\pm}, Z$).  In this case, precision electroweak data 
place strong constraints, requiring that the lightest KK
gauge boson have a mass greater than about 25~TeV.  This value pushes 
the scale on the TeV-brane above 100  TeV, making 
this scenario disfavored in the context of the hierarchy problem.

This bound can be relaxed if the fermions 
also reside in the bulk \cite{bulkph}.  By introducing bulk fermion 5-d masses $m_5$, 
the couplings of the fermion zero 
modes ({\it i.e.}, the Standard Model fermions) to various KK fields
become a function of the bulk mass parameter 
$\nu \equiv m_5/k$.  The parameter $\nu$ 
controls the shape of the fermion zero-mode wavefunction, with
negative (positive) values of $\nu$ serving to localize the wavefunction
near the Planck brane (TeV brane).  Constraints from avoiding flavor-changing
neutral currents, Yukawa coupling blow-up, and the
generation of a new hierarchy result in a rather narrow allowed range of $\nu$.
For some values of $\nu$ in this range, the fermionic couplings of
the KK graviton states essentially vanish, and hence the graviton production
mechanisms discussed above are no longer viable.  In this case, the gravitons
retain a small coupling to the Standard Model gauge bosons, and the most
promising production mechanism \cite{dhr3} is at a photon collider via 
$\gamma\gamma\to G_n\to hh$, with $h$ being the Higgs boson.

\subsection{TeV-scale extra dimensions}

TeV$^{-1}$-sized extra dimensions can naturally arise in some string theory
models \cite {ant}, and in this case, the Standard Model fields may
feel their effects.  
The physics of models with KK excitations of the Standard Model gauge bosons 
arising from TeV-scale extra dimensions has been discussed for some time 
 \cite{antoniadis}.  
The various models in this class of theories 
differ in detail in two regards: ($i$) the placement of the Higgs field(s) 
in the bulk or on the wall(s), and ($ii$) the treatment of the fermion fields. 

If Higgs fields propagate in the bulk, the expectation value of the zero-mode field generates 
electroweak symmetry breaking.  In this case, there is no mixing among the 
various gauge boson KK modes. Thus the KK mass matrix is diagonal, 
with the masses of the excitations given by $[M_0^2+\vec n\cdot \vec n 
M_c^2]^{1/2}$, where $M_0$ is the zero-mode mass, $M_c$ is 
the compactification mass scale and $\vec n$ is a set of integers 
labeling the excitation state. However, if the 
Higgs is a wall field, its expectation value induces off-diagonal elements in the mass matrix 
and thus a mixing  among the gauge KK states.  In this case the mass matrix 
needs to be diagonalized to determine the masses and couplings of the gauge KK 
states.  It is also possible to imagine a more generalized mixed 
scenario with two Higgs fields, one residing in the bulk and one on the wall, 
that 
share the SM symmetry breaking. Clearly, the detailed phenomenology of these possibilities 
will be quite different.  For example, a  small mixing of the gauge KK states may show 
up in precision measurements when $W$ and $Z$ properties are compared with 
Standard Model expectations.

An even more diverse situation arises when one considers the placement of the
Standard Model fermions within the extra dimensions.  There are 
essentially three possibilities: 

$(a)$ The fermions are constrained to 3-branes located at
fixed points. This is the most common 
situation discussed in the literature \cite {bunch} and in this case the 
fermions are not directly affected by the extra dimensions. For models in this 
class, global fits to precision electroweak data place strong lower 
bounds on the value of $M_c$, which corresponds to the mass of the first
gauge KK excitation. Following the analysis of Rizzo and 
Wells \cite {bunch} and employing the most recent data
\cite {moriond_ew}, 
one finds that $M_c>4.4$ TeV when the Higgs field is on the wall;
the bound is 4.6~TeV when the Higgs field is in the bulk.
Such a large mass for gauge KK states is beyond the direct reach of a LC, but 
the KK states can be directly produced as resonances at the LHC in the Drell-Yan channel 
provided that $M_c \lsim 6$ TeV.  This reach at the LHC may be
extended by a TeV or so \cite{leshouches} by examination of the Drell-Yan
line shape at high lepton-pair invariant mass.
However, the LC can indirectly observe the existence of heavy gauge KK states 
via their $s$-channel exchanges 
in the contact interaction limit.  Combining the results from various fermion 
final states in $e^+e^-\to f\bar f$ gives the $95\%$ CL search reach
displayed in Table \ref{tab_tev}.

\begin{table}[htb]
\centering
\begin{tabular}{|c|c|} \hline\hline
 & $M_c$ Reach (TeV) \\ \hline
Tevatron Run II 2 fb$^{-1}$ & 1.1 \\
LHC 100 fb$^{-1}$ & 6.3 ($\sim$ 7.5) \\
LEP II & 3.1 \\
LC $\sqrt s=0.5$ TeV 500 fb$^{-1}$ & 13.0 \\
LC $\sqrt s=1.0$ TeV 500 fb$^{-1}$ & 23.0 \\
LC $\sqrt s=1.5$ TeV 500 fb$^{-1}$ & 31.0 \\ \hline\hline
\end{tabular}
\caption{\label{tab_tev}
$95\%$ CL search reach for the mass of the first KK gauge boson
excitation.  From Rizzo and Wells \cite{bunch}.
The LHC reach is via direct observation of a
resonance, while the LC sensitivities are from indirect effects
as in the case of a search for a new neutral gauge boson.  The
number in parentheses for the LHC is an estimate of the extension
of the complete search reach including indirect effects from contact
interactions.}
\end{table}

If a $\gamma^{(1)}/Z^{(1)}$ KK resonance is observed at the LHC, a $\sqrt 
s = 500$ GeV linear collider can distinguish this state from a new neutral
gauge boson arising from an extended gauge sector by using the Bhabha
scattering channel.  If one attempts to fit the induced
deviations in the Bhabha cross section and polarized asymmetry by
varying the vector and axial-vector couplings of a 
hypothetical non-KK $Z'$, one finds \cite{tgr_muon} that the CL
of the fit is quite poor ($\lsim 10^{-3}$).  This demonstrates
that the assumption that the KK state is a $Z'$ is incorrect.
A separate fit assuming that the resonance is a KK state yields
a good fit.  At the LHC, it is currently unclear whether the
$\gamma^{(1)}/Z^{(1)}$ KK resonance can be distinguished from 
a $Z'$ in a model-independent manner. 

\begin{figure}[b!]
\centerline{
   \psfig{figure=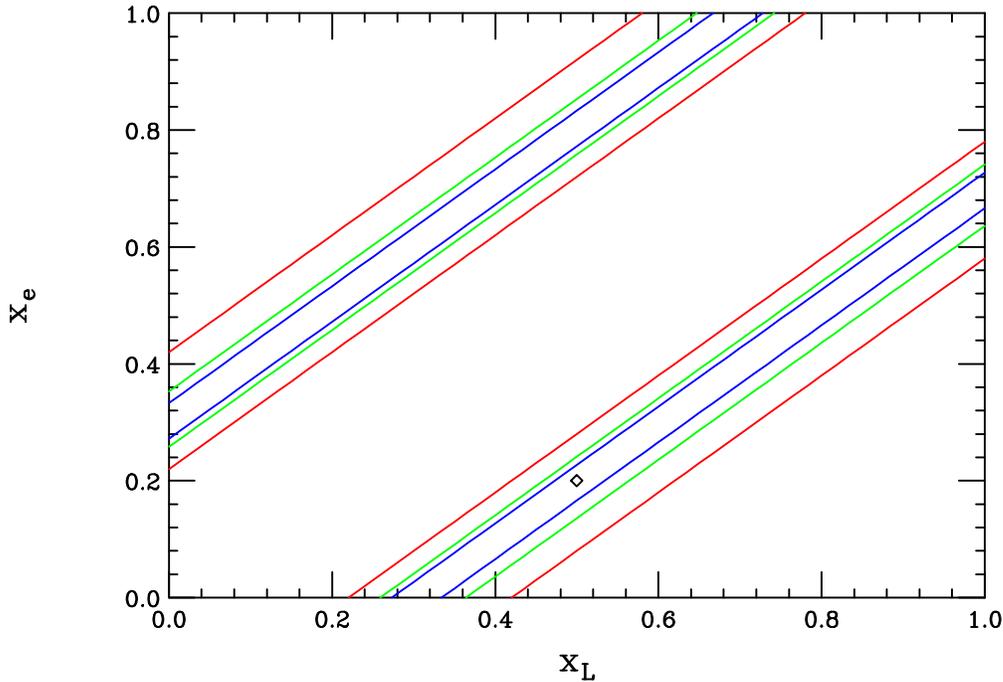,width=9cm,angle=90}}
\caption{\label{ferm_sep}
The ability of a LC to determine the separation in the extra
dimension of right- and left-handed electrons from Bhabha scattering.
The red, green, and blue (outer, middle, and inner) set of curves
correspond to $\sqrt s = $ 500, 1000, 1500 GeV, respectively, with
500~fb$^{-1}$ assumed for each energy.  This case
assumes $M_c=4 $ TeV and that the location of the right- (left-)-handed
electron, $x_{e(L)}$, is given by a Gaussian centered at 
$0.2\ (0.5)\cdot 2\pi r_c$.} 
\end{figure}

($b$) The Standard Model fermions are localized at specific 
points, $x_i$, in the extra TeV dimension, which are not 
necessarily at the orbifold fixed points. Here, the zero- 
and excited-mode fermions obtain narrow Gaussian-like wave functions in the 
extra dimensions with a width $\sigma$ much smaller than the compactification 
scale, {\it i.e.}, $(\sigma/\pi r_c)^2 \ll 1$. 
The placement of SM fermions at different locations and the narrowness of 
their wavefunctions can then suppress \cite{nam} the rates for a number of 
dangerous processes such as proton decay. For the lighter gauge KK modes 
(small values of $n$), the width of the fermion wavefunction centered 
at a given point cannot be resolved, so that the wavefunction 
appears similar to a delta function. Thus the 
coupling of the fermion to these gauge KK states is determined by the 
value of the gauge KK wavefunction evaluated at 
that point. However, when $n\sigma/ \pi r_c$ grows to order unity or 
larger, the KK gauge field can resolve the finite size of the fermion 
wavefunction and the coupling of the fermion becomes exponentially damped.  This
decouples the heavy gauge KK states, providing a means of rendering 
sums over KK towers of gauge bosons  finite in the case of two or more extra 
dimensions \cite {wow}. An analysis of precision electroweak data in this 
case shows that $M_c$ is typically found to be $\geq 3-4$ TeV. Depending upon 
the properties of the compactification manifold, measurements at colliders may 
probe the distance in the extra dimensions 
between two fermions, $|x_i-x_j|$, in $2\to 2$
scattering.  For example, in this case Bhabha scattering can probe the 
distance between the left- and right-handed electrons, as illustrated in
Fig.~\ref{ferm_sep}.  A study of the
cartography of the localized fermions at linear colliders has been
performed in \cite{tgr_cart}.  At very 
large energies, the cross 
section for the polarized version of 
this process will tend rapidly to zero since the two particles completely 
miss each other in the extra dimension \cite{martin}.

($c$) The fermions are fields in the bulk.  This possibility is known as
the `universal extra dimensions' scenario \cite{ACD}.
This case is different in that walls or branes 
are not present and hence momentum is conserved in the additional 
dimensions. The consequence of this is that 
KK number is conserved at all interaction vertices, hence
only pairs of KK gauge bosons couple to the zero-mode fermions. 
In this case, electroweak precision data 
as well as direct searches for KK states lead to a reduced lower bound of 
$M_c\simeq 0.4$ TeV.  Without further 
ingredients, this model may have trouble satisfying
cosmological constraints, since the lightest KK excitations are absolutely 
stable. This may be avoided  if there is any small breaking of
translation invariance in the extra dimensions.  Alternatively, one can imagine the gauge and 
fermion KK fields as confined to a brane of thickness TeV$^{-1}$, {\it i.e.},
a thick brane, embedded in a 
highe-dimensional space that includes gravity. In this case the higher-level
KK modes can 
decay down to the zero modes via graviton emission, but at a rate determined 
by the `form factor' of the brane \cite{deruch}. 
In either case an interesting phenomenology results.  The KK states are
produced in pairs at colliders and then either decay via one of
these two mechanisms or are 
long-lived and appear as tracks in a detector.

\section{Highly non-conventional theories and possible surprises}

So far in this chapter, we have delineated the potential of a linear
collider to explore the new physics that is present in set classes of 
established models.  However, as likely as not, when Nature finally
reveals her mysteries they will be full of surprises that lie
outside the realm of our limited imaginations. 

Along these lines,  we note that some of the most striking recent
developments have occured in string theory.  While it is
currently difficult to relate these theories to experiment, some of
their ingredients, when considered on their own, have interesting 
phenomenological consequences.  Here, we consider two such examples of
this top-down approach, as a demonstration of the potential of the LC 
to discover the unforseen.

\subsection{String resonances}

If the scenario with large extra dimensions discussed in a previous section
is embedded in a string theory, then stringy effects must also appear at
the TeV scale.  Hence, not only the gravitons, but also the Standard Model
fields must have an extended structure.  The exchange of string Regge 
excitations of Standard Model particles in $2\to 2$ scattering may appear 
as contact-like interactions with a strength that overwhelms the 
corresponding graviton exchange.  This is deduced from simple 
coupling-counting arguments.  Yang-Mills bosons live at the end of
open strings, while gravitons correspond to closed string states, which 
require an additional coupling constant factor at the amplitude level.
Hence the exchange of KK graviton states is suppressed by a factor of $g^2$
compared to the exchange of string Regge excitations.

\begin{figure}[htb]
\centerline{
\epsfig{figure=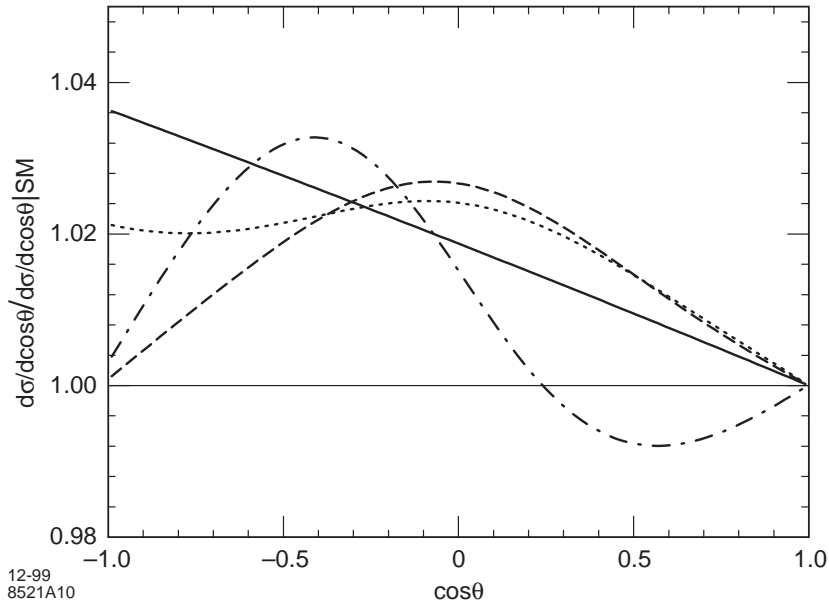,height=8cm,width=11cm,angle=0}}
\caption{\label{michael}
Comparison of deviations from the Standard Model prediction for
Bhabha scattering at 1 TeV due to corrections from higher-dimension
operators \cite{CPP}. The curves correspond to: string model with $M_{str}=3.1$ TeV 
(solid), KK graviton exchange with $M_*/\lambda^{1/4}=6.2$ TeV (dotted),
VV contact interactions with $\Lambda=88$ TeV (dashed), and AA contact
interactions with $\Lambda=62$ TeV (dot-dashed).}
\end{figure}

This has been examined in \cite{CPP}, where an illustrative string model
was assumed.  This model makes use of scattering amplitudes on the 
3-brane of weakly coupled type IIB string theory to describe a string
version of QED.  Electrons and photons then correspond to massless states
of open strings ending on the 3-brane and are characterized by the quantum
theory of fluctuations of an open string with specified boundary conditions.
Within the context of this model, Bhabha scattering
and pair annihilation receive contributions from the string Regge exchanges.
The differential cross section for these processes is modified by a form
factor,
\begin{equation}
{d\sigma\over d\cos\theta}=\left( {d\sigma\over d\cos\theta}\right)_{SM}
\left| {\Gamma (1-s/M^2_{str}) \Gamma (1-t/M^2_{str})\over
\Gamma (1-s/M^2_{str}-t/M^2_{str}) } \right|^2 \,,
\end{equation}
which essentially mirrors the original Veneziano result \cite{venice}.  Here,
$M_{str}$ represents the string scale and can be related to the fundamental
Planck scale in the large extra dimension scenario via $M_*/M_{str}=
\pi^{-1/8}\alpha^{-1/4}$.  Figure \ref{michael} displays the deviation
from Standard Model expectations to Bhabha scattering from these string
exchanges, and compares their effect to those arising from other types
of contact interactions.  The 95\% CL exclusion limits for $\sqrt s=1$ TeV
and 200~fb$^{-1}$ is $M_{str}>3.1$ TeV, which corresponds to 
$M_*/\lambda^{1/4}>9.3$ TeV.

\subsection{Non-commutative field theories}

Recent theoretical results have demonstrated that non-commutative
quantum field theories (NCQFT) naturally appear within the context 
of string theory and M-theory~\cite{seiwit}.  In this case, the usual 
$\delta$-dimensional space associated with commuting space-time 
coordinates is generalized to one that is non-commuting.  In such
a space, the conventional coordinates are represented by operators
that no longer commute,
\begin{equation}
[\hat X_\mu, \hat X_\nu]=i\theta_{\mu\nu}\equiv {i\over\Lambda^2_{NC}}
c_{\mu\nu} \,.
\end{equation}
Here, the effect has been parameterized in terms of an overall scale
$\Lambda_{NC}$, which characterizes the threshold where non-commutative
(NC) effects become important, and a real antisymmetric matrix 
$c_{\mu\nu}$, whose dimensionless elements are presumably of order
unity. The most likely value of $\Lambda_{NC}$ is near the
string scale or the true Planck scale, which could be as low as the  TeV scale.
The matrix $c_{\mu\nu}$ is related to
the Maxwell field-strength tensor $F_{\mu\nu}$ in a straightforward
fashion, since NCQFT arises
in string theory in the presence of background
electromagnetic fields.  The matrix $c_{\mu\nu}$ is identical in all reference
frames, defining a preferred NC direction in space, and hence Lorentz
invariance is violated at energies of order $\Lambda_{NC}$.  The usual
description of Lorentz violation needs to be modified in order to apply
to NCQFT;  present experiments only constrain such effects at
the few-TeV level \cite{carroll}.

\begin{figure}[htb]
\centerline{
\psfig{figure=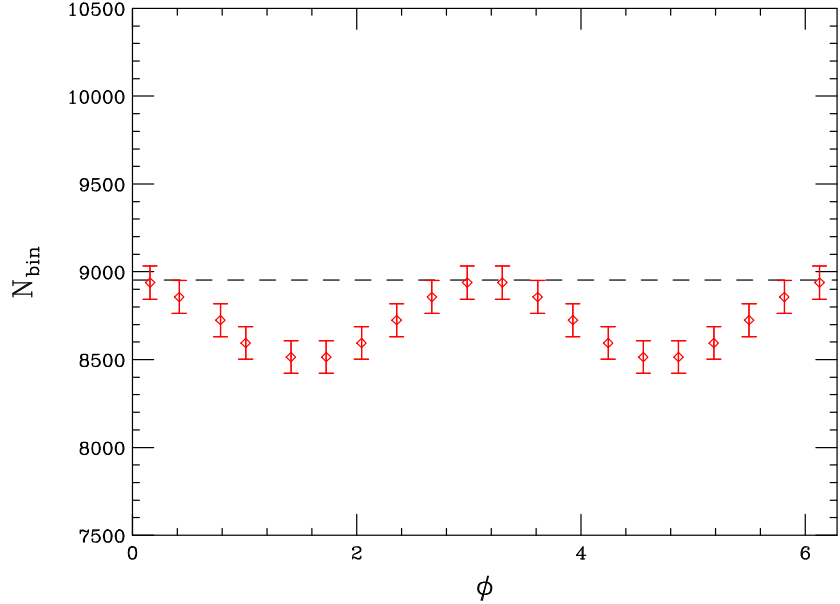,height=8cm,width=11cm,angle=0}}
\caption{\label{azim}
$\phi$ dependence of the 
$e^+ e^- \rightarrow \gamma \gamma$ cross section, taking 
$\Lambda_{NC} = {\sqrt s}= 500$ GeV a luminosity 
of $500 \, {\rm fb}^{-1}$.  A cut of 
$|\cos\theta| < 0.5$ has been employed.  
The dashed line corresponds to the SM 
expectations and the `data' points represent the NCQED results.}
\end{figure}

Caution must be exercised to preserve orderings of the products of
fields when formulating NCQFT.  This is accomplished with the
introduction of the star product, $\phi(\hat X)\phi(\hat X)=\phi(x) * \phi(x)
=\phi(x)e^{[ i\theta^{\mu\nu}\partial_\mu\partial_\nu/2]}\phi(x)$,
which absorbs the effect of the commutation relation via a series
of Fourier transforms.  The NC action for a quantum field theory
is thus obtained from the ordinary one by replacing the products of
fields by star products.  A striking consequence of this is that the
NC version of QED takes on a non-Abelian nature in that both 3-point
and 4-point photon couplings are generated.  In addition, all QED
vertices pick up additional phase factors that are dependent upon
the momenta flowing through the vertex.  We note that propagators,
however, are not modified since quadratic forms remain unchanged
under the properties of the star product.  NC effects thus produce striking
signatures in QED processes at a linear collider.  The modifications
to pair annihilation, Bhabha and M{\o}ller scattering, as well as
$\gamma\gamma\to\gamma\gamma$ have been studied in  \cite{jpr}.
Pair annihilation and $\gamma\gamma$ scattering both receive new
diagrammatic contributions due to the non-Abelian couplings, and
all four processes acquire a phase dependence due to the relative
interference of the vertex kinematic phases.  The lowest-order
correction to the Standard Model in these processes occurs at
dimension 8.  The most striking result is that a $\phi$
dependence is induced in $2\to 2$ scattering processes
because of the existence of the NC preferred direction in space-time.
This azimuthal dependence in pair annihilation is illustrated in 
Fig. \ref{azim} for the case where the NC direction is perpendicular
to the beam axis.  The results of  \cite{jpr} are summarized in
Table \ref{nonctab}, which displays the $95\%$ CL search reach for
the NC scale in these four reactions.  We see that these processes
are complementary in their ability to probe different structures
of non-commuting space-time.

\begin{table}
\centering
\begin{tabular}{|c|c|c|} \hline\hline
Process & Structure Probed  & Bound on $\Lambda_{NC}$   \\ \hline
$e^+e^-\to\gamma\gamma$   & Space-Time & $740-840$ GeV \\
M{\o}ller Scattering & Space-Space &  1700 GeV\\ 
Bhabha Scattering & Space-Time & 1050 GeV \\
$\gamma\gamma\to\gamma\gamma$ & Space-Time & $700-800$ GeV \\ 
   & Space-Space & 500 GeV \\ \hline\hline
\end{tabular}
\caption{\label{nonctab}
Summary of the $95\%$ CL search limits on the NC scale
$\Lambda_{NC}$ from the various processes considered above at
a 500 GeV  linear collider with an integrated luminosity
of 500 fb$^{-1}$.}
\end{table}

\begin{figure}[hb]  
\centerline{  
\epsfig{file=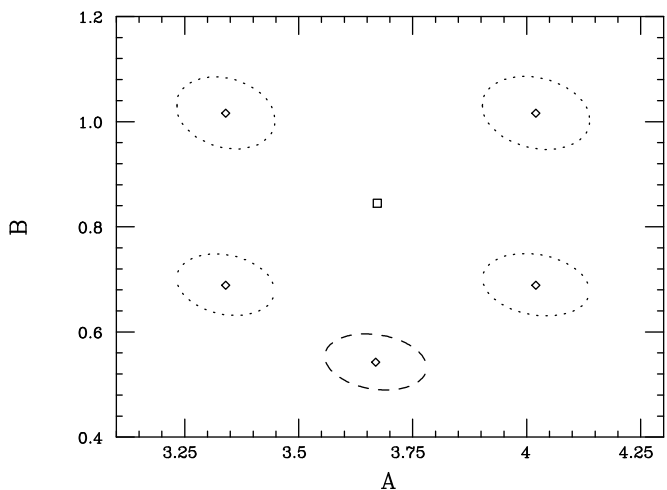,width=8.0cm, angle=-90}} 
\caption{Results of the fit with 
95\% CL contours circled around the fitted values.      
The box in the center corresponds to the Standard Model, 
the dotted ellipses represent
the fit to the four $Z'$ cases 
considered, and the dashed ellipse is for the case
of sneutrino exchange.  The fit was performed taking $\sqrt s=1$ TeV with 150
fb$^{-1}$.
\label{fig:spin_tomr}} 
\end{figure}   

\section{Determining the origin of new physics}  

As demonstrated in this chapter, some reactions at linear colliders may receive 
contributions from many different models. An example of this is $\eeff$,
in which indirect effects of compositeness, extended gauge sectors, extra
dimensions, string resonances, or supersymmetry may be revealed.  Once
a signal for new physics is found, the next step is to unravel the properties
associated with the new phenomena.  If the
mass spectrum of the new particles in these theories is kinematically accessible,
then their properties may be directly measured.  However, if these states
are too heavy, then we must explore their characteristics indirectly.  This
is feasible at a linear collider because of  the precision at which measurements
can be performed.  Here, we give a single example to illustrate our point,
namely, the ability of $e^+e^-$ 
colliders to provide unique information about the spin structure of new objects.
The angular distributions and polarization asymmetries associated with
$\eeff$ are  sensitive probes  of the 
spin of new particles.  An illustration of this was presented in 
Fig. \ref{clee}, which showed the extent to which spin-2 exchange in $\eeff$
is distinguishable from other new physics sources.
This figure showed that deviations induced by spin-2 
graviton exchanges can be distinguished from those due to 
lower spins, such as new vector bosons $Z'$ or a scalar neutrino in 
R-parity-violating models, up to the discovery limit.  In addition,  
discrimination between spin-1 and spin-0 particles  
at a LC was demonstrated \cite{tgr_1999fm}  by studying the angular
distributions induced by the 
exchange of a $Z'$ and of a scalar neutrino, $\tilde \nu$ in $\eeff$. 
A two-parameter fit of a trial 
distribution of the form   $\sim A(1+z)^2+B(1-z)^2$ was performed to the 
observables, with $A,B$  being parameters determined by the fit.  
In the case of the Standard Model and $Z'$,  
the fitted parameters $A,B$ are constant,  while,  in the case of $\tilde \nu$,  
the parameter $B$ depends on $z$. The results of the fit are displayed in 
Fig.~\ref{fig:spin_tomr}. The Standard Model values of $A$ and $B$ are shown 
in the center of the figure 
and are assumed to be known precisely.  The $Z'$ mass was set to 3 TeV and 
four different $Z'$ coupling  values were considered.  The $\tilde \nu$  was 
allowed to mediate  the reaction in both $s$- and $t$-channels. All five regions are 
statistically well separated from each other, and clearly distant from the 
Standard Model solution. 

\section{Conclusions}

In this chapter, we have discussed several classes of motivated models that 
contain
new phenomena,  and we have delineated the ability of a linear collider to explore them.
We have seen that the LHC and the linear collider have a comparable and 
complementary discovery potential.  In many cases, a signal for new physics 
will first be observed at the LHC, and
the linear collider will precisely determine its properties.  
While a 500 GeV linear collider has a large
discovery reach and potential to elucidate the underlying physics, every physics
scenario we have also explored benefits from an upgrade to higher energy.

However, our limited imagination 
does not span the full range of alternatives allowed by present data.
We thus must be prepared to discover the unexpected, which is best accomplished
by exploration of the energy frontier by both $e^+e^-$ and hadron colliders.

\emptyheads
\blankpage \thispagestyle{empty}

\setcounter{chapter}{5}
\fancyheads

\chapter{Top Quark Physics}
\fancyhead[RO]{Top Quark Physics}

\section{Introduction}

The linear collider, operating near the $t\bar t$ production threshold
and at higher energies, can carry out a comprehensive program of top
quark physics. Measurements at the threshold include the determination
of the top quark mass, $m_t$, and width, $\Gamma_t$, as well as the top
quark Yukawa coupling, $g_{tth}$. The quantities 
$m_t$ and $g_{tth}$ can also be
measured at higher energies, 
together with the couplings of the top quark to the
electroweak gauge bosons. In this chapter we present a brief summary of
our current understanding of top quark physics at a linear collider.

 The top is unique among the quarks
in that it decays before nonperturbative 
strong interaction effects can influence it.
Its large mass gives it stronger coupling to many proposed new
physics effects that try to explain electroweak symmetry breaking and/or
the origin of particle masses.  Thus,  precise measurement of
the parameters of the top quark would provide important insights into 
physics beyond the Standard Model.

\section{Physics in the threshold region}

\subsection{Introduction}
\label{sec:yak}

        One of the primary goals of a high-energy $e^+e^-$ linear collider
is the study of sharp features in the cross section for $e^+e^-$
annihilation to hadrons.  The \ttb\ threshold is an excellent example of such a structure.
The cross section for $e^+e^- \to t \bar t$ is expected to rise by an order of
magnitude with only a 5 GeV change in center-of-mass energy around 350 GeV.
Careful study of this \ttb\ threshold structure can precisely measure many
parameters of the top quark, including its mass and
width, and the top quark Yukawa coupling.
In this section we briefly summarize the current status of 
$t\bar t$ threshold studies. More comprehensive discussions can be
found in \cite{Review,Matsui,Frey}. 

\subsection{QCD dynamics and cross section}
It is well known that, because of the large top quark width ($\Gamma_t 
\approx 1.4\GeV \gg \Lambda_{QCD}$), 
a top-antitop pair cannot form narrow toponium resonances.  
Instead, the 
cross section is expected to have a smooth 
line-shape showing only a moderate $1S$ peak.  The dynamics of the top
quark in the threshold region is described by perturbative QCD. The
top quark width serves as an infrared cutoff. As a result, 
nonperturbative QCD effects (as measured, for example, by the influence
of the  gluon condensate) are 
small~\cite{FadinYakovlev}, allowing us, in principle at least,
 to calculate the cross 
section from QCD with high accuracy.

\begin{figure}
\centerline{\resizebox{0.47\textwidth}{!}{\includegraphics{top/oleg_orange2.epsi}}
\hfill%
\resizebox{0.47\textwidth}{!}{\includegraphics{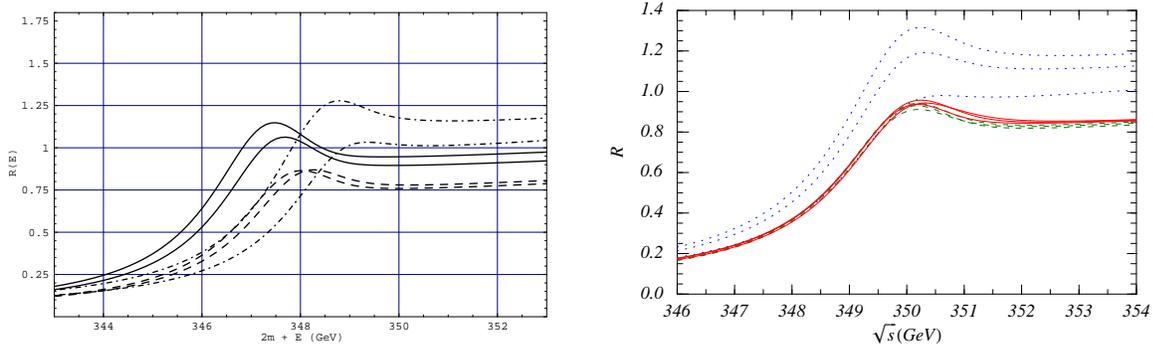}}}
\caption[*]{\label{bfig1} 
The normalized cross section $R_t = \sigma(e^+e^-\to t\bar t)/\sigma(e^+e^-
\to\mu^+\mu^-)$ as a function of  $\sqrt{s}$, computed in QCD perturbation
theory at various levels. These are theoretical curves that do not include
initial state radiation, beamstrahlung, or beam energy spread.
({\bf Left:}) The normalized cross section
computed with the pole mass $m^{\rm pole}_t =175$~GeV, at  LO (dashed-dotted lines),
NLO (dashed lines), and  NNLO (solid lines).   
 Each pair of the curves corresponds to the two different 
soft normalization scales $\mu=30$~GeV (upper curve) and 
$\mu=60$~GeV (lower curve). ({\bf Right:}) The normalized cross section
computed with the $1S$ mass $m_t^{1S}=175$~GeV, 
at LL order (dotted), NLL order (dashed) and NNLL order in QCD (solid).
 The calculation includes the summation of logarithms 
of the top quark velocity, and at each order curves are shown for 
$\nu=0.15, 0.2, 0.4$, where $\nu$ is the so-called subtraction velocity.}
\end{figure}

The convergence of QCD perturbation theory in the threshold region depends
on the quark mass definition used.  The simplest definition of $m_t$ is
the position of the pole in the top quark propagator.  This `pole mass' is 
similar to the kinematic mass observed in top quark pair production above
threshold, and similar to the mass definition used by the CDF and D\O\
experiments in the original papers on the top quark 
discovery \cite{top-CDF,top-DO}.  Unfortunately, with this choice of the
mass definition,  the NNLO corrections are
uncomfortably large~\cite{Review} and shift the $1S$ peak by about 
$0.5\GeV$, spoiling the possibility to extract the
top quark mass with high accuracy.  The threshold cross sections
computed at successive order in QCD are shown in the left-hand 
graph in  Fig.~\ref{bfig1}. 
The instability of this perturbation series is caused by the
fact that the pole mass has a renormalon ambiguity, that is, it obtains
an additive correction from nonperturbative QCD effects.

To remove this difficulty, one can use a different mass definition that 
refers only to short-distance QCD physics.  For example, a possible definition
of the mass,  called the $1S$ mass, 
 is one-half of the  mass of the lowest toponium 
bound state computed in the hypothetical 
limit of zero top quark width~\cite{ht}.  Three other mass 
definitions have been
considered in the literature.
The $\overline{\rm PS}$ mass~\cite{YakovlevGroote} is defined via the
top quark self-energy.  The $LS$ (`low scale') mass is given
in terms of perturbative evaluations of matrix elements of operators in
the heavy quark effective theory that describe the difference between
the pole mass and a fictitious $T$ meson mass~\cite{bigi}. 
Finally, the PS (`potential-subtracted') mass is defined by 
\begin{eqnarray}
m_t^{PS}(\mu)=m_t^{\rm pole}+\frac{1}{2}\int\limits_{|k|<\mu}\frac{d^3k}{(2\pi)^3}
V_C(k)=m_t^{\rm pole}-\frac{4}{3}\frac{\alpha_s}{\pi}\mu+...
\end{eqnarray}
where $\mu$ is the soft renormalization scale. All of these mass
definitions, collectively called `threshold masses' have the property
that they are free of the ${\cal O}(\Lambda_{QCD})$ renormalon 
ambiguity~\cite{Beneke,Scot}.  These masses also have the property 
that they are connected to the $\bar {MS}$ top quark mass by a 
convergent QCD perturbation series.

The position of the $1S$ peak becomes 
much more stable at higher orders of QCD if threshold masses are used.
  The shifts from order
to order are less than  100 MeV. However, a large
theoretical normalization uncertainty of about 10\% remains. The
normalization uncertainty can be reduced to a few percent by
resumming terms logarithmic in the top velocity.  The convergence for
the $1S$ mass definition is shown in the right-hand graph of 
Fig.~\ref{bfig1}~\cite{Stewart}. Simultaneous 
accurate measurements of the top mass and other quantities thus appear
feasible, as discussed further below.

\subsection{Top width}
 The scan of the  $t\bar t$ threshold will allow a direct
measurement of the top quark width, $\Gamma_t$. 
 The cross section at the $1S$ quarkonium bound state
energy is proportional to $1/\Gamma_t$.  
Realistic studies, which include initial state radiation and other
effects, show that $\Gamma_t$ can be measured with an experimental 
precision of
a few percent~\cite{Matsui}, now that higher-order QCD corrections
appear to be under control~\cite{Stewart}. 

$\Gamma_t$ can also be measured using the
forward-backward asymmetry~\cite{Murayama}. 
The $t\bar t$ vector coupling to $\gamma$ and $Z$ produces mainly S-wave
states, while the axial-vector coupling from the $Zt\bar t$ vertex
produces $t\bar t$  in a P state.  The top quark width causes the S and
P states to overlap and allows these states to interfere in the final 
angular distribution.  This produces a forward-backward asymmetry.
 Since the top quark width controls the amount of S-P overlap, the
asymmetry is sensitive to $\Gamma_t$. Realistic studies
are needed to better quantify the experimental sensitivity.

\subsection{Top quark Yukawa coupling}
In addition to the QCD potential, the $t\bar t$ pair interacts via a 
Yukawa potential associated with Higgs boson exchange
\begin{eqnarray}
V_{tth}= - \frac{g_{tth}^2}{4\pi}~\frac{e^{-m_hr}}{r},
\end{eqnarray}
where $m_h$ is the Higgs boson mass and $g_{tth}$ is the Yukawa coupling.
Therefore, top threshold measurements can also
be used to determine $g_{tth}$ if the Higgs boson is light. 
 A SM Higgs boson with a mass of
115~GeV  enhances the normalization of the cross section by 5--8\%
at energies near the threshold.  The theoretical
uncertainty of the cross section in this region is 2--3\% when the 
summation of logarithms of the top quark velocity is taken into
account~\cite{Stewart}. 
A precision measurement of the \ttb\ threshold cross
section thus will be sensitive to the top Yukawa coupling. If we fix all 
other parameters and assume $m_h=115\,{\rm GeV}$,
then varying the SM Yukawa coupling by $\pm 14\%$ gives a $\pm 2\%$
variation in the normalization of the cross section near the $1S$ 
peak~\cite{iain}. For larger values of $m_h$, the sensitivity to
$g_{tth}$ is expected to decrease. Again, realistic
experimental studies that make use of recent theoretical advances
in understanding the threshold cross section are needed.

\subsection{Experimental issues}

        The experimental situation of the \ttb\ threshold is
fairly well understood, and there has not been much progress since the 
experimental methods were reviewed at the 1999
Sitges meeting~\cite{sitgese}.  
It is expected that the top mass can be measured
with a statistical uncertainty of 40 MeV in a modest scan of 10 fb$^{-1}$, 
a small fraction of a year at typical design luminosities.  A longer
scan of about 100 fb$^{-1}$ can determine the top width to 2\%.
A key experimental issue for the threshold study is the measurement of the
\dlde\ spectrum, but many complementary methods have been proposed. The 
issues are similar to and less severe than the measurement of the
\dlde\ spectrum needed for a precision $W$ mass measurement from the 
$W^+W^-$ threshold, discussed in Chapter 8, Section 2.  The
limitations are likely to come from the uncertainty in 
machine-generated backgrounds and from the theoretical understanding of the Bhabha
cross section.  The impact of a precision top quark mass
measurement can be seen 
in~\cite{peskin} and~\cite{sumino}, which show how the current
knowledge of the top mass and precision electroweak measurements
limit the range of the Higgs mass and anomalous $W$ and $Z$ couplings
caused by new physics.

\section{Physics above the top threshold}

\subsection{Determination of the top quark--Higgs Yukawa coupling}

\subsubsection{Introduction}
If there is a light Higgs boson, this particle is likely to be discovered
at the Tevatron or the LHC.  The role of
 a high-energy $e^+e^-$ linear collider is then to test the connection
of this particle to the physics of mass generation by 
accurately
measuring its  mass, width, and  couplings to bosons and fermions. 
The top quark provides a unique opportunity to measure 
the Higgs Yukawa coupling to fermions through the process $e^+e^- \to 
t\bar{t}h$. For a light Higgs boson, the Higgs decays dominantly to $b\bar{b}$.
Assuming BR($t\to W b) = 100\%$, this
leads to multi-jet event topologies involving 4 $b$-jets in the final state.
Therefore, one of the crucial experimental aspects will be flavor 
tagging.

\subsubsection{Basic scenario}

The rate for $e^+e^- \to t\bar{t}h$ has been calculated to 
${\cal O}(\alpha_s)$ and  is less than 1 fb at $\sqrt{s}=500$~GeV.
The total cross section decreases at low $\sqrt{s}$ because of limited phase space and
approaches a constant at high $\sqrt{s}$.  The maximum of the
cross section (for a 100--150~GeV Higgs boson) occurs 
around $\sqrt{s} \simeq 700$--800~GeV.

Since the Yukawa coupling is determined from the cross section measurement,
it is straightforward to estimate the  statistical and some systematic
uncertainties on $g_{tth}$ for a  selection with efficiency $\epsilon$ 
and purity $\rho$, with an integrated luminosity $L$:
\begin{eqnarray}
 \biggl (\frac{\Delta g_{tth}}{g_{tth}} \biggr)_{\rm stat}
 & =& 
\frac{1}{S_{\rm stat}(g_{tth}^2)\sqrt{\epsilon \rho L}}
, \\
 \biggl (\frac{\Delta g_{tth}}{g_{tth}} \biggr)_{\rm syst}
 & = & 
\frac{1}{S_{\rm syst}(g_{tth}^2)} \left [ 
\frac{1-\rho}{\rho}\frac{\Delta \sigma_B^{\rm eff}}{\sigma_B^{\rm eff}} \oplus
\frac{1}{\rho} \frac{\Delta L}{L} \oplus \frac{\Delta 
\epsilon}{\epsilon}\right ] , 
\label{gttherrors}
\end{eqnarray}
\noindent where $(\Delta g_{tth}/g_{tth})_{\rm syst}$ accounts for the 
uncertainties in the effective background cross-section (after
selection), the integrated luminosity
and the selection signal efficiency. $S_{\rm stat}(g_{tth}^2)$ and 
$S_{\rm syst}(g_{tth}^2)$ are defined as:
\begin{equation}
 S_{\rm stat}(g_{tth}^2) = \frac{1}{\sqrt{\sigma_{tth}}}\,
 \Bigg\vert \frac{d\sigma_{tth}}{dg_{tth}^2} \Bigg\vert, \quad
S_{\rm syst}(g_{tth}^2) = \frac{1}{\sigma_{tth}}\,
\Bigg\vert \frac{d\sigma_{tth}}{dg_{tth}^2} \Bigg\vert.
\end{equation}

$S_{\rm stat}$ reaches a `plateau' for
$\sqrt{s} \geq 700$ GeV, whereas $S_{\rm syst}$ is essentially independent 
of $\sqrt{s}$. At $\sqrt{s}=800$ GeV, 
$S_{\rm stat}\simeq 3.09$ fb$^{1/2}$ and $S_{\rm syst}\simeq 1.92$.
Therefore, assuming $\epsilon=5\%$ and $\rho=50\%$, a statistical
precision of around 6.5\% could be achieved in $g_{tth}$
for $\sqrt{s} \geq 700$ GeV and $L = 1000$ fb$^{-1}$.
The case is considerably worse at $\sqrt{s}=500$~GeV where 
$S_{\rm stat}=0.9$~fb$^{1/2}$, leading to a statistical uncertainty
of $22\%$ on the Yukawa coupling measurement (with $\epsilon=5\%$
and $\rho=50\%$). The systematic uncertainty is
dominated by the uncertainty in the background normalization,
if one assumes that both the
signal selection efficiency and integrated luminosity can
be known at the 1\% level or better~\cite{jm}.

\subsubsection{Analysis}

We consider the process $e^+e^-\rightarrow t {\overline t} h \rightarrow
W^+W^- b {\overline b} b {\overline b}$ in both semileptonic and 
fully hadronic $W$ decay channels.
In spite of the apparently clean signature of both channels ($\geq 6$ jets
in the final state, with $\geq 4$  $b$-jets and multi-jet invariant 
mass constraints), the measurement has many difficulties. Among these
are the tiny signal  with backgrounds about 3 orders of magnitude larger, the
limitations of jet-clustering algorithms in properly reconstructing
multi-jets in the final state, and the  degradation of $b$-tagging performance 
due to hard gluon radiation and jet mixing.

\begin{figure}[t]
\centering
\epsfysize=2.6in
\leavevmode\epsffile{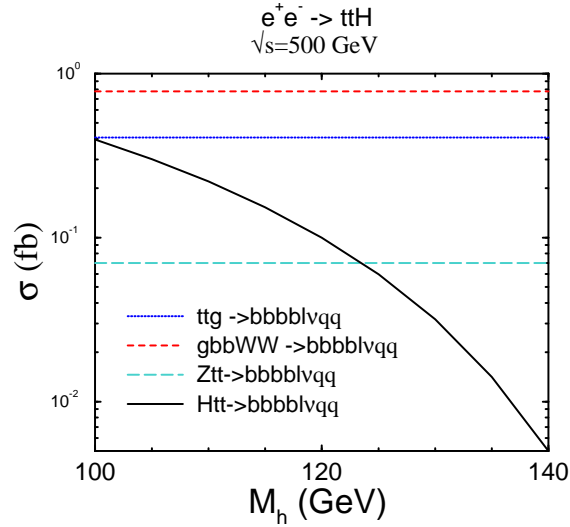}
\caption[]{\label{bkgds}
Parton level signal and backgrounds to
$e^+e^-\rightarrow t {\overline t}h$ at $\sqrt{s}=500$~GeV.}
\end{figure}

The dominant electroweak background to the semi-leptonic decay 
is~\cite{jm,recent,more}:
\begin{eqnarray}
e^+e^- &\rightarrow t {\overline t} Z \rightarrow 
Z W^+W^- b {\overline b} \rightarrow
 b {\overline b} b {\overline b}
\ell^\pm \nu q {\overline q}^\prime\nonumber.
\end{eqnarray}
The largest background is from radiative top quark decays:
\begin{eqnarray}
e^+e^- \rightarrow t {\overline t}  \rightarrow 
g W^+W^- b {\overline b} \rightarrow
 b {\overline b} b {\overline b}
\ell^\pm \nu q {\overline q}^\prime\nonumber.
\end{eqnarray}
This background has been calculated at the parton level~\cite{more} 
and is shown in Fig.~\ref{bkgds}. Since the $b$ jets resulting from 
the gluon splitting are logarithmically enhanced at low energy, cuts on
the jet energy are effective at eliminating this background.
A preliminary study of $e^+e^-\rightarrow t {\overline t}h$ at
$\sqrt{s}=500$~GeV included statistical, but not systematic errors 
and found that the top quark-Higgs Yukawa coupling could be measured 
with $\sim 21\%$ accuracy with perfect $b$-tagging and 
$L=1000$~fb$^{-1}$~\cite{recent}.

The case for a 120~GeV Higgs boson and $\sqrt{s}=800$~GeV with
$L=1000$~fb$^{-1}$ has been considered in \cite{jm}, with 
events  processed through a  simulation  of  a detector for  TESLA. 
In this analysis, the $b$ jets are defined as those four jets
with the lowest probability to originate from the primary vertex. 
The analysis  applies
a standard preselection in order to remove as much background 
as possible while keeping a high efficiency for the signal. 
Then, in order to improve
the statistical sensitivity further, a multivariate analysis using a
Neural Network (NN) is  performed.  
After preselection, the overall effective cross section
for the background is 17.60 fb, while for the signal it is only 0.61 fb.
This translates into such a poor sample purity ($\rho \sim 3.3\%$),
that any uncertainty in the background normalization completely erases 
the significance in the signal. After the NN analysis~\cite{jm}, 
the statistical error is reduced to $5.1\%$, 
and the systematic error to $3.8\%$, leading to an overall uncertainty 
of $6.3\%$ for the Yukawa coupling measurement in the semi-leptonic
channel.
Combining this with the analysis for the hadronic channel gives a total uncertainty
of $5.5\%$.

\subsubsection{Conclusion}
 The reaction $e^+e^- \to t\bar{t}h$ allows a direct determination of 
the top quark-Higgs Yukawa coupling.
For $m_h=120$~GeV and ${\cal L}=1000$~fb$^{-1}$, a total uncertainty of
roughly $5.5\%$ on the top-Higgs Yukawa coupling at $\sqrt{s}=800$~GeV
can be obtained.  Preliminary studies show that the anticipated
precision is about a factor of $4$ worse at $\sqrt{s}=500$~GeV.
The dominant systematic uncertainty is 
from the overall background normalization, pointing to the
importance of a complete $2\rightarrow 8$ background calculation.

\subsection{Top mass reconstruction}

The top quark mass in $e^+e^-$ collisions can not only be measured in 
a threshold scan, but also at
center-of-mass energies above the $t\bar t$ threshold. A recent 
study~\cite{yeh} has shown that a statistical precision of 200~MeV or better
may be reached for the top mass from a full
kinematical reconstruction of $e^+e^-\to t \bar t\to W^+bW^-\bar
b\to\ell^+\nu b\ell^-\bar\nu\bar b$ events. 
It should be noted that the mass measured from final-state shape variables is
the pole mass,  which is subject to a theoretical uncertainty of ${\cal
O}(\Lambda_{QCD})$; this point was  explained in Section 2.2.
Here we give a brief status
report of a new
study that focuses on extracting the top quark mass from the the $b$-$\ell$
invariant mass distribution $d\sigma/dm_{b\ell}$, where $\ell$ is the
lepton from the $W$ decay, and the $b$-quark energy spectrum,
$d\sigma/dE_b$. 

The extraction of the top mass from final-state shape variables is best
done using templates, using a method similar to that described in 
\cite{elinor}. It depends crucially on the modeling of the
multiparton radiation that is associated with the top production and decay 
stages.
Standard Monte Carlo event generators simulate multiple emission in
the soft or collinear approximation and leave empty 
regions of the phase space corresponding to
hard and large-angle gluon radiation (``dead zones''), which can be
populated using the exact matrix element (``matrix-element corrections'').
Matrix-element corrections to top decays $t\to bW(g)$ \cite{mike} have
been implemented in the most recent version of the HERWIG event generator, 
HERWIG~6.2~\cite{herwig}, which is used in the following. These 
corrections were found to 
have a significant effect on jet observables and on the top mass 
measurement at lepton and hadron colliders~\cite{mike,mlm}.

\begin{figure}
\centerline{\resizebox{0.49\textwidth}{!}{\includegraphics{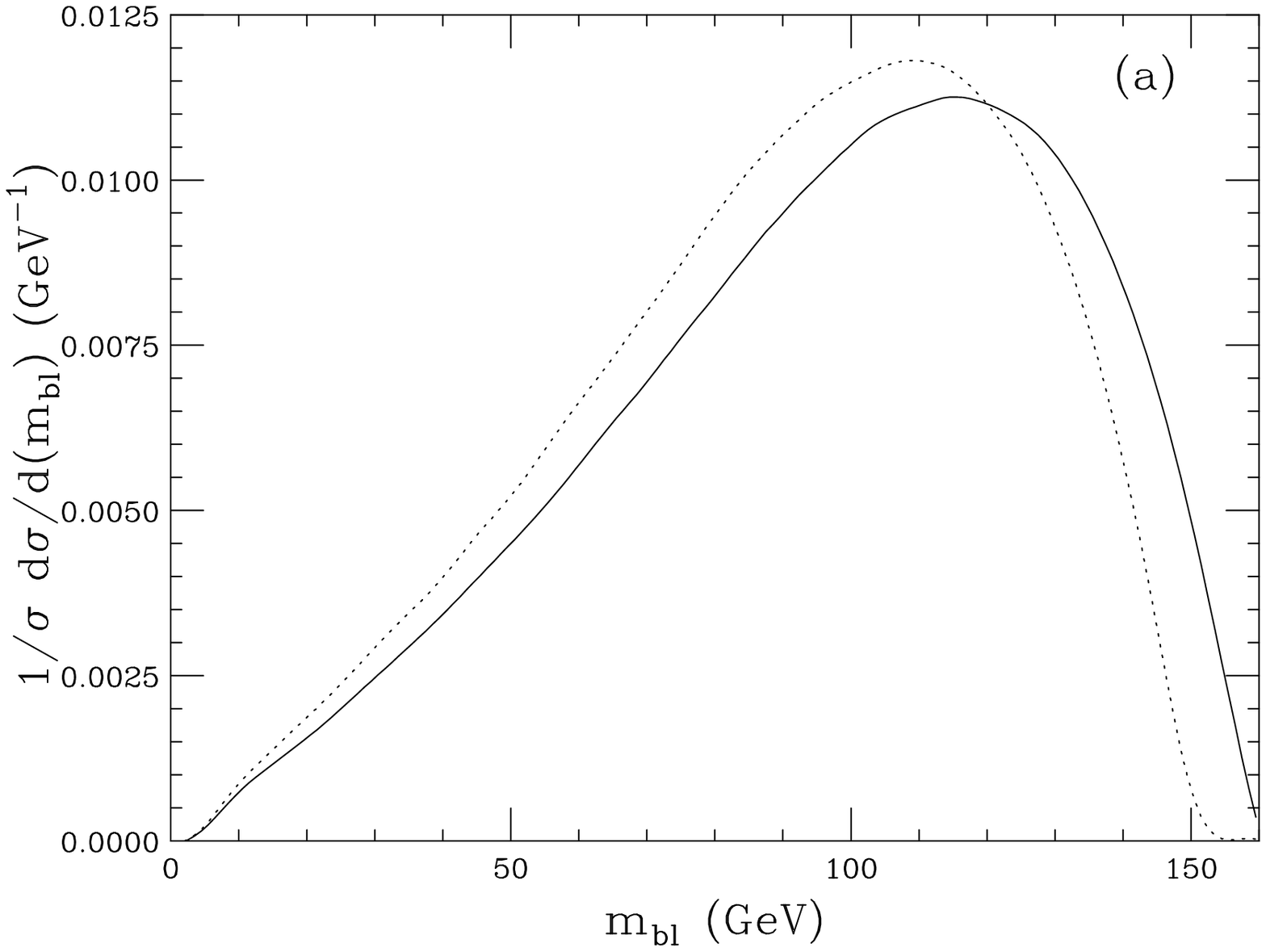}}%
\hfill%
\resizebox{0.49\textwidth}{!}{\includegraphics{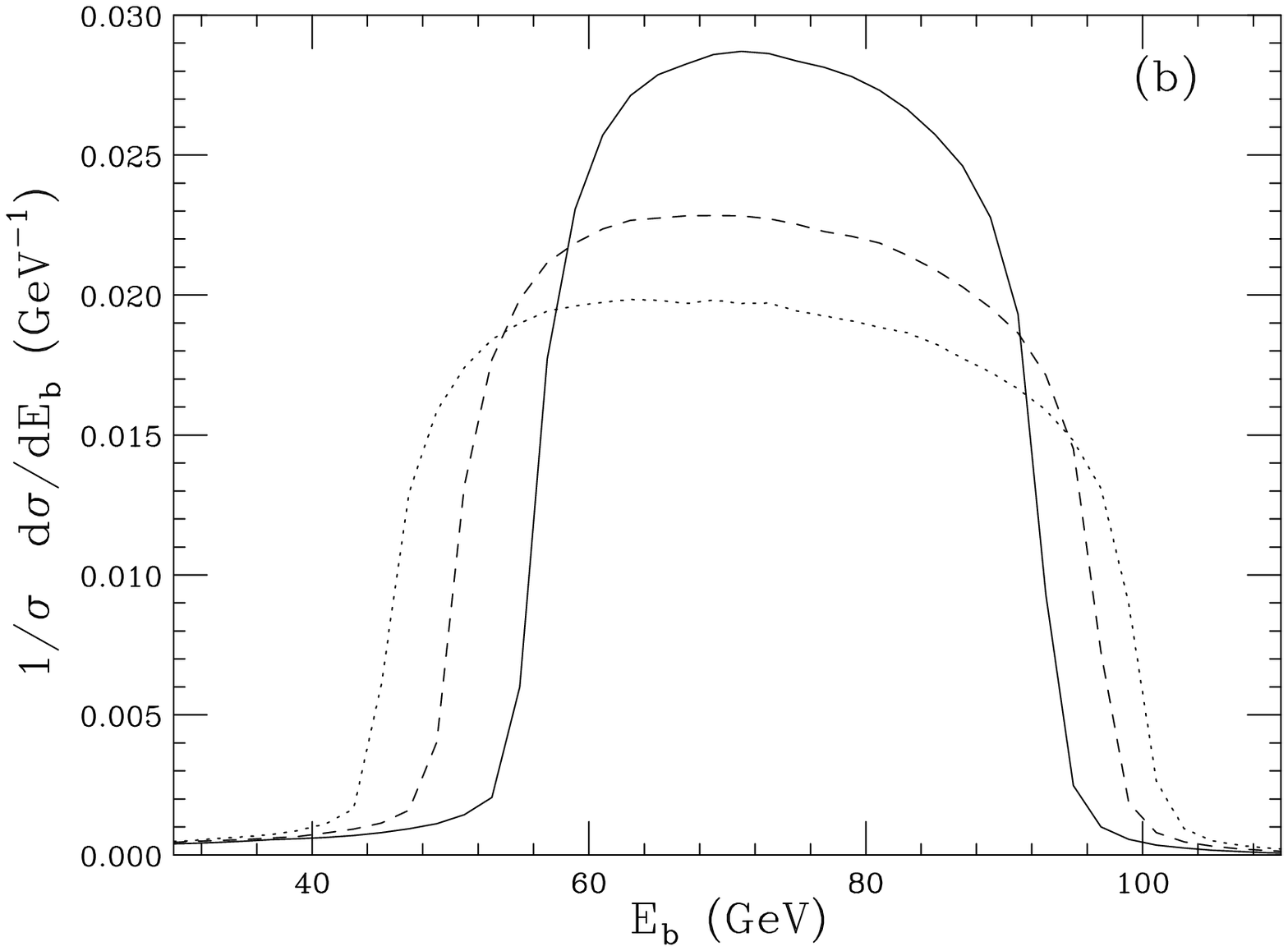}}}
\caption{\label{fig:mbl}
a) Invariant mass 
$m_{b\ell}$ distributions for $m_t=171$~GeV (dotted line) and 
$m_t=179$~GeV (solid line). b) $b$-quark 
energy distribution at $\sqrt{s}=370$~GeV, 
for $m_t=179$~GeV (solid), 175~GeV (dashed) and 171~GeV
(dotted). }
\end{figure}

The $m_{b\ell}$ distribution, within the precision of the Monte Carlo
integration, is independent of the hard-scattering process and of the
center-of-mass energy.
$m_{b\ell}$ is a Lorentz-invariant observable and is therefore 
insensitive to the boost from the top quark rest frame to the laboratory frame.
In Fig.~\ref{fig:mbl}a we plot the $m_{b\ell}$ 
distribution for $m_t=171$ GeV and 179 GeV. As $m_t$ increases, the peak
position of the $m_{b\ell}$ distribution is shifted towards larger
values. The average value $\langle m_{b\ell}\rangle$ is proportional to
the top quark mass. The best fit is:
\begin{equation}
\langle m_{b\ell}\rangle \ = \ 0.756\ m_t-37.761\ {\mathrm{GeV}},\  
\eps=0.002\ {\mathrm{GeV}},
\label{eq:mbl} \end{equation}
where $\eps$ is the mean square deviation in the fit.
Solving Eq.~(\ref{eq:mbl}), one finds 
$\Delta m_t\approx 1.32\ \Delta \langle m_{b\ell}\rangle$, where 
$\Delta\langle m_{b\ell}\rangle$ is the uncertainty on the measurement 
of $\langle m_{b\ell}\rangle$. No detailed study of the precision that
can be achieved with this method has been carried out yet.

In contrast to $m_{b\ell}$, the $b$-quark energy $E_b$ is not a 
Lorentz-invariant observable. One therefore expects that the $E_b$ 
distribution does depend on the boost from the top rest frame
to the laboratory frame, and hence on the center-of-mass energy.
Since the $t\bar t$ pair is produced almost at rest at the $t \bar t$
threshold, the dependence of $E_b$ on the top mass is maximized in this
region. The $E_b$ distribution for $\sqrt{s}=370$~GeV and several values
of $m_t$ is shown in 
Fig.~\ref{fig:mbl}b. For $m_t$ approaching the threshold value of
$\sqrt{s}/2$, the $E_b$ distribution becomes very narrow. 
The half-maximum width $\sigma_b$ therefore shows a strong dependence on
the top mass.
The best polynomial 
fit to express $\sigma_b$ in terms of $m_t$ for $\sqrt{s}=370$~GeV 
is found to be:
\begin{equation}
\sigma_b = \ -0.081\ m_t^2+26.137\ m_t-2048.968\ {\mathrm{GeV}},\ 
\eps=0.393\ {\mathrm{GeV}}.
\label{eq:sigmab}
\end{equation}
For a top quark mass in the range 171~GeV~$\lsim m_t\lsim$~179~GeV, 
the induced uncertainty on $m_t$ is $\Delta m_t\approx 0.35-0.65\ \Delta
\sigma_b$, where $\Delta \sigma_b$ is the uncertainty on the
half-maximum width. $E_b$ thus may be an interesting observable to 
reconstruct the top mass at energies slightly above the $t \bar t$ 
threshold. It is probably less useful at higher energies.

\subsection{Anomalous couplings}

At present, the couplings of the top quark to gluons and the electroweak
gauge bosons are largely untested. A linear collider provides an ideal tool
to probe the couplings of the top quark to the electroweak gauge
bosons. 
It is important to note that the neutral electroweak couplings are 
accessible only at lepton colliders, because top quarks at 
hadron colliders are pair-produced via gluon exchange.
Since the charged electroweak current is involved in the
top decay, $t \bar t$ production in $e^+e^-$ collisions is sensitive to 
both the neutral and charged gauge boson couplings of the top quark.
Because the top quark width,
$\Gamma_t$, is much larger than $\Lambda_{\rm QCD}$, the decay process is
not influenced by fragmentation effects and decay products will provide
useful information. 

The most general $(\gamma,\,Z)t\bar t$ couplings can be written 
as~\cite{Stefano,hioki} 
\begin{equation}
  \Gamma^\mu_{t\overline t \; \gamma,Z} = 
   i \, e \; 
   \left\{ \gamma^\mu \; 
    \left[ F^{\gamma,Z}_{1V} \; + F^{\gamma,Z}_{1A} \, \gamma^5 \right] +
    {(\;p_t^{}-p_{\overline t}^{})^\mu  \over 2 \; m_t^{} } \; 
    \left[ F^{\gamma,Z}_{2V} \; + F^{\gamma,Z}_{2A} \, \gamma^5 \right] \, 
  \right \} \ ,
\end{equation}
where the only form factors different from zero in the SM are
\begin{equation}
F^\gamma_{1V}={2\over 3} \ , \ F^Z_{1V}={1\over 4\sin\theta_W\cos\theta_W} \, 
        \left(1-{8\over 3} \sin^2\theta_W^{} \right)  \ ,\
F^Z_{1A} = -{1 \over 4\sin\theta_W\cos\theta_W }.
\end{equation}
$ \left({e/m_t}\right) \cdot F^\gamma_{2A} $
is the  CP-violating electric dipole moment (EDM) form factor of the top
quark and 
$ \left({e/m_t}\right) \cdot F^Z_{2A} $ is the weak electric dipole moment 
(WDM). $ \left({e/m_t}\right) \cdot F^{\gamma,Z}_{2V} $ are the 
electric and weak magnetic dipole moments (MDM).

In the SM, the EDM and WDM terms violate CP and receive contributions only at
the three-loop level and beyond. The CP-conserving form factors are zero
at tree level but receive non-zero ${\cal O}(\alpha_s)$ QCD
corrections. 

The most general $Wtb$ couplings can be parametrized in the form~\cite{hioki}
\begin{equation} 
  \Gamma^\mu_{tbW} = 
  - {g\over \sqrt 2 } \, V_{tb}^{} \, 
  \left\{ \gamma^\mu \; 
    \left[  f^{L}_{1} \, P_L^{} +
            f^{R}_{1} \, P_R^{} \right] - 
    { i \, \sigma^{\mu\nu }\over M_W^{} } \, (p_t^{}-p_b)_\nu^{} \,
    \left[  f^{L}_{2} \, P_L^{} +  
            f^{R}_{2} \, P_R^{} \right] \,
   \right\} \ ,
\end{equation}
where $P_{R,L}^{} = (1\pm\gamma_5^{})/2$. In the limit $m_b\to 0$,
$f_1^R$ and $f_2^L$ vanish. In the SM, at tree level, 
$f_1^L=1$, and all other form factors are zero. Similarly, the
$W\bar t\bar b$ vertex function can be parametrized in terms of form
factors $\bar f_{1,2}^{L,R}$. If CP is conserved, $\bar f_{1,2}^{L,R}=
f_{1,2}^{L,R}$. 

In Table~\ref{tab:topone}, we present the 
$1\sigma$ sensitivity limits for the real parts of the $(\gamma,\,Z)t \bar t$
form factors obtained from a recent analysis of the process
$e^+e^-\to t \bar t\to\ell^\pm +$~jets at $\sqrt{s}=500$~GeV.
 Only one coupling at a time 
is varied. Top quarks are selected and reconstructed, and $b$
quarks are tagged using the LCD fast simulation package for the L
detector configuration. The combined efficiency is 20\%, and the purity after
selection is 88\%. To extract limits on
$F_{1V}^{\gamma,Z}$ and $F_{1A}^{\gamma,Z}$, the angular distribution of
the reconstructed top quark is used. $F_{1V}^{\gamma,Z}$ and 
$F_{2V}^{\gamma,Z}$ are derived from the left-right polarization
asymmetry, and $F_{2A}^{\gamma,Z}$ from the angular distribution of
the reconstructed top quark and the decay angles of the $t$ and $\bar
t$. 

\begin{table}
\renewcommand{\arraystretch}{1.1}
\begin{center}
\begin{tabular}{|c|c|c|c|c|} \hline\hline
Coupling & LO SM Value & ${\cal P}(e^-)$ & $\int\!{\cal L} dt$ (fb$^{-1}$) &
$1\sigma$ sensitivity \\ \hline 
$F_{1A}^\gamma$ & 0      & $\pm 0.8$ & 100 & 0.011 \\
$F_{1A}^Z$      & $-0.6$ & $-0.8$    & 100 & 0.013 \\
$F_{1V}^\gamma$ & $2/3$  & $\pm 0.8$ & 200 & 0.047 \\
$F_{1V}^Z$      & $0.2$  & $\pm 0.8$ & 200 & 0.012 \\
$F_{2A}^\gamma$ & 0      & $+0.8$    & 100 & 0.014 \\
$F_{2A}^Z$      & 0      & $+0.8$    & 100 & 0.052 \\
$F_{2V}^\gamma$ & 0      & $\pm 0.8$ & 200 & 0.038 \\
$F_{2V}^Z$      & 0      & $\pm 0.8$ & 200 & 0.009 \\
\hline\hline
\end{tabular}
\end{center}
\caption{\label{tab:topone}
The $1\sigma$ statistical uncertainties for the real 
parts of the $(\gamma,\,Z)t \bar t$ form factors obtained from an 
analysis of the process $e^+e^-\to t \bar t\to\ell^\pm +$~jets for
$\sqrt{s}=500$~GeV. Only one coupling at a time is varied. }
\end{table}

The limits shown in Table~\ref{tab:topone} could be strengthened if
positron beam polarization becomes available, mostly from the increased
$t \bar t$ cross section. If ${\cal P}(e^+)=0.5$, the $t \bar t$ cross
section is about a factor 1.45 larger than that obtained with ${\cal
P}(e^+)=0$. This improves the bounds by up to $25\%$.
Increasing the CM energy to
$\sqrt{s}=800$~GeV improves the limits by a factor 1.3--1.5~\cite{breuther}. 

The decay form factor 
$f_2^R$, corresponding to a $(V+A)$ top decay,
 can be measured with a precision of about 0.01 for
$\sqrt{s}=500$~GeV and $\int\!{\cal L}dt=500~{\rm fb}^{-1}$ if electron and
positron beam polarization are available~\cite{hioki}. This quantity can
also be measured at the LHC, though the expected limit is a factor three to
eight weaker than the limit we project for a linear collider~\cite{toplhc}. 

Many models predict anomalous top quark couplings.
In technicolor models and other models with a strongly-coupled Higgs
sector, the CP-conserving couplings may be induced
at the 5--10\% level~\cite{top-TC1,top-TC2,top-TC3}.  In supersymmetric
and multi-Higgs models, the CP-violating 
couplings $F_{2V,A}^{\gamma,Z}$ may be induced at the one-loop level,
with predictions in the range
$F_{2V,A}^{\gamma,Z}={\cal O}(10^{-3}-10^{-2})$~\cite{sumino}. A 
measurement of the $(\gamma,Z)t \bar t$ couplings at a linear collider will
thus be sensitive to interesting sources of non-SM physics.

\subsection{QCD and electroweak radiative corrections}

For $\sqrt{s}=500$~GeV and an integrated luminosity of 500~fb$^{-1}$,
the statistical error of the $e^+e^-\to t \bar t\to\ell\nu jj\bar bb$
cross section is well below 1\%. In order to match this experimental 
accuracy with robust theoretical predictions, precision calculations 
beyond tree level are required. Such theoretical accuracy is needed both
when top itself is 
the subject of study and when top is a background to other physics of
interest.

QCD corrections can have important effects in top events.  
Jets from radiated gluons can be indistinguishable from quark jets, 
complicating identification of top quark events from the reconstruction of
the top  decay products.  In addition, real
emission may occur either in the top production or decay
processes, so that radiated gluons may or may not themselves be 
products of the decay.  Subsequent  mass measurements can 
be degraded, not only from misidentification of jets but also
from subtle effects such as jet broadening when gluons are 
emitted near other partons.  Virtual corrections must also be included 
to predict correct overall rates.

Most calculations of QCD corrections in $e^+e^-\to t \bar t$ to date have 
been performed for on-shell top quarks.
In this approximation, corrections to the production and decay processes
can be computed separately.  A calculation of the QCD corrections to 
the production process $e^+e^- \to t \bar t$,
which includes real gluon emission from the $t$ and $\bar t$ and virtual 
gluon exchange between the $t$ and $\bar t$ 
has been presented in~\cite{prod}.  A discussion of the QCD corrections
to the decay $t\to Wb$ can be found in \cite{decay}; QCD corrections are
found to reduce the tree-level 
width of 1.55 GeV to $\Gamma_t^{{\cal O}(\alpha_s)}=1.42\ {\rm GeV}$
after all the known QCD and EW corrections are taken into account.

Because of the large width of the top quark and the fact that it does not
hadronize before decaying \cite{lifetime}, it is necessary to compute 
corrections to the entire production and decay process, including
off-shell effects.  
In the soft gluon approximation, real gluon corrections for the process
$e^+e^- \to t \bar t \to WWbb$ 
with the top allowed to be off-shell were calculated in \cite{soft}.
Interference effects of gluons radiated in
the production and decay stages were found to be sensitive to the top 
width $\Gamma_t$,
with the effects being largest for gluon energies comparable to
$\Gamma_t$.
Similarly, real gluon radiation in top production and decay is
sensitive to top width effects~\cite{hardglu}.

Since the process observed experimentally is 
\begin{equation} \label{proc1}
 e^+ e^- \rightarrow\ b\ W^+\ \bar{b}\ W^-\, ,
\end{equation}
 it is desirable to take into account all Feynman diagrams that 
contribute to~(\ref{proc1}). This has not been done yet. At 
next-to-leading order,
it is sufficient to take into account only the QCD corrections to the 
diagrams containing an intermediate top and antitop quark, as
has been done in the computations discussed here.
This approach  uses  the 
double pole approximation (DPA), in which  only the double resonant
terms (due to top and antitop propagators) are kept.
Work done in this area follows closely the treatment of the
$W$ pair production process at LEP II \cite{WW}. 

Radiative corrections to  
$e^+ e^- \rightarrow\ t\bar{t}\rightarrow b W^+\bar{b}W^-$ are 
usually split into two classes: corrections to particular subprocesses
(production and decay), also called factorizable corrections, and 
corrections involving interference between these subprocesses (non-factorizable
corrections). In most approaches, the factorizable corrections are computed 
using the on-shell approximation for the top quarks; either using
the on-shell phase space, or making an on-shell projection from 
the exact phase space \cite{top-MY1996,BBC_top}. In the latter the on-shell
projection restricts the  effect of the 
off-shell particles to the interference terms. 
These interference terms are computed in DPA, for virtual as well
as for  real gluons. As a consequence, interference terms do not
contribute to the total cross section.

In \cite{mcos}, a different approach is used. Instead of starting with the 
on-shell computation and adding the nonfactorizable corrections, the
starting point is the exact amplitudes for the off-shell process from
which terms that are not 
doubly resonant are dropped. Also, the real gluon contributions are
treated exactly (as in \cite{hardglu});
as a consequence, the cancellation between virtual
gluon and real gluon interference is no longer complete.
Table~\ref{Tcr_sec} summarizes the total cross section results. The 
QCD corrections
are found to increase the $t \bar t$ production cross section by
up to a factor two near the threshold, and by about 11--13\% in the
continuum. 

\begin{table}[t] 
\begin{center}
\begin{tabular}{|c|c|c|c|}
\hline\hline
 $2E_{beam}$    & 360 GeV & 500 GeV & 1000 GeV \\
\hline
 $\sigma_0 $             & 0.386 pb & 0.565 pb& 0.172 pb\\
 $\sigma_1^{on-shell}$   & 0.737 pb& 0.666 pb& 0.186 pb \\
 $\sigma_1^{DPA}$        & 0.644 pb& 0.652 pb& 0.191 pb \\
\hline\hline
\end{tabular}
\end{center}
\caption{\label{Tcr_sec}
Cross sections (tree level, on-shell NLO and DPA NLO)
for top production and decay at a linear collider~\cite{mcos};
results do not include ISR, beamstrahlung or beam energy spread. }
\end{table}

Electroweak ${\cal O}(\alpha)$ corrections for top processes at 
linear colliders have also been computed so far 
only to on-shell $t \bar t$ production and top decay.
The electroweak ${\cal O}(\alpha)$ corrections can be naturally subdivided
into two gauge-invariant subclasses, QED and weak corrections.  The QED 
corrections depend on the cuts imposed
on the photon phase space and thus on the experimental setup.  
As discussed in \cite{Beenakker:1991ca}, initial-state ${\cal
O}(\alpha)$ QED corrections can significantly reduce the cross
section because of large logarithms of the form $\alpha/
\pi \ln(s/m_e^2)$ with $s \gg m_e^2$.  These terms arise when photons are
radiated off in the direction of the incoming electrons. Thus, the
inclusion of higher-order initial-state radiation (ISR) has to be
considered. The leading-log initial-state QED corrections are
universal and can be calculated using the so-called structure function
approach~\cite{Beenakker:1996kt}. 

The model-dependent contributions to corrections to top pair 
production are contained in the weak corrections. The numerical 
impact of the weak one-loop corrections is discussed in detail in
\cite{Beenakker:1991ca}.  Close to the $t\bar t$ threshold, 
the weak corrections to $\sigma_{t \bar t}$ are found to 
be quite sensitive to the Higgs boson mass.  
An updated analysis of the weak corrections
to $\sigma_{t\bar t}$, using the current value of the top-quark mass, is
presented in~\cite{Hollik:1999md}. The weak corrections are found
to reduce the Born cross section (expressed in terms of
$G_\mu$) near threshold by about $7 \%$, which is mainly
due to the box diagrams.

The complete electroweak ${\cal O}(\alpha)$ corrections to $\Gamma_t$ 
are calculated in~\cite{Denner:1991ns}.  When using $G_{\mu}$ and 
$M_W$ to parametrize
the lowest-order top decay width, the electroweak corrections 
amount to typically 1-2 \% with no significant dependence on $m_h$.
  
Ultimately it will be necessary to combine the QCD and electroweak 
corrections to
top processes.  This has been done for $e^+e^-\to t \bar t$ 
in \cite{combine}, and work is in progress to combine both types
of correction for the entire production and decay process \cite{mow}.

\section{Conclusions}

\begin{table}[t]
\vskip 2.mm
\begin{center}{\small
\begin{tabular}{ccccc}
\hline\hline \\[-3.mm]
Observable & Precision & $\int\!{\cal L}dt$ (fb$^{-1}$) & $\sqrt{s}$
(GeV) & Comment \\[1.mm] \hline 
$m_t$ & $<100$~MeV & 10 & $350$ & theory dominated \\[1.mm]
$m_t$ & 200~MeV & 50 & 500 & not fully explored \\[1.mm]
$\Gamma_t$ & ${\cal O}(30$~MeV) & 100 & $350$ & not fully explored
\\[1.mm]
$g_{tth}$ & ${\cal O}(10\%)$ & 100 & $350$ & need realistic study
\\[1.mm] 
$g_{tth}$ & $21\%$ & 1000 & 500 & stat. uncert. only \\[1.mm]
$g_{tth}$ & 5.5\% & 1000 & 800 & need improved bgd. estimate \\[1.mm]
$F_{iV,A}^{\gamma,Z}$, $f_2^R$ & $0.01 - 0.2$ & 500 &
500 & polarized beams essential \\[1.mm]
\hline\hline
\end{tabular}
}
\end{center}
\caption{\label{toptab:three}
Summary of top quark-related measurements at a linear $e^+e^-$ 
collider.}
\end{table}

Remarkable progress has been made in the last two years in our
theoretical understanding of $t \bar t$ production in $e^+e^-$ collisions
at the
threshold. Problems associated with defining the top quark mass in a way
that removes QCD ambiguities have been solved. The remaining
theoretical uncertainties are sufficiently small to allow a
simultaneous measurement of $m_t$ (to 100~MeV),
$\Gamma_t$ (to  a few percent) and $g_{tth}$. The top
quark mass can also be measured with a precision of 200~MeV or better 
at higher energies, using a variety of kinematic variables. Not all
interesting variables have been fully explored yet. An ideal process to
determine the top quark Yukawa
coupling at energies above the $t \bar t$ threshold is $t \bar
th$ production in $e^+e^-$ collisions. However, to fully exploit 
this process, energies
significantly larger than $\sqrt{s}=500$~GeV are necessary. On the other
hand, a center-of-mass energy of 500~GeV is sufficient to measure 
the top quark couplings to the electroweak gauge bosons with a 
precision of ${\cal O}(1-10\%)$. 
Polarized electron and positron beams are essential to
disentangle the various couplings. We have summarized
the  estimated precision on the
various quantities in Table~\ref{toptab:three}.
Finally, we have given a brief overview of
the status of calculations of the QCD and electroweak corrections to
$e^+e^-\to t \bar t$. The potential for precision studies of top quark
physics at a linear collider requires a detailed understanding of these
corrections.

\emptyheads
\blankpage \thispagestyle{empty}

\setcounter{chapter}{6}

\fancyheads

\chapter{QCD and Two-Photon Physics}
\fancyhead[RO]{QCD and Two-Photon Physics}

\section{Introduction}

A relatively clean environment and well-understood initial-state parton
content render $ e^+e^- $ colliding beam experiments ideal for both
the qualitative confirmation and quantitative testing of Quantum
Chromodynamics (QCD).
Through the years, a number of seminal discoveries and measurements
performed at $e^+e^-$ colliding beam facilities have served to
establish the SU(3) color gauge theory  QCD
as the accepted dynamical model of the strong nuclear interaction.
Highlights unique to the $e^+e^-$ QCD program include the discovery of
the gluon at PETRA in 1979, the confirmation of the
SU(3) gauge structure of quark-gluon and gluon-gluon vertices at
LEP in the early 1990s, and the precise measurement of the strong
coupling constant $\alphas$ from hadronic observables and from the $Z$ and $\tau$
decay widths.

The study of QCD, and the dynamics of the strong force in general, is
expected to provide a significant contribution to the physics program
at a high-energy $e^+e^-$ colliding beam facility. The highlights of
this program include
\smallskip

\noindent
$\bullet$
the precise determination of the strong coupling constant $\alphas$;

\noindent
$\bullet$
the search for anomalous strong couplings of the top quark;

\noindent
$\bullet$
the study of photon structure; and

\noindent
$\bullet$
the study of strong-interaction dynamics at high $\sqrt{s}$ and fixed $t$.

\smallskip
\noindent
Together, these measurements probe some of the most important topics in the
study of strong force dynamics, in ways that are often superior
to measurements at hadron colliders.

\section{QCD from annihilation processes}

\boldmath
\subsection{The precise determination of $\alphas$}
\unboldmath

As the single free parameter of the SU(3) gauge theory of the strong 
interaction,
the strong coupling constant $\alphas$ should be measured to the
 highest available
precision. Renormalization group extrapolations of the U(1), SU(2) and SU(3)
coupling strengths constrain physics scenarios at the GUT
scale. The current constraints are limited by the few-percent relative 
precision~\cite{Sigalpha}
of the value of \alpmz. The value of $\alphas$ should also be determined
with comparable accuracy over as large a range of scales as possible in 
order to measure the renormalization-group running of $\alphas$ and to 
reveal potential anomalous running in the strength of the strong interaction.
In this article, as a matter of convention, measurements of $\alphas$
performed at other scales will be  evolved to the scale $Q^2 = M_Z^2$
according to Standard Model renormalization group equations
and quoted in terms of their implied value of \alpmz.

\subsubsection{Event observables in $\ee$ annihilation}

The determination of \alpmz\ from the process
$e^+e^- \rightarrow Z/\gamma \rightarrow q{\bar q}(g)$, using `shape'
observables that are
sensitive to the underlying parton content, has been
pursued for two decades and is generally well understood~\cite{philalp}.
In this method one usually forms a differential distribution,
makes corrections for detector and hadronization effects,
and fits a perturbative QCD prediction to the data, allowing \alpmz\  to vary.
Examples of such observables are thrust, jet masses and jet rates.

The latest generation of such \alpmz\  measurements, from SLC and LEP, has
shown that statistical errors below the 1\% level can be obtained
with samples of a few tens of thousands of hadronic events.
With the current linear collider design luminosity of
$2.2\times10^{34}$ cm$^{-2}$s$^{-1}$, at $\sqrt{s}$ = 500 GeV,
hundreds of thousands of $e^+e^-$~$\rightarrow$~$q\overline{q}$
events would be produced each year,
and a statistical error on \alpmz\  below 0.5\% would
be achieved.

At energies far above the $Z$ pole, the electron-positron collision cross
section is dominated by $t$-channel processes such as $ZZ$ and $W^+W^-$
production. In addition, because of the substantial mass of the $t$ quark, 
the inclusive characteristics of $e^+ e^-$~$\rightarrow$~$t{\bar t}$
events tend to mimic those of lighter quark events with hard gluon radiation.
A prescription for the elimination of these backgrounds was developed for
the 1996 Snowmass workshop~\cite{Bethke,Schumm}.  This prescription
 makes use of electron beam
polarization and precise tracking to reduce the effects of these backgrounds
on the measured three-jet rate to less than 5\%, with the 
corresponding systematic
uncertainty on the extraction of \alpmz\  expected to be 
substantially less than 1\%. 
The
sizable initial-state and beamstrahlung radiation associated with linear
collider energies will act to smear the CM energy of the $e^+e^-$ annihilation
process, as well as to boost the particle flow into the forward regions
of the detector. A PYTHIA study~\cite{schumm-truitt}, including the full
effects of ISR, has shown that these considerations can
be accurately taken into account in the measurement of \alpmz.

Hadronization effects, which lead to corrections of order 10\% at the
$Z^0$ pole, are expected to fall at least as fast as $1/\sqrt{s}$, leading
to corrections of order 1\% at $\sqrt{s} \ge$  500 GeV~\cite{pyth-had}.
The corresponding
systematic error on the extraction of \alpmz\  is thus expected to be 
substantially
below 1\%. Detector systematics, due primarily to limited acceptance and
resolution smearing, and which are observable-dependent, are
found to contribute at the level of
$\delta$\alpmz = $\pm 1$--4\% at LEP-II~\cite{OPALAS}. The greater hermeticity
and $\cos\theta$ coverage anticipated for linear collider detectors are
again expected to reduce this substantially.

Currently,  perturbative calculations of event shapes
are complete only
up to $O(\alpha_s^2)$, although resummed calculations are available for some
observables~\cite{resum}.
One must therefore
estimate the possible bias inherent in measuring
\alpmz\  using the truncated QCD series.
Though not universally accepted, it is customary to estimate this from
the dependence of the fitted \alpmz\  value on the QCD renormalization scale,
yielding a large and dominant uncertainty of about $\Delta$\alpmz $\simeq$
$\pm$6\%~\cite{philalp}.
Therefore, although a $\pm 1\%$-level \alpmz\  measurement is 
possible experimentally,
it will not be realized until $O(\alpha_s^3)$ contributions are
completed. There is a reasonable expectation that this will be achieved within
the next three years~\cite{zvi,Gehrmann:2001zt}.

\subsubsection{The $t {\bar t} (g)$ system}

The dependence of the $e^+e^- \rightarrow t {\bar t}$ cross section
on $m_t$ and \alpmz\  is presented in Chapter 6, Section 2.
As  discussed there,
next-to-next-to-leading-order calculations
of the $t {\bar t}$ cross section in the resonance region show 
convergence to the few-percent level for an appropriate definition of
$m_t$, if logarithms of the top quark velocity are resummed. 
This is good news for the extraction of $m_t$; however, we will
probably not obtain a competitive value of  \alpmz\ from this
system.

\subsubsection{A high-luminosity run at the $Z^0$ resonance}

A sample of $10{^9}$ $Z^0$ decays offers two additional options for
the determination of 
\alpmz\ via measurements of the inclusive ratios
$\Gamma^{\rm had}_Z/\Gamma_Z^{\rm lept}$ and
$\Gamma^{\rm had}_{\tau}/\Gamma_{\tau}^{\rm lept}$. 
In both cases, $\alphas$ enters in through the QCD radiative correction; thus,
both observables require
a very large event sample for
a precise measurement. For example, the current LEP data sample of 16M
$Z^0$ decays yields an error of $\pm 2.5\%$ on \alpmz\  from
$\Gamma^{\rm had}_Z/\Gamma_Z^{\rm lept}$, with an experimental 
systematic of $\pm 1\%$.
With a Giga-Z sample, the statistical error would
be pushed to below $\Delta$\alpmz = 0.4\%. Even with no improvement
in experimental systematics, this would be a precise and reliable measurement.
In the case of
$\Gamma^{\rm had}_{\tau}/\Gamma_{\tau}^{\rm lept}$ the experimental precision
from LEP and CLEO is already at the 1\% level on \alpmz. However,
there has been considerable debate about the size of the theoretical
uncertainties, with estimates as large as 5\% \cite{tau}.
If this situation is clarified, and the theoretical uncertainty is small,
$\Gamma^{\rm had}_{\tau}/\Gamma_{\tau}^{\rm lept}$ may offer a further
1\%-level \alpmz\  measurement.

\boldmath
\subsection{$Q^2$ evolution of $\alphas$}
\unboldmath

In the preceding sections we discussed the expected precision on the
measurement of the benchmark parameter
\alpmz. Translation of the measurements of $\alphas(Q^2)$
($Q^2 \neq M_Z^2$) to \alpmz\  requires the assumption that the `running' of
the coupling is determined by the QCD $\beta$ function.
However, since the logarithmic decrease of $\alphas$ with $Q^2$ is a telling
prediction of QCD, reflecting the underlying non-Abelian dynamics,
it is essential to test this $Q^2$ dependence explicitly.
In particular, such a test would be sensitive to new colored
degrees of freedom with mass below the limit for pair production
at the highest explored scale.
For this measurement of the $Q^2$-dependence of $\alphas$, rather
than its overall magnitude, many common systematic effects would be
expected to cancel.
Hence it would be desirable to measure $\alphas$ in the same detector,
with the same technique, and by applying the same treatment to the
data at a series of different $Q^2$ scales, so as to maximize the
lever-arm for constraining the running.

\begin{figure}[htb]
\begin{center}
\epsfig{file=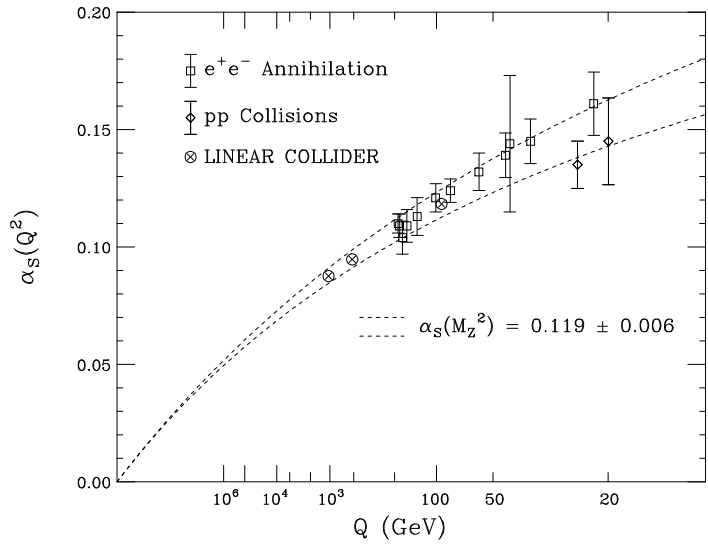,height=2.8in}
\caption{\label{fig:alpha_s_running}
Linear collider measurements of \alpmz, in comparison to existing
measurements from $e^+e^-$ and $p{\overline p}$ collisions, as a function of
interaction scale.}
\end{center}
\end{figure}

Proposed linear collider
measurements of $\alphas(Q^2)$ at $\sqrt{s}$ = 91, 500 and 1000 GeV
are shown in Fig.~\ref{fig:alpha_s_running}, together with
existing measurements which span the range $20\leq \sqrt{s} \leq 200$ GeV.
The linear collider
point at $\sqrt{s}$ = 91 GeV can be obtained either from jet rates
or from the $\Gamma_Z^{\rm had}/\Gamma_Z^{\rm lept}$
technique, while those at 500 and 1000 GeV are based on
jet rates. A theoretical uncertainty of $\pm 1\%$ is 
assumed for all LC points.

\begin{figure}[htb]
\begin{center}
\epsfig{file=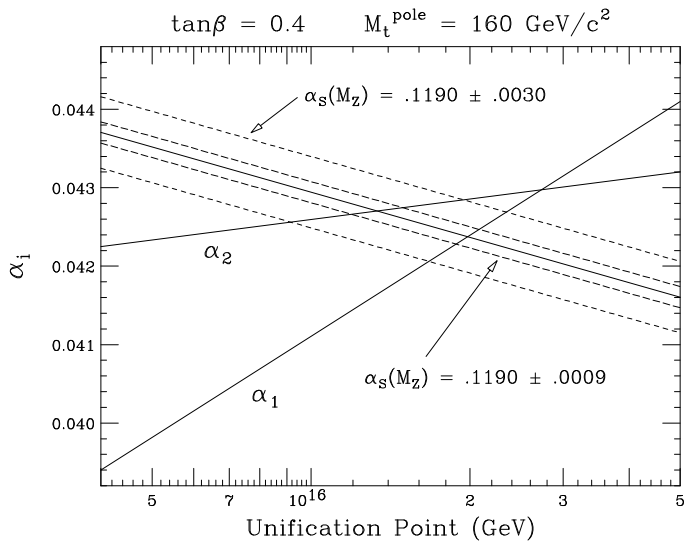,height=2.8in}
\caption{\label{fig:gutscale_as}
Improvement in the GUT scale constraint, assuming a $\pm 1\%$
measurement of \alpmz\  at the linear collider.
Renormalization group trajectories assume the
MSSM with $\tan\beta$ = 0.4 and $m_t^{\rm pole}$ = 
160 GeV~\cite{lang_un}.}
\end{center}
\end{figure}

The linear collider data would add significantly to the
lever-arm in $Q^2$, and would allow a substantially improved extrapolation
to the GUT scale. Consider, for example, making 
a simultaneous fit for \alpmz\  and for $\beta_0$, the
leading term in the expansion of the QCD $\beta$-function which establishes the
rate at which the strong coupling constant runs. (This term is
 expected to be about 0.61 in
the SM.)  The linear collider data alone would give a precision on these
quantities of 
$\pm 0.0018$ and $\pm 0.034$, respectively. Including
accurate measurements at low $Q^2$ (particularly from $e$ and $\mu$ deep
inelastic scattering),
the existing constraints are $\pm 0.0030$ and $\pm 0.042$, respectively.
Combining existing data with that available from the LC would yield
constraints of $\pm 0.0009$ and $\pm 0.016$, providing a substantial
improvement on the measurement of the running of \alpmz, as well as the
extrapolation to the GUT scale (see Fig.~\ref{fig:gutscale_as}). Note that,
unlike the determination of $\beta_0$, the accuracy of the GUT-scale
extrapolation is not dependent upon future running at the $Z^0$.

\subsection{Top quark strong moments}

The very large mass of the recently discovered top quark suggests the
possibility that top plays a central role in physics beyond the
Standard Model. If this is the case, it is likely that this new
physics will manifest itself via anomalous top-quark moments, which
represent the low-energy manifestation of effective higher-dimensional
couplings.  The measurement of the electroweak anomalous moments of the 
top quark is discussed in Chapter 6, Section 3.3.

In the case of the strong interactions of top, the lowest-dimensional 
gauge-invariant and CP-conserving extension to SM top quark couplings 
is the anomalous chromomagnetic moment, which we can parameterize via
a dimensionless quantity $\kappa$. The corresponding chromoelectric
moment, parameterized by $\tilde{\kappa}$, violates CP and arises from an
operator of the same dimension. The resulting generalized three-point
$t\bar{t}g$ vertex takes the form
\beqa
              L  = g_s \bar{t} T_a \left(\gamma_{\mu} +
   {i \over 2m_t} \sigma_{\mu\nu}(\kappa - i\tilde{\kappa}\gamma_5)q^{\nu}\right)
   tG^{\mu}_a,
\eeqan
where $g_s$ is the SU(3) gauge coupling parameter, $m_t$ is the top quark mass,
$T_a$ are the SU(3) color generators, $G^{\mu}_a$ are the vector gluon fields,
and $q$ is the outgoing gluon four-momentum.

\begin{figure}[htb]
\begin{center}
\epsfig{file=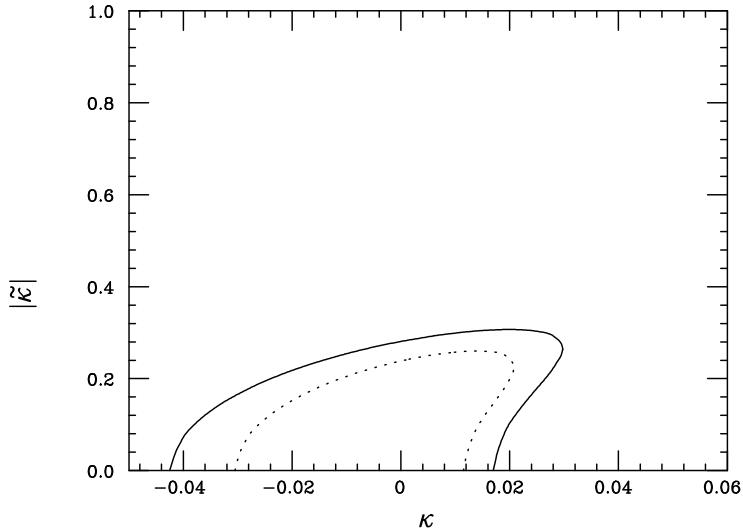,height=2.7in}
\caption{\label{top_mom}
Constraints on anomalous strong moments of the top quark, derived from
a LC sample of 100 fb$^{-1}$ (solid) and 200 fb$^{-1}$ (dotted) for
$\sqrt{s} = 1$ TeV.}
\end{center}
\end{figure}

This interaction leads to a substantially different spectrum of gluon radiation
for $e^+e^- \rightarrow t \bar{t}$ events above threshold than for the 
pure vector
interaction case corresponding to $\kappa = \tilde{\kappa} = 0$. Fits to this
spectrum thus provide limits on the values of $\kappa$ and $\tilde{\kappa}$.
Figure~\ref{top_mom}, from Ref.~\cite{rizzo_top},
shows the limits in the $\kappa$-$\tilde{\kappa}$ plane that
can be achieved with an integrated luminosity of 100 and 200 fb$^{-1}$
at $\sqrt{s} = 1$ TeV. Similar studies for the Tevatron and
 LHC~\cite{had_top_mom}
indicate that the corresponding sensitivities at hadron colliders will be
substantially weaker, in particular for the case of $\kappa$, for which
sensitivities of $|\kappa| < 0.1$ will be difficult to achieve.
In~\cite{kra_top}, the authors offer a technicolor model for which the
unique capability of the LC to measure strong moments of top precisely
would be a critical asset.

\section{Two-photon physics}

At a future $e^+ e^-$ linear collider, we will be able to study the
two-photon processes $e^+ e^- \rightarrow e^+ e^- + \gamma^{(*)}
\gamma^{(*)} \rightarrow e^+ e^- + {\rm hadrons}$ for all combinations
of real ($\gamma$) and virtual 
($\gamma^*$) photons.  Reactions of real photons can also be studied 
by using a dedicated backscattered-laser 
photon beam, as described in Chapter 13.
These reactions test
QCD in photon structure measurements and in the dynamics of parton
distribution function evolution.  Direct measurement of the photon
structure function $F_2^{\gamma} (x,Q^2)$ in $\gamma \gamma^*$ collisions
pushes into currently
unattainable regimes of 
lower $x$ and higher $Q^2$, testing scaling
behavior and $Q^2$ evolution.  Extending the measurement of the total
$\gamma \gamma$ cross section to higher $\sqrt{s}$ tests whether
QCD-based models of parton emission describe photon interactions.  By
colliding two virtual photons, QCD dynamics can be studied in a
relatively background-free environment.  No other planned or
anticipated future collider will be able to compete with an $e^+ e^-$
linear collider in these areas.

We now present a 
 comprehensive plan for the study of photon structure through $e
\gamma$ deep inelastic scattering (DIS) and $\gamma \gamma$ scattering,
and through the study of QCD dynamics through $\gamma^* \gamma^*$ scattering.
We discuss the  relative merits of employing photons produced
by bremsstrahlung and laser backscattering and the utility of having
well-defined photon  polarization.	

\subsection{Experimental requirements}
Experimental issues related to two-photon physics are mainly
concerned with instrumentation of the forward parts of the interaction
region (IR), particularly inside the conical shielding masks.
The cases in which the initial photons are produced by bremsstrahlung from 
$\ee$ and from laser backscattering have some  differences, but also 
many similarities.

\subsection{Bremsstrahlung photon beam}

In an IR designed for $\ee$ collisions, the study of two-photon 
processes requires 
small-angle-tagging electromagnetic 
calorimeters in the forward regions.  Some physics topics also require
hadronic calorimetry from beampipe to beampipe.

Virtual photons are produced when, in the bremsstrahlung process, an
$e^+$ or $e^-$ transfers a significant amount of 4-momentum to the
radiated photon.  The virtuality, $Q^2$, of the ``tagged'' photon is
determined by measuring the energy and angle of the scattered lepton
in an electromagnetic calorimeter via the relation
\beq
Q^2 = 2 E_e E_e^{\prime} (1 - \cos \theta) \ ,
\eeqn
where $E_e$ is
the incoming lepton beam energy, and $E_e^{\prime}$ and $\theta$ are the
scattered lepton energy and angle, respectively.  Since some
physics analyses require that the measurement of $Q^2$ be as small as
possible, the electromagnetic tagging calorimeters must be positioned
as closely as possible to the outgoing beampipes on both sides of the
interaction region and inside the shielding cone in order make the
minimum measurable scattered lepton angle as small as possible,
leading to the requirement of a compact design.  
Also, since $Q^2 \simeq E_e E_e^{\prime} \theta^2$ at small angles, radial
position resolution is an important consideration in $Q^2$ reconstruction,
requiring fine-grained readout in the radial direction~\cite{srm1}.  
Fine-grained
sampling calorimeters with these properties have been successfully used in
photon-tagging experiments at LEP~\cite{lepsiw}. 

Almost-real photons ($Q^2 \simeq 0$) from the 
bremsstrahlung process are defined by anti-tags in the forward
electromagnetic tagging calorimeters.  For
example, a single tag on one side of the IR, combined with an anti-tag
on the other side with hadronic activity in the main detector, signals
a $\gamma^* \gamma$ interaction  ($e \gamma$ DIS).  Double anti-tags
signal $\gamma \gamma$ interactions in which both interacting photons
are almost real.   It is important to note that the energy spectrum of
bremsstrahlung-produced photons is dominated by low-energy photons.  
Furthermore, since the untagged photon energy is not known, 
it is important to have hadronic energy and angle measurement in the
forward IR, to as small an angle as possible, in order to determine the 
kinematics
of the interaction.

\subsubsection{Backscattered laser beam}

It would be  desirable to create a beam of high-energy real photons
by Compton backscattering  of a high-power, 
high-repetition-rate laser from the electron beams.  The technology for 
achieving this backscattered-laser photon beam is described in 
Chapter 13.  To prepare the  Compton-backscattered beam,
1 eV laser photons backscatter from the 
incoming 250 GeV
$e^-$ beam, producing a beam of photons carrying about 75\% of the 
electron beam energy with an energy spread of
5--10\%.
 Since the resulting photon
beam energy spread is small, the kinematics of the high-energy photon
interactions can be determined from
the known photon energy.  Also, since these are high-energy photons
at nearly the
incoming lepton beam energy, the mass of the two-photon system 
$W_{\gamma* \gamma}$ is much 
larger than that obtained
from bremsstrahlung-produced photons, leading to the possibility
of reaching very low $x$ in
$e \gamma$ DIS.

In addition, the polarization state of the interacting 
photons and/or leptons can 
have a big effect on the physics impact of a measurement.  For example, by
combining the circular polarizations of the incoming leptons and 
the laser photons in an optimal way, the
energy spread of the resulting backscattered photon beam can 
be reduced by almost a factor of 2.

\subsection{Photon structure}
A real photon can interact both as a point-like particle, or as a
collection of quarks and gluons, {\it i.e.},  like a hadron.  The structure of the
photon is determined not by the traditional valence quark
distributions as in a proton, but by fluctuations of the point-like
photon into a collection of partons.  As such, the scaling behavior of
the photon structure function, $d F_2^{\gamma}/d \ln Q^2$, is always positive.
Single-tag and double-anti-tag events can be used to measure
$F_2^{\gamma}$ directly and to constrain the relative quark/gluon
fractions in the photon, testing predictions for this content and its behavior.

\subsubsection{$\gamma^* \gamma$ scattering---{\lowercase{e}}$\gamma$ DIS}
Direct measurement of the photon structure function $F_2^{\gamma} (x,Q^2)$
in $e \gamma$ DIS is accomplished by tagging a single virtual photon
probe, anti-tagging an almost-real or real target photon, and requiring
hadronic activity anywhere in the detector.

If the anti-tagged target photon is produced by bremsstrahlung from an
incoming lepton, it has very small virtuality, 
$\VEV{Q^2} \simeq 10^{-4}$ GeV$^2$, 
and low
energy, neither of which is known.  In order to determine the longitudinal
momentum fraction, $x$, the
mass $W_{\gamma^* \gamma}$ of the $\gamma^* \gamma$ system must be
measured, which requires hadronic calorimetry to measure the energy and angle
of all hadrons.  The best measurements of $F_2^{\gamma}$ using
bremsstrahlung photons as the target are done at relatively low
$W_{\gamma^* \gamma}$ where it is well-measured away from the forward
IR, which in kinematic space is at the high end of the $x,Q^2$ range.
Physics topics that can best be addressed in this region are the
scaling behavior of $F_2^{\gamma}$ as $x \rightarrow 1$ and its
 evolution with $Q^2$.

\begin{figure}[hbt]
\begin{center}
\epsfig{file=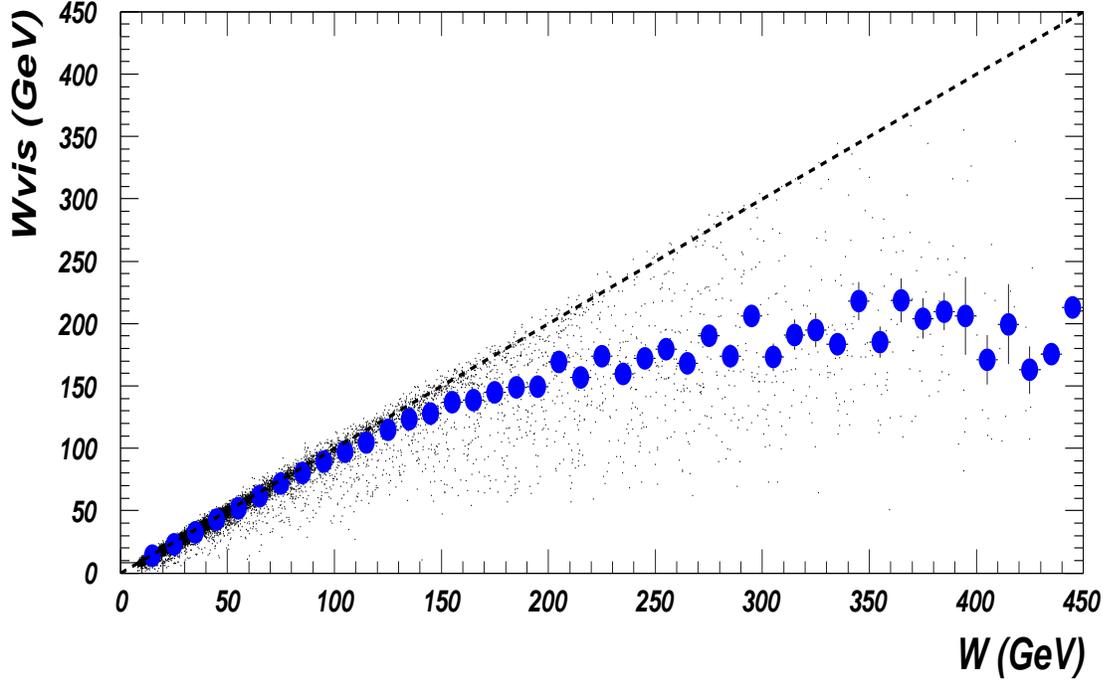,width=6in,height=4in}
\caption{\label{oranwvw}
Comparison of $W_{\rm vis}$ with $W_{\gamma^* \gamma}$ 
from PYTHIA~\cite{pythw} for a typical LC detector,
including the average
value (profile plot).}
\end{center}
\end{figure}

As $W_{\gamma^* \gamma}$ increases (towards low $x$),
increasingly more of the hadronic mass escapes undetected in the beam
direction and the mass of the observed hadrons, usually referred 
to as $W_{\rm vis}$, begins to differ substantially 
from the true hadronic mass.  
Figure~\ref{oranwvw} illustrates this effect by comparing 
$W_{\rm vis}$ with the true mass, $W_{\gamma^* \gamma}$.

Monte Carlo simulations of the fragmentation of the $\gamma^* \gamma$ system
are used to correct $W_{\rm vis}$
for this loss until the uncertainty in the correction begins
to dominate the measurement.  Eventually, this limits the low-$x$
range of the $F_2^{\gamma}$ measurement.

However, if the target photon is produced by laser backscattering,
two advantages are realized: 1) the high $W_{\gamma^* \gamma}$ 
(low-$x$) region 
is enhanced since the real photon energy
is high; and 2) the energy spread of the real photons is small enough
that the error on $x$ caused by assuming a monochromatic photon
does not dominate the systematics.

\begin{figure}[hbt]
\begin{center}
\epsfig{file=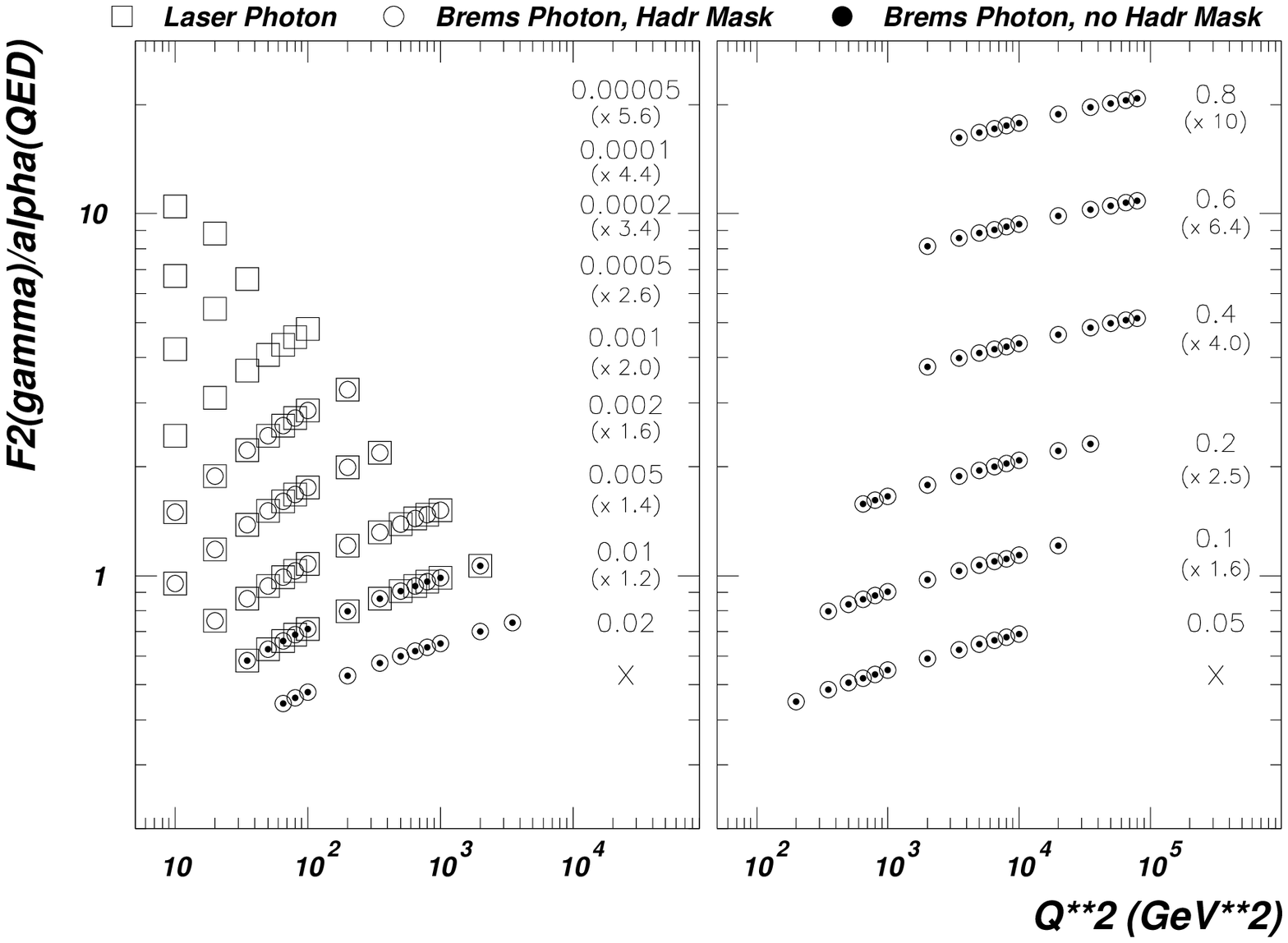,width=6.0in}
\caption{\label{f2fig}
$F_2^{\gamma}/\alpha$ versus $Q^2$ in $x$ bins.  Open squares:
real photon
target from laser backscattering; open circles: almost-real
photon target from
bremsstrahlung with small-angle hadronic calorimetry; solid dots:
almost-real photon target from bremsstrahlung with hadronic calorimetry 
outside mask.}
\end{center}
\end{figure}

Figure~\ref{f2fig} shows $F_2^{\gamma}$ versus $Q^2$ for various $x$
bins from possible measurements at a future $e^+e^-$ 
linear collider~\cite{srm2}.
The various points are differentiated according to the measurement method. 
The open squares represent the very low-$x$ region accessible only with photons
produced by laser backscattering; open
circles represent measurements with
target photons from bremsstrahlung and with hadronic calorimetry built into a
shielding mask down to $30$~mrad; solid dots represent measurements 
with
bremsstrahlung photons and with hadronic calorimetry only outside the mask.
Note that there is enough overlap between the methods to 
provide cross-checks on
the various measurements and experimental conditions. 

With known polarization of both the target photon and the tagged
virtual photon, polarized photon structure functions can be measured
for the first time.  The `BFKL' terms involving $\ln(1/x)$ in the 
unpolarized structure functions enter in polarized 
scattering as $\ln^2(1/x)$. These effects are then enhanced at low $x$ over 
the unpolarized case. Thus, in polarized $e \gamma$ DIS,
forward particle and jet measurements, such as have been performed at
HERA~\cite{forwj}, 
can be done at a future $e^+ e^-$ linear collider with increased sensitivity to
any BFKL effects. 

In addition to the $F_2^{\gamma}$ structure function, $e \gamma$ DIS
can be used to test QCD in other ways.  For example, dijet production
in DIS can be used to extract the strong coupling parameter,
$\alpha_s$, as is done at HERA~\cite{zeus1}.  
At a future $e^+ e^-$ linear collider,
$\alpha_s$ from $e^+ e^-$ event shapes and from dijets in DIS 
can be compared using the same detector.

\boldmath
\subsection{$\gamma \gamma$ scattering---total cross section}
\unboldmath

\begin{figure}[hbt]
\begin{center}
\epsfig{file=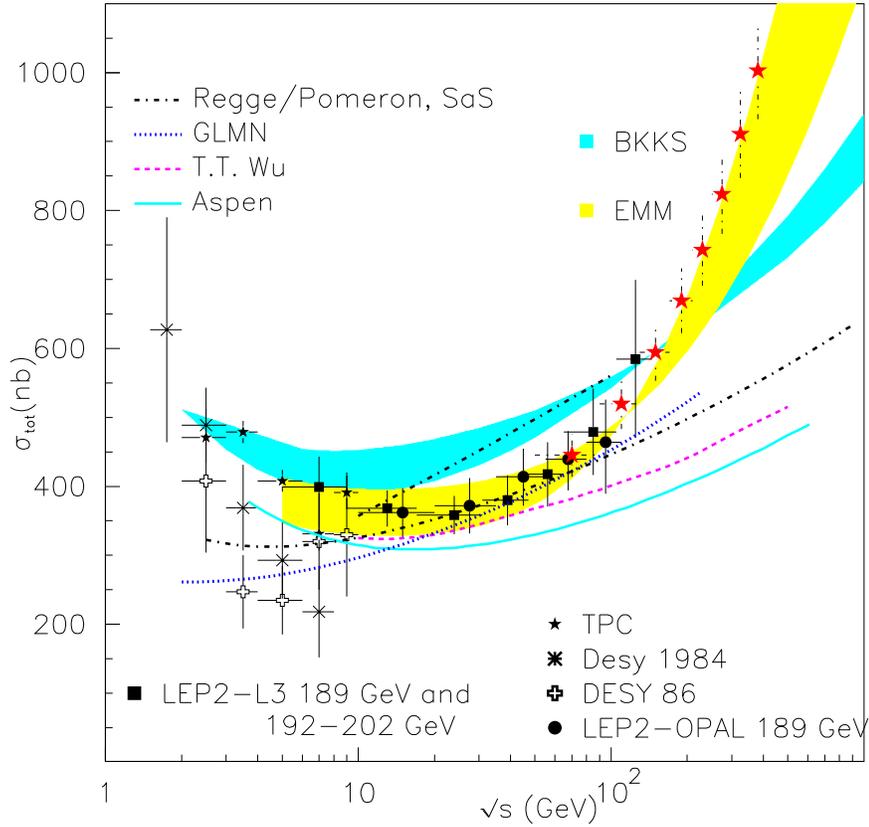,width=5in}
\caption{\label{orangg}
$\sigma_{\rm tot}$ versus $\sqrt{s}$ at a LC (large stars) compared to 
existing data and various models.}
\end{center}
\end{figure}

Various models have been developed to describe the rise with energy of
the total $\gamma \gamma$ cross section.  These give either 
a fast rise driven by
QCD effects such as minijets, or a slower rise based on reggeon
exchange.  To get to the highest $\sqrt s$ and $W_{\gamma \gamma}$,
real photons from the laser backscattering process are required.
Studies show that a precision of $\sim 20 \%$ on the total cross
section will enable adequate discrimination of model types for
energies up to $1$ TeV~\cite{rohini}.  Figure~\ref{orangg} shows possible
$\sigma_{\rm tot}$ measurements at a 500  GeV linear 
collider (large stars) compared
to existing measurements at lower $\sqrt{s}$ and to various models.

Using dijets from $\gamma \gamma$ scattering, the relative quark/gluon
structure of the photon can be determined.  Interactions between the
almost-real photons produced by bremsstrahlung are determined primarily by 
interacting gluons in the ratio
of approximately $70 \%$ gluons to  $30 \%$ quarks.
At higher $\sqrt s$, the gluon component should be more predominant.  
Thus, if real photons from laser backscattering are used, we expect to find 
an almost pure gluon-constituted photon ($90 \% g / 10 \% q$)~\cite{adr}.

\boldmath
\subsection{$\gamma^* \gamma^*$ scattering---QCD dynamics}
\unboldmath

Double-tagged virtual photon scattering completes the study of the
photon at the linear collider by allowing the evolution of photon
structure to be studied in an almost background-free environment.  The
$Q^2$ of each of the scattered leptons 
(denoted $Q_1^2$ and $Q_2^2$)
is measured in the forward electromagnetic
tagging calorimeters.
By requiring the
ratio $Q_1^2/Q_2^2 \sim 1$, production of hadrons in the region between the two
virtual photons through traditional
DGLAP evolution is suppressed. This suppression grows stronger as the
rapidity separation, $Y$, between the two virtual photons increases.
At large values of $Y$, any signal above the small DGLAP background
points to alternative forms of structure function evolution, \eg,  to
the $\ln (1/x)$ evolution of BFKL~\cite{bfkl}.  
Virtual photon scattering at a linear 
collider provides perhaps the cleanest environment in which to study
BFKL physics~\cite{brodskyetal,deroeck}.  

With total center-of-mass energy $\sqrt s$ and
photon virtuality $Q^2$, 
BFKL effects are expected in the kinematic region where the
square of the photon-photon invariant mass (or, equivalently, the
hadronic final-state system) is large, and
$$s \gg Q^2 \gg \Lambda_{QCD}^2.$$
At fixed order in QCD, the dominant process is four-quark production with
$t$-channel gluon exchange.  Each photon couples to a quark box, and 
the  quark boxes are connected via the gluon.
The corresponding BFKL
contribution arises from diagrams in which the 
$t$-channel gluon becomes a gluon ladder.
  At lepton-hadron or hadron-hadron colliders, the 
presence of hadrons in the initial state can complicate or even mask 
BFKL effects.

\begin{figure}[hbt]
\begin{center}
\epsfig{file=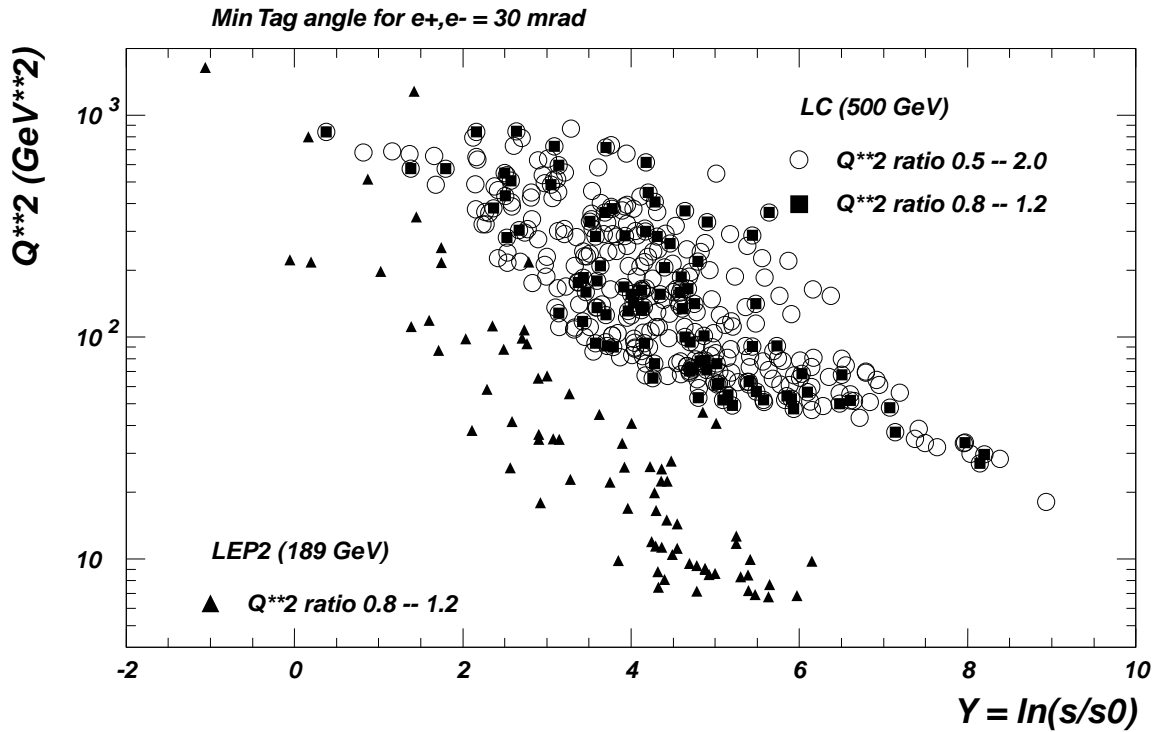,width=6.0in}
\caption{\label{orangsgs}
$Q^2$ versus $Y$ for a 500  GeV LC compared to LEP2.}
\end{center}
\end{figure}

The largest values of $Y$ are
obtained at low $Q_{1,2}^2$, again emphasizing the need for the
electromagnetic tagging calorimeters to be positioned as close to
the beampipe as possible.   Figure~\ref{orangsgs} shows the substantially
greater reach in $Y$ available to the 500 GeV LC relative to that of 
LEP2 running
at $189$ GeV.

Experiments at LEP have looked for BFKL effects in virtual photon 
scattering~\cite{LEP}.  The data tend to lie between the predictions of
fixed-order QCD and analytic solutions to the BFKL equation 
(asymptotic full-order QCD).
However,
the data were compared to the asymptotic QCD prediction in
a non-asymptotic regime~\cite{os}, so the disagreement with QCD is not 
surprising.  In contrast, a linear collider will be expected to reach closer
to 
the asymptotic regime, providing a more definitive test of BFKL evolution.
Improved predictions are also on the way with the development of
BFKL Monte Carlo programs that incorporate kinematic constraints, such
as~\cite{bfklmc}.  On the more theoretical front, next-to-leading log
corrections have been calculated and found to be large, but 
the source of the large corrections is understood and they
are being brought under control; see~\cite{salam} for a review and references.

\subsection{Summary of two-photon physics}

The study of 
two-photon physics from $e^+ e^-$ collisions has grown tremendously in
the past several years of higher-energy LEP2 running and will continue
to provide a wealth of precision measurements at a future $e^+ e^-$ linear
collider.  Using combinations of tagged and untagged bremsstrahlung
photons, aspects of real and virtual photon structure will be
addressed, especially $F_2^{\gamma}$ at high $Q^2$, the relative
quark/gluon content of the photon from dijets, and possible BFKL
effects in QCD evolution.

With laser-backscattered real photons, the highest energies available at
the linear collider can be fully exploited.  $F_2^{\gamma}$ can be
measured at very low $x$, which in combination with high $Q^2$
measurements from bremsstrahlung photons, will map out a kinematic
region in photon structure as extensive as that known for the proton.
The total $\gamma \gamma$ cross section will also be measured at the
highest $\sqrt{s}$ available at the linear collider, leading to
understanding of the dominant mechanisms responsible for this interaction.

Finally, with combinations of lepton and photon polarization, BFKL
effects can be enhanced and the first measurements of polarized
structure functions of the photon can be made.

\section{Overall summary and conclusions}

The high-energy linear collider offers a unique program of QCD and related
two-photon studies.
The strong
coupling constant $\alphas$ can be measured at high $Q^2$ to a precision
approaching $\pm 1\%$, free of the initial-state ambiguities that make the
corresponding determination at a hadron collider substantially less precise,
and allowing for substantial improvements in the determination of the
running of the QCD coupling strength, as well as its extrapolation to the GUT scale.
Constraints on the strong coupling properties of the top, providing sensitivity
to a number of new physics scenarios inspired by the large 
mass of the top quark,
can be made as much as an order of magnitude more stringent at an $e^+e^-$
collider than at a proton collider of equivalent reach.

In two-photon reactions, the precisely defined state of the incoming electron
and positron beams permits the kinematic properties of the interacting virtual
and nearly on-shell photons to be inferred from the properties of the recoiling
electrons. This in turn allows for a unique program of photon structure and
strong-force dynamics which cannot be emulated by any other proposed facility.
In addition, the possibility of precisely controlled real 
photon beams from the
Compton backscattering of polarized laser light opens up further vistas
in the exploration of photon structure, and may allow the resolution of
long-standing questions regarding the energy evolution of the photon-photon
total cross section. Again, these studies are only possible within the larger
context of an $e^+e^-$ linear collider program.

Together, these physics topics present a unique and compelling program of
strong-interaction studies at a high-energy linear collider, one that
adds substantial weight to the promise of the proposed linear
collider physics program.

\setcounter{chapter}{7}

\chapter{Precision Studies at the \boldmath $Z$ and the $WW$ \unboldmath Threshold}
\fancyhead[RO]{Precision Studies at the $Z$ and $WW$ Threshold}

A high-precision program of electroweak and heavy-quark physics provides
a natural complement to the direct searches for the Higgs boson and
other new particles.  The study of loop corrections to the electroweak
parameters measured at the $Z$, in $p\bar{p}$ collisions and in neutrino
experiments made impressive indirect predictions for the top quark mass,
and constrains the mass range for a Standard Model Higgs.  Limits on
${\cal B}(B\to X_s\gamma)$ provide the tightest mass limits on type II
Higgs doublets.  Because the new particles appear virtually in loops,
the sensitivity extends over a much higher mass range than can be
obtained in direct searches, though generally at the expense of some
model--dependence.

While the physics program at 500 GeV has the potential to be very rich,
it is also possible that at this center-of-mass energy there is only one
Higgs-like particle seen, or no such particle at all.  Under either
scenario, the constraints from the electroweak and heavy-quark studies
can be powerful.  In the case that we do see a plethora of new
particles, the full spectrum of states predicted by any model must
satisfy the rules dictated by the precision measurements.  In the case
that very little is seen directly, the precision low-energy measurements
have a good chance of showing deviations from the Standard Model.  These
deviations will indicate the direction that future studies must take.

There remain open issues with respect to implementing a low-energy
program at a linear collider.  If only the basic electroweak program is
undertaken, the goals may be met by devoting a modest amount of running
time at low energy.  A single facility for both the high-energy and the
$Z$ running, however, requires incorporation of this capability into the
design of the accelerator.  For a broader program,
including running at \ww threshold and extended running at the $Z$ pole
for heavy flavor physics, a low-energy facility that can operate in
parallel with the high-energy may be required.

\section{Electroweak observables on the  $Z$  resonance}

In principle, all measurements done at LEP and SLC can be repeated at
the linear collider with much higher statistics.  In about 100 days of
running, it is possible to collect a sample of $10^9$ $Z$ decays
(`Giga-Z'), about 100 times the LEP or 1000 times the SLC statistics.  A
high degree of electron polarization seems certain and $\pmi = 80\%$
will be assumed in the following.  Positron polarization is 
desirable and the R\&D to achieve it is under way.  Both options, with
and without positron polarization, will be discussed.  The issue of
positron polarization is discussed further in Chapter 12.

\subsection{Machine issues}

In the present designs, the linear collider can deliver a luminosity
${\cal L} \sim 5 \times 10^{33} {\rm cm}^{-2} {\rm s}^{-1}$ at the $Z$
resonance.  The energy loss due to beamstrahlung for colliding particles
is around $0.05\%-0.1\%$ and the depolarization in the interaction
region is negligible.  By sacrificing some luminosity, beamstrahlung can
be reduced substantially, for example, by a factor three for a
luminosity loss of a factor two \cite{peter_mike}.

Apart from the beamstrahlung there are several other effects that
influence the precision of the measurements:
\begin{itemize}
\renewcommand{\itemsep}{0pt}
\item The mean energies of the two
  beams have to be measured very precisely.
  A precision of $10^{-5}$
relative to the $Z$ mass might be needed to
  relate $A_{LR}$ to $\sweff$ with the desired precision.
\item The beam energy spread of the machine plays a crucial role in
  the measurement of the total width of the $Z$. If the shape of the
  distribution is known, the width can be measured from the acolinearity of
  Bhabha events in the forward region as long as the energies of the
  two colliding particles are not strongly correlated.
\item With the high luminosities planned, the $Z$ multiplicity in a train
  becomes high. This can influence $Z$ flavor tagging or even $Z$ counting.
\item With positron polarization, the positron source must be able to switch
  polarizations on a time scale commensurate with the stability of the beam
  conditions.
\end{itemize}
The two main designs, X-band and superconducting, differ in some aspects
relevant for $Z$ running.  For the X-band design a bunch train contains
190 bunches with 1.4~ns bunch spacing, for which over half of the $Z$
bosons are produced in the same train as at least one other $Z$.
Typical event separation is about 150~ns, but the experimental
consequences merit some study.  A TESLA bunch contains 2800 bunches with
280~ns bunch spacing.  In this case bunch separation is not a problem,
but data acquisition system requirements are higher.  The smaller
wakefields in the superconducting machine should reduce the beam energy
spread.  The larger bunch spacing may allow sufficient time for energy
feedback, resulting in a smaller energy difference between
the bunches in a train.

The LC design must accommodate the needs of the precision electroweak
program in advance for the program to be viable.  Suitable space in the
beam delivery system for precise beam energy measurement and for
polarimetry must be provided, or the beam energy measurement must be
directly incorporated into the Final Focus magnet system.  A measurement
of these quantities behind the IP is also desirable, though it is difficult.  
A nonzero crossing angle might  be needed.

\subsection{Electroweak observables}
There are three classes of electroweak observables that can be
measured during $Z$-running at a linear collider:
\begin{itemize}
\renewcommand{\itemsep}{0pt}
\item observables related to the partial widths of the $Z$, measured in
  a $Z$ resonance scan;
\item observables sensitive to the effective weak mixing angle;
\item observables using quark flavor tagging.
\end{itemize}
Table \ref{tab:line} summarizes the present precision and the
expectations for the linear collider for these quantities.

\begin{table}
\begin{center}
\begin{tabular}[c]{ccc}
\hline \hline
 & LEP/SLC/Tev \cite{ref:lepew} & LC \\
\hline
$\sweff$ & $0.23146 \pm 0.00017$ & $\pm 0.000013$\\
\hline
\multicolumn{3}{l}{lineshape observables:}\\
\hline
$\mz$ & {$ 91.1875 \pm 0.0021 \GeV$} & {$ \pm 0.0021 \GeV$} \\
$\alpha_s(\mz^2)$ & {$ 0.1183 \pm 0.0027 $} & {$ \pm 0.0009$} \\
$\Delta \rho_\ell$ & {$ (0.55 \pm 0.10 ) \times 10^{-2}$}
& {$ \pm 0.05\times 10^{-2}$}  \\
$N_\nu$ & {$ 2.984 \pm 0.008 $} & {$ \pm 0.004 $} \\
\hline
\multicolumn{3}{l}{heavy flavors:}\\
\hline
$\cAb$ & $0.898 \pm 0.015$   &$\pm 0.001$ \\
$\Rb^0$ &$0.21653 \pm 0.00069$ & $\pm 0.00014$  \\
\hline\hline
\end{tabular}
\end{center}
\caption{\label{tab:line}
Possible improvement in the electroweak physics quantities
for $10^9$ $Z$'s collected at a linear collider.
$N_\nu=3$ is assumed for $\alpha_s$ and $\Delta \rho_\ell$.}
\end{table}

\subsubsection{Observables from the  $Z$  resonance line scan}

From a scan of the $Z$ resonance curve the following quantities are measured:
\begin{itemize}
\renewcommand{\itemsep}{0pt}
\item the mass of the $Z$ ($\mz$);
\item the total width of the $Z$ ($\gz$);
\item the hadronic pole cross section
  ($\so = (12\pi/\mz^2)\cdot (\Gamma_e \Ghad/\Gamma_Z^2) $);
\item the ratio of the hadronic to the leptonic width of the $Z$
($R_\ell = \frac{\Ghad}{\Gamma_l} $).
\end{itemize}
From these parameters, two
interesting physics quantities can be derived: the radiative correction parameter
$\Delta \rho_\ell$ that normalizes the $Z$ leptonic width, and the strong
coupling constant $\alpha_s$.

The LEP measurements are already systematics-limited, so statistical
improvement is not the issue.  From LEP, $\mz$ is known to $2\times
10^{-5}$, and the other three parameters are all known to $10^{-3}$.  To
improve on $\alpha_s$ and especially on $\Delta \rho_\ell$, all three
measured parameters must be improved.  This requires one to understand
the beam energy and the beam energy spread for $\gz$, the hadronic and
leptonic selection efficiencies for $R_\ell$, and the absolute
luminosity for $\so$.  With the better detectors and the higher
statistics available for cross checks, the errors on the selection
efficiency and on the luminosity might be improved by a factor of three
relative to the best LEP experiment \cite{ref:marc}.  It is not clear whether
the theory error on the luminosity can be improved beyond its present
value of $0.05\%$.  These errors would improve the precision on $R_\ell$
by a factor of four and that on $\so$ by 30\%.

With a M{\o}ller spectrometer, one could possibly obtain a
precision of $10^{-5}$ in the beam energy relative to $\mz$.  This would
give a potential improvement of a factor of two in $\gz$.  However,
because the second derivative of a Breit-Wigner curve at the maximum is
rather large, $\gz$ and $\so$ are significantly modified by
beamstrahlung and beam energy spread.  For illustration, the fitted
$\gz$ is increased by about $60 \MeV$ and $\so$ is decreased by 1.8\%
for the TESLA parameters.  The energy spread dominates the effect, so 
this particularly needs to be understood to about 2\% to avoid limiting the
precision on $\gz$ and $\Delta \rho_\ell$.  There is a potential to
achieve this precision with the acolinearity measurement of Bhabha
events \cite{ref:bsmeas} or to extend the scan to five scan points and
fit for the energy spread, but both options need further study.

\subsubsection{The effective weak mixing angle}
If polarized beams are available, the most sensitive quantity by far to
the weak mixing angle is the left-right asymmetry:
\begin{Eqnarray}
\ALR & = & \frac{1}{{\cal P}}\frac{\sigma_L-\sigma_R}{\sigma_L+\sigma_R}\CR
     & = &   \cAe \CR
     & = & \frac{2 v_e a_e}{v_e^2 +a_e^2} \CR
  {v_e}/{a_e} & = & 1 - 4 \sweff.
\label{eq:alrdef}
\end{Eqnarray}
$\ALR$ is independent of the final state.

The $\ALR$ measurement has been analyzed for the linear collider
environment in \cite{kmogigaz,peter_mike}.  With $10^9$ $Z$'s, an
electron polarization of 80\% and no positron polarization, the
statistical error is $\Delta \ALR = 4 \times 10^{-5}$.  The error from
the polarization measurement is $\Delta \ALR/\ALR = \Delta {\cal
P}/{\cal P}$.  At SLC, $\Delta {\cal P}/{\cal P}=0.5\%$ has been reached
\cite{alrsld}.  With some optimism a factor two improvement in $\Delta
{\cal P}/{\cal P}$ is possible \cite{peter_mike}.  In combination with the improved
statistics, this leads to $\Delta \ALR = 3.8 \times 10^{-4}$.  This
precision is already more than a factor of five improvement over the
final SLD result for $\sweff$ and almost a factor of four over the
combined LEP/SLD average.

\begin{figure}[t]
\centering
    \epsfig{figure=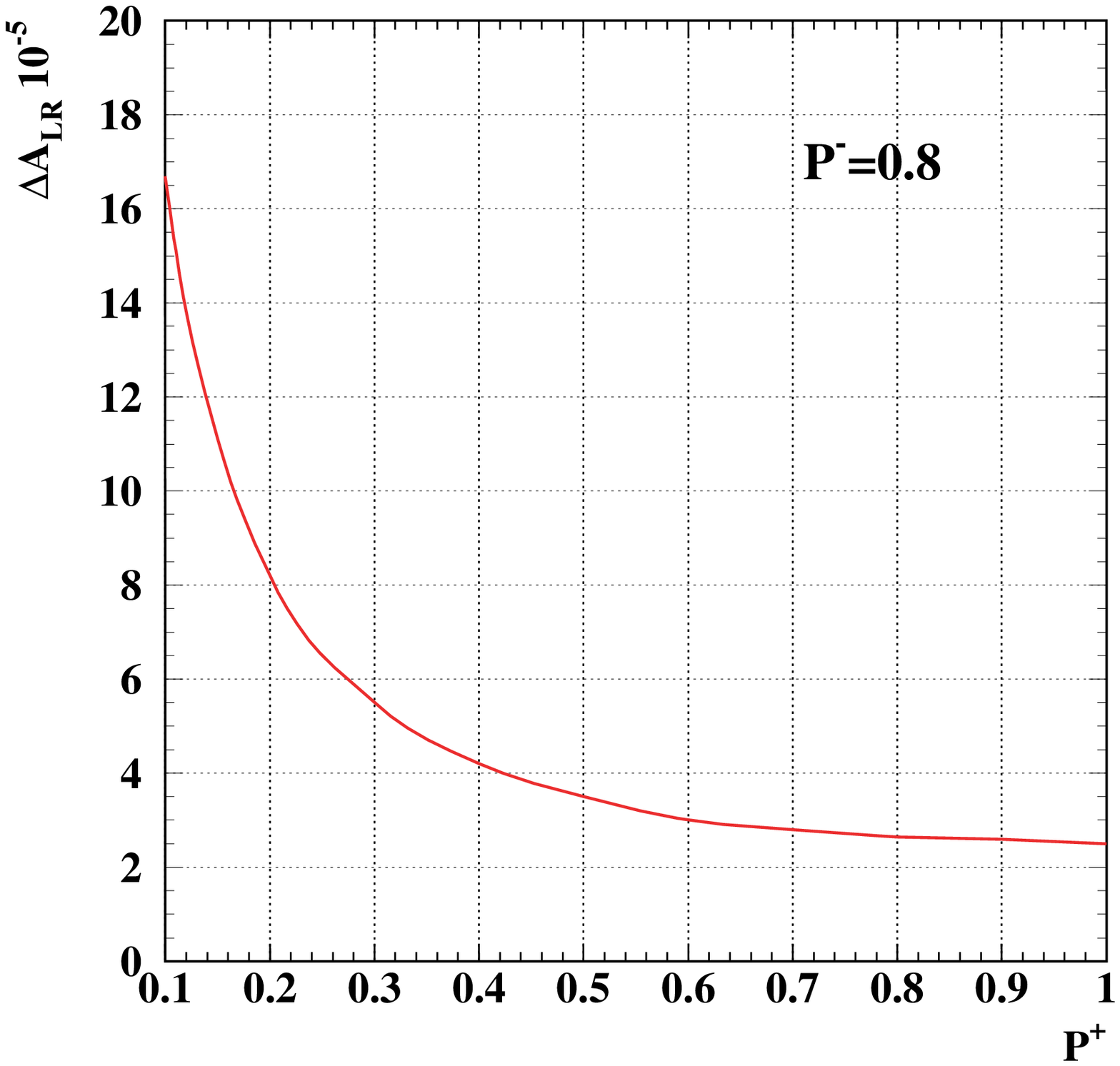, height=6cm}
    \hspace{3mm}
    \epsfig{figure=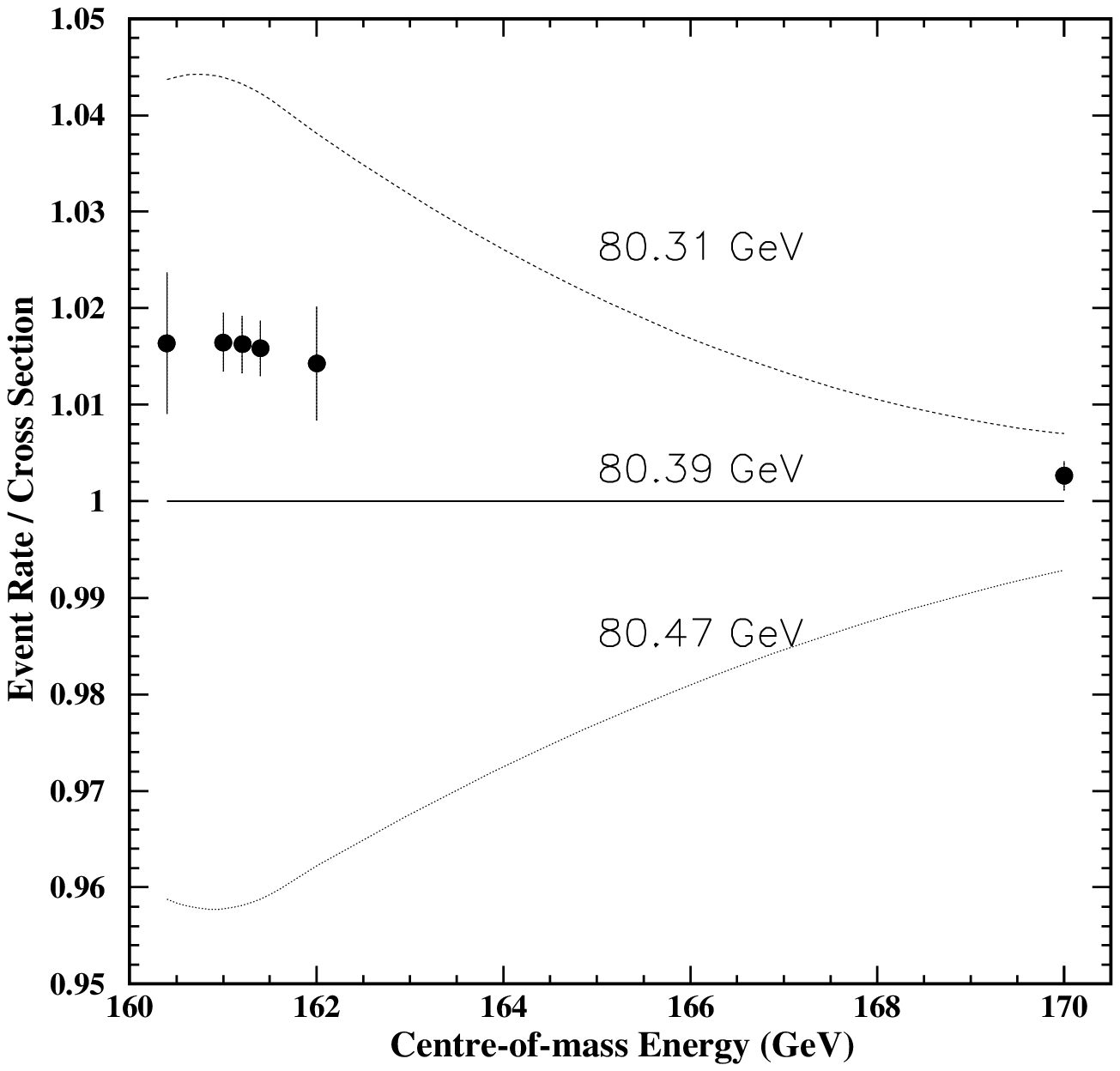, height=6cm}
  \caption[*]{ \label{fig:polerr}
Left: Error of $\ALR$ as a function of the positron
    polarization for a luminosity corresponding to $10^9$ unpolarized $Z$'s.
   The errors assume that switching of the positron polarization can be done
    on a time scale over which the beam conditions are suitably stable.
    Right: The ratio of the measured \ww cross section to
    the predicted cross section for $M_W=80.39$ GeV (see
    Section~\ref{sec:wwscan}).  The data were generated using
    $M_W=80.36$ GeV.  The upper (lower) curves show the ratio of
    the predicted cross section for $M_W=80.31$ GeV ($M_W=80.47$ GeV)
    to that for $M_W=80.39$ GeV.}
 \end{figure}

If positron polarization is available,  there is the potential to go
much further using the `Blondel scheme' \cite{ref:alain}. This method of
polarization measurement, and the associated techniques for obtaining
polarized positrons, are described in more detail in Chapter 12.  To
summarize the results,
the total cross section with both beams polarized is given as
  $\sigma \, = \, \sigma_u \left[ 1 - \ppl \pmi + \ALR (\ppl - \pmi) \right]$,
where $\sigma_u$ is the unpolarized cross section.
If all four helicity combinations are measured, $\ALR$ can be determined
without polarization measurement as
\[
\ALR \, = \, \sqrt{\frac{
    ( \sigma_{++}+\sigma_{-+}-\sigma_{+-}-\sigma_{--})
    (-\sigma_{++}+\sigma_{-+}-\sigma_{+-}+\sigma_{--})}{
    ( \sigma_{++}+\sigma_{-+}+\sigma_{+-}+\sigma_{--})
    (-\sigma_{++}+\sigma_{-+}+\sigma_{+-}-\sigma_{--})}} \ .
\]
Figure \ref{fig:polerr} shows the error on $\ALR$ as a function of the
positron polarization.  For $\ppl > 50\%$ the dependence is relatively weak.
For $10^9$ $Z$'s, the Blondel scheme with a positron polarization of 20\%
gives a better result than a
polarization measurement of $0.1\%$ and electron polarization only.

Polarimeters are still needed to resolve one remaining question.  There
could potentially be a difference between the absolute values of the
polarization in the left- and right-handed states.  If the two
polarization values for electrons and positrons are written as $\ppm =
\pm |\ppm| + \delta \ppm$, the dependence on this difference is $\rm{d}
\ALR / \rm{d} \delta \ppm \approx 0.5$.  One therefore needs to
understand $\delta \ppm $ to $< 10^{-4}$.  If polarimeters with at least
two channels are available, $\delta P$ can be measured together with other
systematic effects intrinsic to the polarimeters in a way that does not
increase the statistical error from the Blondel scheme.

Because of  $\gamma$--$Z$ interference, the dependence of $\ALR$ on the beam
energy is $\rm{d} \ALR / \rm{d} \sqrt{s}$ $ = 2 \times 10^{-2}/\GeV$.  The
difference $\sqrt{s} - \mz$ thus needs to be known to about $10 \MeV$ to
match the measurement with electron polarization only, and to about
$1\MeV$ if polarized positrons are available.  For the same reason
beamstrahlung shifts $\ALR$. The shift is $9\times 10^{-4}$ for TESLA and is
larger for NLC/JLC \cite{peter_mike}.  The uncertainty can only be a few percent.  If
beamstrahlung in the $\ALR$ running is identical to that in the $Z$ scan
used to calibrate the beam energy, the effect is absorbed into the mean
energy measured in the calibration.  In that case, practically no
correction would be needed for $\ALR$.  How well the beam parameters can
be kept constant during the scan and how well the beamstrahlung can be
measured still need further study.  However, for $\ALR$, only the
beamstrahlung and not the energy spread matters.  If the beamstrahlung
cannot be understood to the required level in the normal running mode
one can still go to a mode with lower beamstrahlung at the expense of
lower luminosity.  The cost is an increase in the statistical error or
the running time.

Finally, the rate at which the positron polarization must be switched,
and the switching rates that are achievable are still unknown.

For the interpretation of the data it will be assumed that $\Delta \ALR
=10^{-4}$ is possible.  This leads to $\Delta \sweff = 0.000013$.  It
must be kept in mind that this error will increase by a factor of four
if no positron polarization is available.

\subsubsection{Observables with tagged quarks}

By the use of quark tagging in addition to the observables discussed
above, the partial widths and forward-backward asymmetries for $b$  and
$c$ quarks can be measured.  These observables are sensitive to vertex
corrections at the $Zqq$ vertex and to new Born-level effects that alter
the SM relations between quarks and leptons.  The $Zbb$ vertex is
particularly interesting, since the $b$ is the partner of the heavy top
quark, and since the vertex corrections are naturally enhanced with the
quark mass.

To date, only the improvement to the $b$-quark observables has been
estimated \cite{kmogigaz}.  For the ratio $\Rb$ of the $Z$ partial
widths to $b$ quarks and to hadrons, an improvement of a factor five to
the LEP/SLD average is possible.  This improvement is due to the much
better $b$ tagging than at LEP.  The improved tagging results in a
higher purity (over 99\% for a 30\% efficiency) and a smaller
energy dependence, which in turn reduces the hemisphere correlations.

The forward-backward asymmetry with unpolarized beams measures the
product of the coupling parameters for the initial-state electrons and
the final-state quarks:  $A_{\rm FB}^q = \frac{3}{4} \cAe \cAq$, while
the left-right forward-backward asymmetry with polarized beams measures
the quark couplings directly:  $A_{\rm LR,FB}^q = \frac{3}{4} {\cal P}
\cAq$.  For this reason a factor 15 improvement on $\cAb$ relative to
the LEP/SLC result is possible if polarized positrons are available, and
if other systematic effects are relatively small.
With polarized electrons only, the improvement is limited by the
polarization error to a factor of six.  For control of systematics, the
improved $b$-tagging capabilities are essential here as well.

Though the SM predicts that $Z$ decays to quarks are flavor-diagonal
to a very good approximation, loop effects of new physics can induce
flavor-violating rare decays~\cite{Buchalla:2001sk}.
These could be searched for at a high-luminosity $Z$ factory.
For $Z \to b \bar{s}$ decays, the SM predicts a branching ratio
of ${\cal{B}}(Z \to b \bar{s}) \simeq 1.4 \cdot 10^{-8}$.  To date,
the direct experimental bound on this process is relatively weak, at the
level of about $10^{-3}$~\cite{gudrun-L3}, though bounds from rare
$b$ decays such as
$b \to s \ell^+ \ell^-$ and $b \to s \nu \bar{\nu}$ lead to a bound
${\cal{B}}(Z \to b \bar{s}) \lsim 5 \cdot 10^{-7}$ \cite{Buchalla:2001sk}.
Still, there is room for a new physics contribution that might be revealed
in a large sample of $Z$ decays.

\section{$m_W$ from $WW$ threshold running  \label{sec:wwscan}}

The mass $\mw$ of the $W$ boson plays a fundamental role in constraints
on the Standard Model via comparison of direct measurement with the
prediction based on other electroweak parameters.  The electroweak
measurements from LEP1 and Giga-Z---combined with the Higgs boson and
top quark mass measurements from the linear collider---allow $\mw$ to be
predicted to about 3 MeV within the SM.  Measurements at the Tevatron
and at LEP2 combine to give an $\mw$ precision of 34 MeV
\cite{bb:lepewwg}.  The LEP2 experiments hope to reach a combined
precision of 35 MeV. With Run II at the Tevatron, 30 MeV per experiment
appears feasible with 2 $\textrm{fb}^{-1}$, though systematics,
correlated between experiments, will dominate~\cite{Signore98}.  The LHC
experiments hope to reach an uncertainty of 20 MeV each, for perhaps an overall
uncertainty of 15 MeV~\cite{bb:atlas_lhc_mw}.  Unfortunately, these
uncertainties remain significantly larger than that expected for the
indirect determination and would limit the power of the electroweak
constraints.

A high-luminosity linear collider presents an opportunity to measure
$\mw$ with a much higher precision.  The two potential approaches
\cite{bb:mw_working} are a \ww threshold scan and kinematic fitting of
events with $W^+W^-$ production.  With expected linear collider
luminosities, one could obtain 100 \invf\ in one year ($10^7$ s) at
$W^+W^-$ threshold and about 1000 \invf\ at $\sqrt{s} = 500$ GeV in
several years.  The threshold scan requires precise determination of the
absolute average beam energy and of the distortion of the luminosity
spectrum by beamstrahlung.  The kinematic fitting method also requires
precise knowledge of the beam energy, since it relies on a beam energy
constraint.  The uncertainty from this parameter will grow with energy,
since beam calibration will likely refer back to the $Z$ peak.
Furthermore, the energy spread from beamstrahlung grows approximately as
the square of the beam energy.

The four-quark ($4q$) channel (46\% of the rate) cannot be used in the
kinematic analysis because of theoretical uncertainties associated with
final-state interactions between the decay products of the $W^+$ and the
$W^-$.  This uncertainty contributes an error of 40--90 MeV for the
current LEP $4q$ measurements
\cite{bb:aleph_mw,bb:delphi_mw,bb:opal_mw,bb:l3_mw}.  Scaling of the
LEP2 statistical precision for the remaining channels results in a 5 MeV
$\mw$ precision at 500 GeV.  However, significant reductions in
systematics will be needed.  In particular, the difficulties in beam
energy calibration disfavor the direct reconstruction method.

\subsection{$\mw$ from a polarized threshold scan}

The extraction of $\mw$ from a threshold scan requires an accurate
theoretical description of the cross-section dependence on $m_W$.  The
main corrections to the Born approximation near threshold come from QED.
Fortunately, the dominant Coulomb correction (about 6\%) is already
known to all orders \cite{bb:CoulCorr}.  To keep the theoretical
uncertainty down to 2 MeV, however, the electroweak and QCD corrections
to the cross section must be known to 0.12\% (about the size of the
second-order Coulomb contribution).  While work is needed, this goal
appears attainable.

Recent studies \cite{bb:graham_writeup,bb:graham_talk} indicate that
experimental systematics can be controlled to obtain a 5 MeV $\mw$
measurement with 100 \invf of data {\em if} a polarization of 60\% for
the positron beam can be achieved.  The strategy capitalizes on the
domination of the \ww cross section near threshold by the $t$--channel
$\nu_e$ exchange process, which couples only to the $e_R^+e_L^-$
helicity combination.  The correct $e_R^+e_L^-$ beam polarization
enhances \ww production relative to the background, while the
$e_L^+e_R^-$ polarization has almost negligible \ww production and so
can constrain the background levels.

A sample scan is illustrated in Fig.~\ref{fig:polerr}.  This study
assumes that the absolute luminosity and the reconstruction efficiencies
can be determined with a relative (point\-to-point) accuracy of 0.25\%.
This is four times looser than that achieved for the LEP1 $Z$ line-shape
scan.  Beam polarizations are assumed known to 0.25\%, and are further
constrained at each scan point by exploring various polarization
combinations.  About 90\% of the luminosity is given to the main
$e_R^+e_L^-$ to $e_R^+e_L^-$ configurations, in a 5:1 ratio, with the
10\% devoted to the remaining configurations to determine the beam
polarization.  LEP signal efficiencies and background rates
\cite{bb:oxford97} are assumed; this should be conservative for a linear
collider detector.  The $W$ width $\Gamma_W$ is assumed to have the SM
value.  Under these assumptions, a precision on $\mw$ of 4.9 MeV is
predicted for 100 \invf of data.

To reduce the dependence of the $\mw$ precision on the absolute beam
polarization determination, `radiative return' ($e^+e^-\to \gamma + Z$)
events can be incorporated into the analysis.  They are sufficiently
numerous---$10^7$ in 100 fb$^{-1}$---that the Blondel scheme described
in the previous section can be employed to measure the polarization.
After fine tuning of the luminosity distribution among various helicity
configurations, a scan can still determine $\mw$ to 5 MeV without the
0.25\% polarization calibration.

The background from $e^+e^-\to q\bar{q}$ and its polarization asymmetry
is neglected in this analysis.  It is possible that the polarization
asymmetry of the sample of background events that pass the $WW$ event
selection cuts will be poorly known.  In this case, the scan strategy
above may not be optimal for control of the systematics.  While further
study is warranted, incorporation of a scan point below threshold should
control the uncertainties without significantly degrading precision on
$\mw$.

The beam-energy and beamstrahlung uncertainties of a \ww threshold scan
must be controlled to a few MeV to achieve the desired $\mw$ precision.
One method~\cite{bb:graham_beam_proposals} provides a direct measurement
of the average $\sqrt{s}$ via reconstruction of $e^+e^-\to \gamma + Z$,
$Z\to e^+e^-/\mu^+\mu^-$.  This measurement includes the average
beamstrahlung effect.  A precision of 2.5 MeV may be possible for 100
\invf.  Absolute alignment of the detector polar angle to $10^{-5}$ and
knowledge of the radiative corrections will be needed.  One could also
calibrate a precise beam spectrometer using the $Z$ line shape and
extrapolate to the \ww threshold.  The uncertainty from the LEP1 $\mz$
measurement will cancel in the $\mw/\mz$ ratio.
Beamstrahlung both reduces the effective \ww cross section at threshold
and distorts the shape.  To limit the effects to 2 MeV, the absolute
induced distortion must be known to 0.1\%.  Mapping of the distortion to
this accuracy appears feasible by measurement of the distribution in the
acolinearity angle in Bhabha scattering at forward
angles \cite{bb:desy_97_048}.  All of these aspects of the precision
energy determination will be challenging if one wishes to achieve a
2 MeV error from this source.

\subsection{Conclusion}

The experimental systematics for an $\mw$ measurement near \ww threshold
appear to be under control at the few-MeV level.  Issues related to beam
energy and beamstrahlung deserve further attention, but cautious
optimism is appropriate.  Certainly the $\mw$ issues should be
considered in the accelerator and interaction region design.  Given the
one year of running required to reach the order 5 MeV accuracy in $\mw$,
consideration of a dedicated low-energy facility seems appropriate.  The
feasibility of the measurement without positron polarization needs
examination.  A much longer running period would be necessary just to
make up the loss in \ww production.  The impact on control of the
background level is currently unknown.

\section{Electroweak tests of the Standard Model}

The physics program outlined above opens new opportunities for 
high-precision physics in the electroweak sector.  For reference,
\refta{tab:precallcoll}~\cite{gigaz2} summarizes the present and
anticipated precisions for the most relevant electroweak observables at
the Teva\-tron---Run II (2~fb$^{-1}$) and TeV33 (30~fb$^{-1}$), the LHC,
and a future linear collider without (LC) and with (Giga-Z) a low-energy
program.

\begin{table}[t]
\BC
\begin{tabular}{ccr@{~}lr@{~}lr@{~}lr@{~}lr@{~}l} \hline\hline
            & now & \multicolumn{2}{c}{Run II}
                  & \multicolumn{2}{c}{TeV33}
                  & \multicolumn{2}{c}{LHC}
                  & \multicolumn{2}{c}{LC}
                  & \multicolumn{2}{c}{Giga-Z} \\ \hline
$\delta\sweff(\times 10^5)$& 17 & 50 & \cite{Baur96} &
                                  13 & \cite{Baur96} &
                                  21 & \cite{Baur96,EWrep} &
                                 (6) & \cite{Baur96} &
                                 1.3 & \cite{kmogigaz}\\
$\delta\mw$ [MeV] & 37          & 30 & \cite{Signore98} &
                                  15 & \cite{bb:atlas_lhc_mw} &
                                  15 & \cite{bb:atlas_lhc_mw,EWrep} &
                                  15 & \cite{Haber97} &
                                   6 & \cite{wwthreshold}\\
$\delta\mt$ [GeV] & 5.1         & 4.0 & \cite{Baur96} &
                                  2.0 & \cite{Baur96} &
                                  2.0 & \cite{Baur96,Toprep} &
                                  0.2 & \cite{Frey97} &
                                  0.2 & \\
$\delta\mh$ [MeV] & ---         & --- & &
                               2000 & \cite{Gunion97} &
                               100 & \cite{Gunion97} &
                                50 & \cite{Gunion97} &
                                50 & \cite{Gunion97}\\ \hline\hline
\end{tabular}
\caption{\label{tab:precallcoll}
The expected experimental precision from various collider programs
for $\sweff$, $\mw$, $\mt$  and
the Higgs boson mass, $\mh$, assuming $\mh = 110 \GeV$.
For the LC entry in parentheses, a fixed-target polarized M\o ller
scattering experiment using the $e^-$ beam has been  assumed.
The present uncertainty on $\mw$ will be improved with the final
analysis of the LEP2 data.
}
\EC
\end{table}

The SM predictions for the electroweak precision observables
are affected via loop corrections by contributions from
the top quark mass, $\mt$, and the Higgs boson mass, $\mh$.
The prediction for the $W$~boson mass is obtained from
\beq
\mw = \frac{\mz}{\wz} \,
      \sqrt{1 + \sqrt{ \frac{4\,\pi\,\alpha}{\wz\,\GF\,\mz^2}
                       (1 + \Delta r)}} ,
\eeq{defMW}
where the loop corrections are contained in $\Delta r$~\cite{deltar}.
Beyond  one-loop order, the QCD corrections are known at
\order{\alpha\alphas}~\cite{deltaralals} and
\order{\alpha\alphas^2}~\cite{deltaralals2}. The electroweak \twol\
corrections have recently been extended to include the complete fermionic
contribution at \order{\alpha^2}~\cite{pdeltaral2}.

The effective leptonic weak mixing angle, $\sweff$, is defined through the
effective couplings $g_V^f$ and $g_A^f$ of the $Z$~boson to fermions at
the $Z$~resonance,
\beq
\sweff = \ed{4\,Q_f} \KL 1 - \frac{\re\, g_V^f}{\re\, g_A^f} \KR ,
\eeq{defsw2eff}
where the loop corrections enter through $g_{V,A}^f$.
The radiative corrections entering the relations (\ref{defMW}) and
(\ref{defsw2eff}) depend quadratically on $\mt$, while the leading
dependence on $\mh$ is only logarithmic.

The current theoretical
uncertainties~\cite{PCP} are dominated by the uncertainties
in the input parameters $\mt$ and $m_h$, and in the value of the
running electromagnetic coupling constant evaluated at the scale $\mz$.
Let $\Delta\alpha = \alpha(\mz) - \alpha(0)$.  This difference results from
electromagnetic vacuum polarization corrections due to the charged leptons
and light quarks.
The hadronic contributions to $\Delta\alpha$ currently
give rise to an uncertainty
$\delta\Delta\alpha \approx \pm 2 \times 10^{-4}$~\cite{delalpha}.
If future low-energy \epem\ experiments can measure the
hadronic total cross section up to the $J/\psi$ to 1\%, it is possible to
reduce this uncertainty to about
$\delta\Delta\alpha = \pm 7 \times 10^{-5}$~\cite{delalphajegerlehner}.
As an estimate for the future theoretical uncertainties in the prediction
of $\mw$ and $\sweff$ from unknown higher-order
corrections (including the uncertainties from $\delta\Delta\alpha$) we use
\beq
\delta\mw(\mbox{theory}) = \pm 3 \MeV, \quad
\delta\sweff(\mbox{theory}) = \pm 3 \times 10^{-5} \quad
\mbox{(future)} .
\eeq{eq:futureunc}
The experimental  error on $\mz$ ($\delta\mz = \pm 2.1 \MeV$~\cite{bb:lepewwg})
leads to an uncertainty in $\sweff$ of $\delta\sweff = \pm 1.4 \times 10^{-5}$.
While this uncertainty can currently be neglected, it will
have non-negligible impact
given the precision obtainable at Giga-Z.
The future experimental error in the top quark mass,
$\delta\mt = \pm 130 \MeV$,
induces further uncertainties $\delta\mw = \pm 0.8 \MeV$ and
$\delta\sweff = \pm 0.4 \times 10^{-5}$.

\begin{table}[t]
\BC
\begin{tabular}{lrrr} \hline\hline
                                       & $\mw$ & $\sweff$ & all  \\  \hline
now                                    &200\% & 62\%    & 60\% \\
Run II                                 & 77\% & 46\%    & 41\% \\
TeV33                                  & 39\% & 28\%    & 26\% \\
LHC                                    & 28\% & 24\%    & 21\% \\
LC                                     & 18\% & 20\%    & 15\% \\
Giga-Z                                  & 12\% &  7\%    &  7\% \\ \hline\hline
\end{tabular}
\caption{
\label{tab:indirectmh}
{Cumulative expected precisions for the indirect determination
of the Higgs boson mass, $\delta\mh/\mh$, taking into account the error
projections in \refta{tab:precallcoll} and the theoretical uncertainties
of $\mw$ and $\sweff$. The first two columns use $\mw$ and $\sweff$
constraints alone, while the last column uses the
full set of precision observables.}   }
\EC
\end{table}

Comparison of an indirect determination of the SM Higgs boson mass,
which would be significantly improved by Giga-Z
\cite{gigaz1,gigaz2,pgigaz4,kmogigaz}, with a future direct measurement
will provide a sensitive test of the SM.
Table~\ref{tab:indirectmh}~\cite{gigaz2} summarizes both today's
accuracy for the indirect prediction of $\mh$ and the accuracy available
from the prospective improvements at forthcoming colliders listed in
Table~\ref{tab:precallcoll}.  The current accuracies assume
$\delta\Delta\alpha = \pm 2 \times 10^{-4}$~\cite{delalpha}, while the
future cases assume $\delta\Delta\alpha = \pm 7 \times
10^{-5}$~\cite{delalphajegerlehner}.  The Giga-Z scenario allows an
indirect determination of $\mh$ with an uncertainty of $\delta \mh/\mh =
\pm 7\%$ (about the level of the current indirect $\mt$ determination).
This represents a factor of three improvement over the EW constraints
that could be made using LHC measurements, while a linear collider
running solely at high energy would provide only a modest gain.

\begin{figure}[t]
\vspace{1em}
\begin{center}
\mbox{
\epsfig{figure=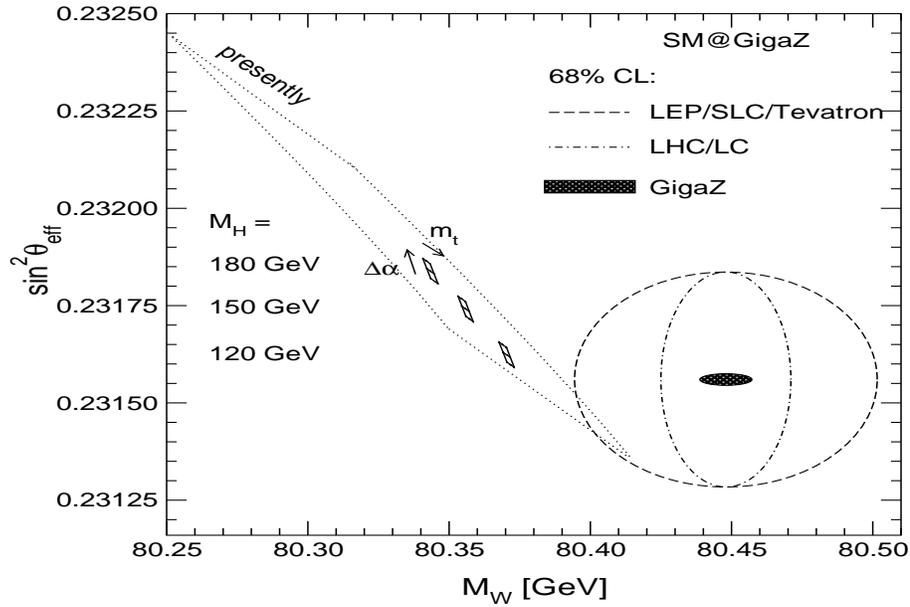,width=12cm,height=8cm}}
\end{center}
\caption[]{\label{fig:SWMW}
The present and prospective future theoretical predictions in the SM (for
three $\mh$ values) are compared with the
current experimental accuracies and those expected from LHC and Giga-Z
(see \refta{tab:precallcoll}).
The future theoretical uncertainties arising from
$\delta\Delta\alpha = \pm 7 \times 10^{-5}$ and $\delta\mt = \pm 200 \MeV$ are
indicated.
}
\end{figure}

Figure~\ref{fig:SWMW} compares the potential of Giga-Z for testing the
electroweak theory with the present status from both theoretical and
experimental standpoints.  The SM prediction corresponds to an allowed
$\mh$ interval of $113 \GeV \leq \mh \leq 400 \GeV$ and to an allowed
$\mt$ interval within its measured uncertainty.  The theoretical
prediction assumes that the Higgs boson has been found, with masses of
120, 150 and 180~GeV considered.  The uncertainty induced assuming
$\delta\mt = \pm 200$~MeV and $\delta \Delta\alpha = \pm 7 \times
10^{-5}$ is indicated.  The figure illustrates that the improved
experimental accuracy at Giga-Z will allow tests of the internal
consistency of the SM at an unprecedented level.


\subsection{Parameterizations of deviations from the Standard Model}

The precision achievable at Giga-Z allows for the exploration of
possible effects of new physics with great sensitivity.  This section is
devoted to more general parameterizations of physics beyond the SM
through the specific example of the $S$, $T$, $U$
parameters~\cite{Peskin90}.  While these parameters are widely used,
considerable confusion exists concerning their meaning and range of
applicability.  Because it is important to understand precisely how the
effects of new physics can be probed in a sensible way given the
potential Giga-Z accuracies, we briefly summarize the main points.

By definition, the $S$, $T$, $U$ parameters describe only the
effects of new physics contributions that enter via vacuum-polarization
effects
(\ie, self-energy corrections) to the vector-boson propagators of the
SM. That is, the new physics contributions
are assumed to have negligible couplings to SM fermions.
The parameters can be computed in
new models as certain combinations of one-loop
self-energies. Experimentally, their values are determined by
comparing the measurements, ${\cal A}_i^{\mathrm{exp}}$, of a number of
observables with their SM predictions, ${\cal A}_i^{\mathrm{SM}}$,
\begin{equation}
\label{eq:STUgen}
{\cal A}_i^{\mathrm{exp}} = {\cal A}_i^{\mathrm{SM}} +
f^{\mathrm{NP}}_i(S, T, U).
\end{equation}
Here ${\cal A}_i^{\mathrm{SM}}$ contains all known
radiative corrections in the SM evaluated at reference values of
$\mt$ and $\mh$.
The (linear) function $f^{\mathrm{NP}}_i(S, T, U)$ describes the
contributions of new
physics.  For most
precision observables, the corrections caused by a variation of $\mt$ and
$\mh$ at one-loop order can also be absorbed into $S$, $T$,
and $U$. A non-zero result for $S$, $T$, $U$ determined in this way
indicates non-vanishing contributions of new physics (with respect to
the SM reference value).

The $S$, $T$, $U$
parameters can only be applied for parameterizing effects of physics
{\em beyond} the SM.  To compute the SM predictions to which these
parameters provide corrections, one must take into account the full
contributions, which also contain vertex and box corrections, since
these effects cannot be consistently absorbed into the $S$, $T$,
$U$ parameters. For a more detailed discussion of this point, see
\cite{ringbg}.  Because the $S$, $T$, $U$ parameters are
restricted to the leading-order contributions of new physics, they should
only be applied for {\em small} deviations
from the SM pre\-dic\-tions. Their application to cases with large
deviations from the SM, like extensions of the SM with a very heavy Higgs
boson in the range of several TeV, is questionable.
The current experimental values~\cite{Erler99A} (assuming
$\mt = 173.4 \GeV$ and $\mh = 100 \GeV$) are
\beq
S = -0.07 \pm 0.11, \quad
T = -0.10 \pm 0.14, \quad
U = ~0.10 \pm 0.15. \quad
\eeq{eq:STUnow}

Other parameterizations, defined via linear combinations of
various observables without reference to the SM contribution,
have been suggested (see, \eg, \cite{eps,schild}).  While
any new physics model can be explored, it is not in all cases obvious
that studying parameters is of advantage compared to studying the
observables themselves.  For this reason and for brevity, we restrict our
discussion to the $S$, $T$, $U$ parameters.

Examples of new physics contributions that can be described in the
framework of the $S$, $T$, $U$ parameters are contributions from a
fourth generation of heavy fermions or effects from scalar quark loops
(see \refse{sec:susy}). A counterexample going beyond the
$S$, $T$, $U$ framework is given by corrections of the kind that could
bring the prediction for the anomalous magnetic moment of the muon in
agreement with the experimental value~\cite{g2exp,czarnmarc}.

While many SM extensions result in a vanishing or small contribution to
the $U$ parameter (see ~\citere{Erler99A} and references therein), sizable
contributions to $S$ and $T$ can be expected
from a number of models. For instance, the contribution of a heavy Higgs
boson with $\mh = 1 \TeV$ gives rise to a contribution in $S$ and $T$ of about
$S \approx 0.1$, $T \approx -0.3$~\cite{PeskinWells01} (see however
the discussion above). In technicolor models one typically expects $S$
and $T$ to be positive and of order 1~\cite{PeskinWells01}.
Peskin and Wells~\cite{PeskinWells01}
have also examined the `topcolor seesaw' model of Dobrescu and
Hill~\cite{bb:DobHill}, which predicts little or no new physics observable
at the LHC or LC. The Giga-Z scenario, however, would reveal a significant
departure in the $(S,T)$ plane from the minimal SM with a light Higgs
boson.

These additional contributions to the $S,T,U$ parameters have to be
compared with the errors with which these
parameters can be extracted at Giga-Z~\cite{gigaz2}:
\beq
\Delta S = \pm 0.05, \quad \Delta T = \pm 0.06, \quad \Delta U = \pm
0.04. \quad
\eeq{eq:deltas}
These parameters are strongly correlated. Assuming $U =  0$, as justified
above, the anticipated errors in $S$ and $T$ would decrease to
about
\beq
\Delta S = \pm 0.02, \quad \Delta T = \pm 0.02. \quad
\eeq{eq:newdeltas}
The increased precision, compared to the present
situation given in Eq.~\leqn{eq:STUnow}, will constrain or
exclude of many possible extensions of the SM.


\subsection{Tests of supersymmetry}
\label{sec:susy}

We now explore the utility of the precision electroweak observables
in a scenario with direct observation of new particles, by examining
a specific example. Suppose that particles compatible with
a MSSM Higgs boson and a light scalar top quark $\Stope$
have been discovered at the Tevatron or the LHC, and
further explored at an \epem\ linear collider. With the luminosity expected 
at a linear collider,  the $\Stope$ mass,
$\mste$, and the mixing angle in the stop sector, $\costt$,
can be measured in the process $e^+e^- \to \Stope \tilde{t}_1^*$ to a
level below 1\%~\cite{susyattesla,plcstop}.

The precision electroweak variables provide several constraints.
First, the measurements and predictions for $\mw$ and $\sweff$
provide an indirect test of the MSSM, as they do for the SM.
Comparison of the predicted
to the measured value of the lightest CP-even
MSSM Higgs boson mass, $\mh$, provides a further constraint.  In the
MSSM,  $\mh$ is not a
free parameter as in the SM; it is
calculable from the other SUSY parameters.
Furthermore, because $\mw$, $\sweff$ and $\mh$ are
particularly sensitive to the SUSY parameters of the scalar top and
bottom sector and of the Higgs sector, they provide an
indirect probe of the masses of supersymmetric particles
that might not be seen at the LHC or LC.
In particular, the heavier scalar top quark, $\Stopz$, and the heavy Higgs
bosons $A$, $H$ and $H^\pm$ could be outside the kinematic reach of the
initial-stage
LC, and background problems could preclude their observation at the LHC.
Reference~\cite{gigaz2} explores this scenario and demonstrates that
upper bounds on $\MA$ could be established through the
SUSY contributions to $\mw$ and $\sweff$, just as the Higgs
boson mass can be bounded in the SM.

Finally, we examine the indirect information on the mass
of the heavier scalar top quark, $\mstz$, that can be
obtained by requiring consistency of the
MSSM with measurements of $\mw$, $\sweff$, and $\mh$ in addition
to those of $\mste$ and $\costt$.
The SUSY contributions to $\mw$ and $\sweff$ include
the complete \onel\ results in the MSSM~\cite{deltarsusy} as well as
the leading higher-order QCD corrections~\cite{mssm2lqcd}.
The prediction for
$\mh$ is obtained with the program \fh~\cite{feynhiggs}, based on the
Feynman-diagrammatic two-loop result of \citere{mhiggs2l}.
A future uncertainty in
the theoretical prediction of $\mh$ of $\pm 0.5 \GeV$ is assumed.

We examine the scenarios
for a LC with and without the Giga-Z option and for the
LHC (see \refta{tab:precallcoll}),
taking $\mste = 180 \pm 1.25 \GeV$ for LC/Giga-Z,
and $180 \pm 18 \GeV$ for the LHC. The other parameters
have been chosen according to the mSUGRA reference
scenario~2 specified in \citere{msugrapoints}, with the following
accuracies: $\MA = 257 \pm 10$~GeV, $\mu = 263 \pm 1$~GeV,
$M_2 = 150 \pm 1$~GeV, $\mgl = 496 \pm 10$~GeV.  For $\tb$ a lower bound
of $\tan\beta > 10$
has been taken.  The central values for $\mw$ and $\sweff$ have been chosen in accordance
with a non-zero contribution to the precision observables from SUSY
loops.

\begin{figure}[ht!]
\begin{center}
\mbox{
\epsfig{figure=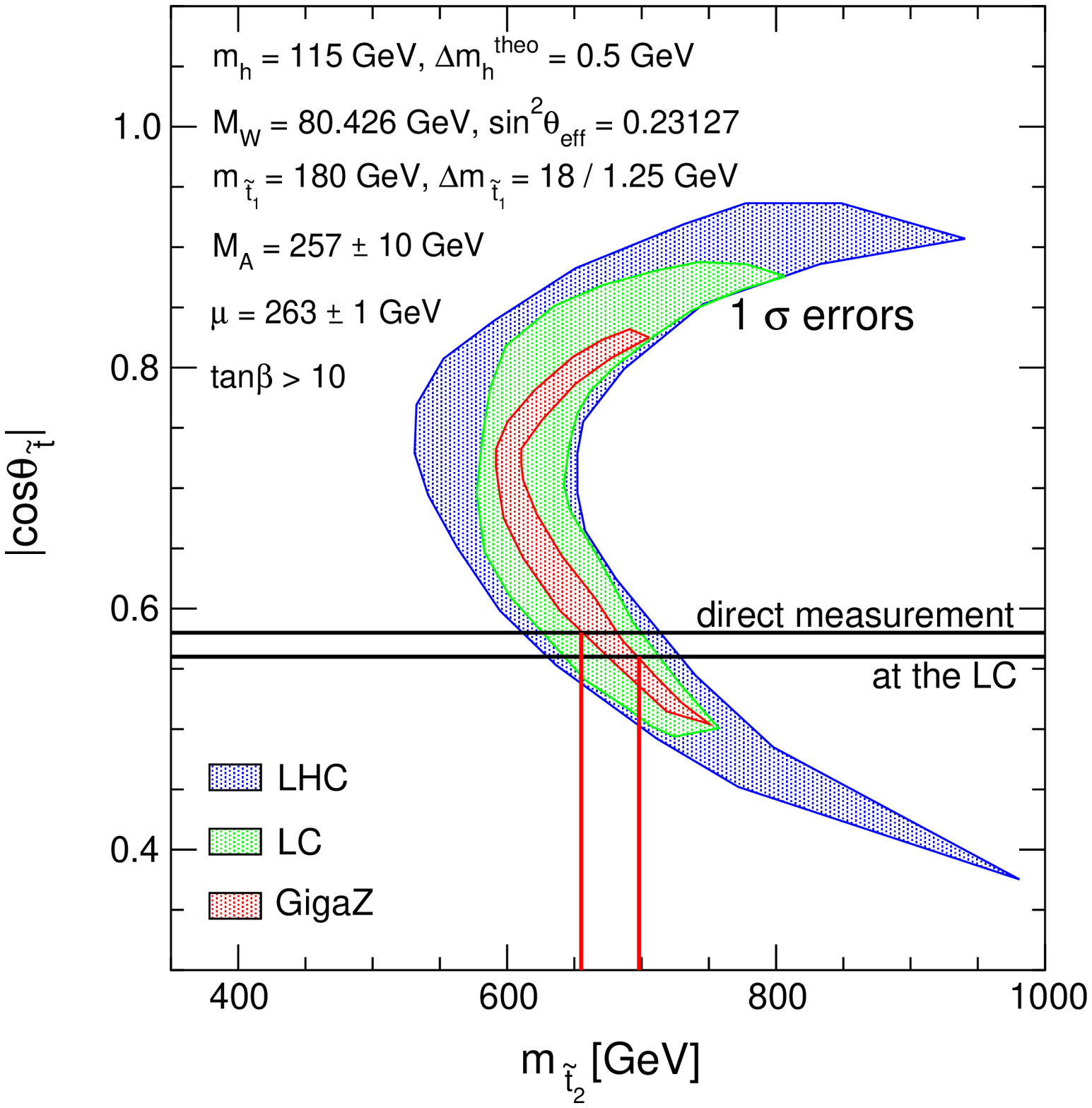,width=10cm, height=8cm}}
\end{center}
\vspace{-1em}
\caption[]{\label{fig:LCvGiga-Z}
Indirect constraints on the MSSM parameter space in the
$\mstz$--$|\costt|$ plane from measurements of
$\mh$, $\mw$, $\sweff$, $\mt$ and $\mste$ at a LC with and
without the Giga-Z option and at the LHC.  The solid lines indicate the
direct information on the
mixing angle from a measurement at the LC and the
corresponding indirect determination of $\mstz$.
}
\end{figure}

As one can see in Fig.~\ref{fig:LCvGiga-Z}, the allowed parameter space in the
$\mstz$--$|\costt|$ plane is significantly reduced in the Giga-Z scenario
relative to the others. Using
the direct information on $|\costt|$ from \citere{plcstop}
allows an indirect determination of $\mstz$ with
a precision of better than 5\% in the Giga-Z case. By comparing this indirect
prediction for $\mstz$ with direct experimental information on
the mass of this particle, the MSSM could be tested at its quantum level
in a sensitive and highly non-trivial way.

\section{Heavy flavor physics}

The $Z$ pole has already been established as an excellent laboratory
for the study of $b$ physics.  The large boost and resulting detached
vertices for the $b$ decays have amply compensated the relatively modest
statistics of the LEP experiments, allowing them to make many competitive
and important measurements.  SLD, with much smaller statistics, has benefitted
greatly from the SLC's beam polarization in the $b$ studies that require
production tagging and has produced measurements competitive with LEP.  The
hadronic
experiments, LHC-b and BTeV, will be faced with large backgrounds, with typical
signal-to-noise ratios of $S/N\approx 5 \times 10^{-3}$ compared
to $S/N \approx 0.21$ at the $Z$ (albeit with $10^4$ to $10^5$ more $b$'s
produced).

The $Z$-pole running will result in a very powerful $b$ experiment. With 80\%
and 60\% polarizations for the electron and positron beams, respectively,
production
flavor tags that include the forward--backward production asymmetry should
 reach a
signal $\times$ purity
$\varepsilon D^2$ approaching 0.6.  (With 80\% electron polarization and
no positron polarization, one finds about half of this value.)
For comparison, the
$B$ factories have achieved $\varepsilon
D^2 \approx 0.25$ \cite{bb:babar_sin2beta}
while the hadronic facilities will have rather lower values.  Coupled with the
excellent resolution expected from the vertex detector for the linear collider,
a reach in $\delta m_s$ of $40\  \textrm{ps}^{-1}$ is possible with
$10^9$ $Z$'s, with a resolution
limit of around $80\  \textrm{ps}^{-1}$.

The scenario in which $2\times 10^9$ $Z$ decays are produced, yielding
about $6 \times 10^8$ $b$ hadrons,
has been investigated.
This sample should be compared to the $\Upsilon(4S)$ and hadronic $b$ samples
that will be available in the same time period~\cite{kmogigaz,MANNEL}.
This section is largely based on a
review of such work in~\cite{MANNEL}.
With these statistics, $b$ studies at the $Z$ offer a number of
measurements that
are of fundamental importance for the comprehensive
$b$-physics program that is being undertaken worldwide,
but which cannot be addressed adequately at other existing or planned
facilities.   A longer running period at the
$Z$ ($10^{10}$ $Z$'s) is necessary to improve upon the sensitivity
for the `canonical' measurements planned at other $b$ facilities,
despite the combined
advantages of tagging, boost and purity.  Such a facility would be quite
competitive. A precision on $\sin 2\beta$ of about 0.01 would be obtainable,
similar to that obtainable from LHC-b and BTeV.
If one translates the studies of $B\to\pi\pi$ to an effective value of
$\sin 2\alpha$, the
uncertainty would be about 0.02, approaching that of BTeV and
somewhat better than that
expected from LHC-b.

The topics unique to a polarized $Z$ facility are the following:
\begin{enumerate}
\renewcommand{\itemsep}{0pt}
\item
The quark-level transition
\begin{equation}
b \to q + \nu \bar \nu
\label{bqnunu}
\end{equation}
could well be affected significantly by new physics in ways
quite different from
$b \to q + l^+l^-$.
Searching for $b \to q \nu \bar \nu$ in hadronic colliders appears
hopeless. The searches also pose quite a challenge for an
$\Upsilon (4S)$ experiment because of the intermingling of the decay
productions from the two $B$ decays~\cite{bb:alexander_privcomm}.
\item
The CKM elements $|V(cb)|$ and $|V(ub)|$, determined in semileptonic $B$ decay,
suffer from a potentially
considerable source of uncertainty due to limitations in the validity of quark-hadron
duality, of which at present little is known for certain. Detailed
comparisons of semileptonic
$B_s$ and $B_{u,d}$ decays  would be invaluable in this respect.
The $\Upsilon(4S)$ machines will not have $B_s$ samples, while the hadronic
machines will have difficulty providing precise inclusive measurements.
\item
The availability of polarized beams will allow production of a huge sample of
highly polarized beauty baryons whose weak decays can be analyzed. In
this way a determination of the handedness of a quark transition becomes
feasible.
\end{enumerate}
The canonical measurements for which $2\times 10^{9}$ $Z$'s may be competitive
include
\begin{enumerate}
\renewcommand{\itemsep}{0pt}
\item
The transition $b \to \tau\nu$ contains multiple neutrinos in the final
state, with
an experimental situation similar to that for $b \to q + \nu \bar \nu$.  This
measurement determines the product $F_B|V_{ub}|$, and would play a fundamental
role in constraints of the CKM matrix.  The reach at Giga-Z has not yet been
studied.
\item
The production flavor tagging from the $Z$ running might offer the most
precise
measurements of ${\cal B}(B^0\to \pi^0 \pi^0)$ and
${\cal B}(\bar{B}^0\to \pi^0 \pi^0)$, which are of
great significance for extracting the angle $\phi_2$  or $\alpha$
from the measured CP asymmetry in $B^0 \to \pi ^+\pi^-$.
\end{enumerate}
The following subsections elaborate on these points.

\subsection{Measurement prospects for   ${\cal B}(B\to \pi^0\pi^0)$}

One of the promising strategies for measuring the CKM angle $\alpha$ is the
study of the CP asymmetry in the decay $\rm B^0\rightarrow\pi^+\pi^-$.
The presence of significant `penguin' contributions to
$\rm B\rightarrow\pi^+\pi^-$ complicates the extraction of $\alpha$ from
the measured time\-dependent CP asymmetry.
The penguin and tree contributions can be separated by measuring the
branching ratios  ${\cal B}(\rm B^0\rightarrow\pi^+\pi^-)$,
${\cal B}(\rm B^+\rightarrow\pi^+\pi^0)$ and
${\cal B}(\rm B^0\rightarrow\pi^0\pi^0)$
and the charge conjugate modes~\cite{bb:london}.
The first can be measured as a by-product of the CP-asymmetry analysis, but
the other two are more difficult. The need to reconstruct $\pi^0$s makes
them extremely challenging for hadron machines. The expected
branching ratios are also very small, of order $10^{-6}$, with experimental
upper limits of $12.7\times 10^{-6}$  ($\pi^+\pi^0$) \cite{cleorarea}
and $9.3\times 10^{-6}$ ($\pi^0\pi^0$) \cite{cleorareb}.

The feasibility of measuring these branching ratios at a linear collider
was studied~\cite{kmogigaz} using the fast Monte Carlo simulation
SIMDET \cite{simdet}.
The reconstructed $B$ mass resolutions were found to be 150\,MeV ($\pi^0\pi^0$)
and 120\,MeV ($\pi^+\pi^0$), dominated by the calorimeter angular resolution.
 Assuming signal
branching ratios of a few $10^{-6}$ gives signal samples  of about 200 events
 for $2\times 10^9\
\rm Z^0$ decays, on top of several hundred events of combinatorial background.
This would allow a flavor-independent measurement comparable to that of BABAR
 or BELLE with about
$200\,\rm fb^{-1}$ \cite{kmogigaz}. For the separate $B$ versus the $\bar{B}$
 branching
fractions,  which are needed for the $\alpha$ determination,
the factor of two  or more improvement in $\varepsilon D^2$ at the $Z$
 relative to
that for the $B$ factories makes these measurements with $10^9$ $Z$'s
 competitive with, if not
better than, those obtainable at the $B$ factories.
It should
be emphasized that this study was performed with a very crude calorimeter
simulation and further background rejection may certainly be possible
after more detailed studies.

\subsection{$B \to X_q \nu \bar \nu$}

The large backgrounds at hadronic machines make  measurement of
 $B \to X_q \nu \bar \nu$ impossible there.  In an $e^+e^-$ threshold
machine, such
transitions could be found only at the cost of reconstructing
one $B$ more or less fully.  At Giga-Z, however,
the relative cleanliness of
the $Z$, the
hemispheric separation of the $b$ quarks, and the well-defined initial
state provide powerful tools for discovering and
actually measuring properties of
such transitions at the $Z$. This is illustrated by the
fact that the current upper limit on this decay mode comes from LEP1:
\begin{equation}
{\rm BR}(B \to X_s \nu \bar \nu) \leq 7.7 \times 10^{-4}
\; \; \; {\rm (ALEPH)} \ .
\end{equation}

New physics can affect $b \to q l^+l^-$ and
$b \to q \nu \bar \nu$ in quite different way for various
reasons \cite{ALI}.  For example, new contributions to an
effective $bsZ$ vertex would enhance $b \to q \nu \bar \nu$
relative to $b \to q l^+l^-$ by a large factor, and
study of $b \to q \nu \bar \nu$ (with contributions from
$b \to q \nu_\tau \bar \nu_\tau$) in addition to
$b \to q e^+e^-$ and $b \to q \mu^+\mu^-$ can help disentangle
new physics scenarios with generation--dependent couplings .

At the $Z$, the statistics will be high enough to  make meaningful searches
for $B \to X_s \nu \bar \nu$.
With an inclusive branching fraction in
the standard model of about $4 \times 10^{-5}$, and exclusive branching
 fractions
to $K$ and $K^*$ of order $10^{-5}$ \cite{ALI}, one can expect a few
times $10^3$ events in
exclusive channels and about $10^4$ inclusively.  The expected reach, including
control of backgrounds such as $b\to \tau \nu$, is not known at this time,
but warrants study.

\subsection{Semileptonic  $B_s$ decays}

The CKM parameters $\Vcb$ and $\Vub$  play a central
role in the prediction of various CP asymmetries in $B$ and $K$ decays.
With precision measurements, constraints on new physics scenarios
would be obtained by comparison of the predictions with direct measurements.
It is crucial for this program to have
reliable determination of $\Vcb$ and $\Vub$,
obtained from semileptonic $B$ decays through observables
in exclusive and inclusive modes.

Inclusive measurements play an important role in these
determinations. The known uncertainties are estimated at the
5\% level for $\Vcb$ and at the (10--15)\% level for $\Vub$.
However, there may be an additional
significant source of systematic uncertainty, the validity of
 quark-hadron
duality, which underlies almost all
applications of the $1/m_Q$ expansions.
A large
body of folkloric or circumstantial evidence suggests that duality is a
useful and meaningful concept. But for a full evaluation of the
data from beauty physics it is essential to know with {\em tested}
confidence whether the deviations from exact duality in semileptonic
transitions arise at the 10\%, the 5\%, or the 1\% level.
It is quite unlikely that this question can be answered
by theoretical means alone.

Experimentally, one can probe duality via an independent
extraction of
$|V_{cb}|$ in $B_s$ decays through measurement of $\Gamma _{SL}(B_s)$.
One could also
determine the rate for $B_s \to l \nu D^*_s $, extrapolate
to zero recoil,  and extract the product
$|V(cb)F_{B_s \to D_s^*}(0)|$. The form factor can be obtained from the
result of the
Heavy Quark Expansion
\begin{equation}
|F_{B_s \to D_s^*}(0)| \simeq |F_{B \to D^*}(0)|
\end{equation}
up to $SU(3)$ breaking corrections, which can be estimated.

The physical origin of duality violation would be the
accidental presence of a nearby
hadronic resonance with appropriate quantum numbers to affect the decay
pattern for one of the $B$ mesons.  On one hand, this resonance may
affect
$B_d \to l \nu X_c$ and $B_u \to l \nu X_c$, but not
$B_s \to l \nu X_c$; conversely, it may affect $B_s$
transitions while having no impact on $B_{u,d}$ channels.
If the same value emerged for $|V(cb)|$ in both cases, we would
have verified the validity of duality in this case at least. If not, we
would not know which, if any, of the values is the correct one, but
we would be aware of a serious problem.

Duality violation could exhibit a different pattern in
$B\to l \nu X_u$ channels. Here theory also calls for a detailed comparison of
$B_d$ and $B_u$ modes, since one expects a difference in the endpoint region of
$B_d$ and $B_u$ semileptonic decays \cite{WA}. Hadronic
resonances could affect $B_d\to l \nu X_u$ and
$B_u \to l \nu X_u$ quite differently. In addition, measurements of
$B_s \to l \nu X_u$, both inclusive and exclusive,
would provide crucial cross checks.

\subsection{Weak decays of polarized beauty baryons}

\begin{table}
\begin{center}
\begin{tabular}{lcc}
\hline \hline
Mode & Branching Ratio & Number of Events \\
\hline
$\Lambda_b \to \Lambda_c \ell \bar{\nu}_\ell$
           & $8 \times 10^{-2}$ & $5 \times 10^{6}$\\
$\Lambda_b \to p \ell \bar{\nu}_\ell$
           & $8 \times 10^{-4}$ & $5 \times 10^{4}$\\
$\Lambda_b \to X_s \gamma$ & $3 \times 10^{-4}$ & $11 000$ \\
$\Lambda_b \to \Lambda \gamma$ &  $5 \times 10^{-5}$ & $ 1400 $ \\
$\Lambda_b \to \Lambda \ell \ell$ & $ 1 \times 10^{-6}$ & 50 \\
\hline \hline
\end{tabular}
\end{center}
\caption{\label{tab:2}
Expected numbers of events for $\Lambda_b$ decays, based
         on the Standard Model estimates.}
\end{table}

The large polarization asymmetry for $Z$ decay to $b$ quarks implies that
beauty baryons produced in $Z$ decays are highly polarized.
From $2\times10^9$ $Z$s,
one expects about  $3 \times 10^7$ polarized $b$-flavored baryons.
The study of the
weak decays of these particles offers
a whole new field of dynamical information.
The  existence of initial-state polarization in
$\Lambda _b$ decays allows one to analyze the chirality of the
quark coupling {\em directly}; it also leads to
a new program of studying
observables revealing direct CP violation. Charmed baryons also merit
study.

A generic analysis of $b \to s \gamma$
results in two transition operators, mediating the decays
\begin{equation}
b_R \to s_L \gamma \; , \; b_L \to s_R \gamma \; .
\end{equation}
While the second operator
 is highly suppressed in the SM, by a factor $m_s/m_b$, these operators
could be of comparable size in new physics scenarios,
for example, in Left-Right Symmetric
models or the MSSM.
While the decays of mesons realistically
cannot distinguish between these two transitions, a study of the
$\Lambda$ polarization in
the decay $\Lambda_b \to \Lambda \gamma$ with polarized $\Lambda_b$
could probe the SM prediction that the ratio of
left- to right-handed couplings is
$r \lsim 0.04$.
One measures the asymmetry in the angular distribution
defined between the $\Lambda_b$
spin and the photon in the parent baryon rest frame.
Based on the statistics of
Table~\ref{tab:2},
corresponding to roughly 750 fully reconstructed events,
the measurement would be sensitive to values of  $r$
between 0.5 and 1.9 at the $5\sigma$ level.  For
comparison,  the sensitivity extends from 0.2 and 4.1 with $10^{10}$ $Z$'s
 \cite{HK01}.
It should be noted that the angular asymmetry
is a theoretically very clean observable
and the extraction of $r$ is essentially limited only by statistics.

A significant non-vanishing contribution of
$b_L \to s_R \gamma$ would signal the
intervention of new physics. One can actually undertake an
{\em inclusive} polarization study of $\Lambda _b \to \Lambda  \gamma +X$
with large statistics; the clean environment of the $Z$ is
crucial here. Corresponding studies can be performed with
$\Lambda _b\to l^+l^-X$ with smaller statistics.

Although theoretically less clean, similar angular asymmetries in rare
hadronic 2-body decays such as $\Lambda_{\rm{b}} \to \Lambda \phi$
offer a unique opportunity to probe for new physics contributions to
four-quark penguin operators with chiralities opposite to those in the
SM~\cite{HK01}.

As an advantage over experiments with unpolarized $\Lambda_b$ baryons,
spin correlations between the spin of
the $\Lambda_b$ and the daughter baryon are fully accessible.
It is possible, for example, to distinguish between
pseudoscalar and vector transition form factors~\cite{pakvasaetal}.
This allows for novel, powerful
consistency checks of the Standard Model including its CP
and chirality properties.

Semileptonic decays of polarized $\Lambda _b$ allow testing of
the $V-A$ character of $b$ quarks with unprecedented accuracy
and searches for CP asymmetries in the
decay spectra.  For example,
comparison of
\begin{equation}
\Lambda _b \to l^- (p+X)_{\rm no\; charm}
\; \; \; \textrm{vs.} \; \; \;
\bar \Lambda _b \to l^+(\bar p+X)_{\rm no\; charm},
\end{equation}
might reveal CP violation from new physics.
In final states with at least three particles
($\Lambda _b \to ABC$), one can also form  $T$-odd correlations such as
\begin{equation}
C_T \equiv \langle \vec \sigma _{\Lambda _b} \cdot
(\vec p_A \times \vec p_B)\rangle
\end{equation}
with $\vec p_A$, $\vec p_B$ denoting the momenta of $A$ and $B$, respectively,
and $\vec \sigma _{\Lambda _b}$ the $\Lambda _b$ polarization.
A nonzero value of $C_T$
can be due either to $T$ violation or to final-state interactions.
Measurement of  $\bar C_T$ in the CP-conjugate process resolves the
ambiguity. If $C_T \neq \bar C_T $, one has a signature of
direct CP violation.
Since these effects are typically quite suppressed in the
Standard Model, such studies represent largely a search for
new physics. They can be performed in nonleptonic modes
\begin{equation}
\Lambda ^0_b \to \Lambda ^+_c \pi ^- \pi ^0, p \pi ^- \pi ^0,
\Lambda K^+ \pi ^-
\end{equation}
as well as in semileptonic channels containing a
$\tau$ lepton, since the effect is proportional to the
lepton mass \cite{BENSON}.

\section{Summary}

A sample of order $10^9$ $Z$'s will provide important and unique
tools in the search for and constraint of physics beyond the
Standard Model.  The program available with polarized positron
beams in particular provides dramatic improvement in the measurement
precision of the electroweak observables at the $Z$.  This
improvement leads to markedly more powerful constraints
on Standard Model and new physics scenarios. The polarized
$b$-baryon program offers a unique window of exploration for
new right--handed couplings.  With the statistics and $b$-tagging
capabilities available with two polarized beams, running for
several years ($10^{10}$ $Z$'s) could provide a $b$ physics program
rivaling the proposed hadronic experiments in some fundamental
CKM measurements.

Without positron polarization, significant gains can still be
made.  Much of the $b$ physics would suffer only from a decrease
in statistics.  Impact on the $\Lambda_b$ asymmetry measurements
needs to be evaluated.  The improvement in $\Delta\ALR$ is still
significant and useful.  The most damaging aspect could be
the loss of the $\mw$
determination from threshold running, for which it is unclear that
a 5--6 MeV determination would be realistic without positron
polarization.  This impact still needs study.

\emptyheads

\end{document}